\documentclass{aa}  

\usepackage{graphicx}
\usepackage{txfonts}

\usepackage{lscape}
\usepackage{amsfonts,amsmath,natbib,xcolor,subfigure,gensymb,longtable,ulem}
\usepackage{hyperref}
\hypersetup{breaklinks,colorlinks,citecolor=blue,linkcolor=blue,urlcolor=red}

\def\swift{{\it Swift}}

\begin{document} 

\title{Radio data challenge the broadband modelling of GRB\,160131A afterglow}

\author{M.~Marongiu\thanks{marco.marongiu@unife.it}\inst{1,2,3}
    \and C.~Guidorzi\inst{2,4,5}
    \and G.~Stratta\inst{6}
    \and A.~Gomboc\inst{7}
    \and N.~Jordana-Mitjans\inst{8}
    \and S.~Dichiara\inst{9,10,11}
    \and S.~Kobayashi\inst{12}
    \and D.~Kopa{\v c}\inst{13}
    \and{C.~G.~Mundell}\inst{8}
       }

\institute{INAF -- Osservatorio Astronomico di Cagliari - via della Scienza 5 - I-09047 Selargius, Italy
    \and Department of Physics and Earth Science, University of Ferrara, via Saragat 1, I--44122, Ferrara, Italy
    \and ICRANet, Piazzale della Repubblica 10, I--65122, Pescara, Italy
    \and INFN -- Sezione di Ferrara, via Saragat 1, I--44122, Ferrara, Italy
    \and INAF -- Osservatorio di Astrofisica e Scienza dello Spazio di Bologna, Via Piero Gobetti 101, I-40129 Bologna, Italy
    \and INAF -- Istituto di Astrofisica e Planetologia Spaziali, via Fosso del Cavaliere 100, I-00133 Rome, Italy
    \and Center for Astrophysics and Cosmology, University of Nova Gorica, Vipavska 13, 5000 Nova Gorica, Slovenia
    \and Department of Physics, University of Bath, Claverton Down, Bath, BA2 7AY
    \and Department of Astronomy, University of Maryland, College Park, MD 20742–4111, USA
    \and Astrophysics Science Division, NASA Goddard Space Flight Center, 8800 Greenbelt Rd, Greenbelt, MD 20771, USA
    \and Department of Astronomy and Astrophysics, The Pennsylvania State University, 525 Davey Lab, University Park, PA 16802, USA
    \and Astrophysics Research Institute, Liverpool John Moores University, IC2, Liverpool Science Park, 146 Brownlow Hill, Liverpool L3 5RF, UK
    \and Faculty of Mathematics and Physics, University of Ljubljana, Jadranska 19, Ljubljana SI-1000, Slovenia
    }

\date{Received January 20, 2021; accepted -}

\abstract
{Gamma--ray burst (GRB) afterglows originate from the interaction between the relativistic ejecta and the surrounding medium. Consequently, their properties depend on several aspects: radiation mechanisms, relativistic shock micro-physics, circumburst environment, and the structure and geometry of the relativistic jet. While the standard afterglow model accounts for the overall spectral and temporal evolution for a number of GRBs, its validity limits emerge when the data set is particularly rich and constraining, especially in the radio band.}
{We aimed to model the afterglow of the long GRB\,160131A (redshift $z=0.972$), for which we collected a rich, broadband, and accurate data set, spanning from $6\times10^{8}$~Hz to $7\times10^{17}$~Hz in frequency, and from 330~s to 160~days post burst in time.}
{We modelled the spectral and temporal evolution of this GRB afterglow through two approaches: (1) the adoption of empirical functions to model optical/X-rays data set, later assessing their compatibility with the radio domain; (2) the inclusion of the entire multi-frequency data set simultaneously through the Python package named {\sc sAGa} (Software for AfterGlow Analysis), to come up with an exhaustive and self-consistent description of the micro-physics, geometry, and dynamics of the afterglow.
}
{From deep broadband analysis (from radio to X-ray frequencies) of the afterglow light curves, GRB\,160131A outflow shows evidence of jetted emission.
Moreover, we observe dust extinction in the optical spectra, and energy injection in the optical/X-ray data.
Finally, radio spectra are characterised by several peaks, that could be due to either interstellar scintillation (ISS) effects or a multi-component structure.}
{The inclusion of radio data in the broadband set of GRB\,160131A makes a self-consistent modelling hardly attainable within the standard model of GRB afterglows.}


\maketitle

%

\section{Introduction}
\label{par:intro}

Gamma-ray bursts (GRBs) consist in short and intense pulses of gamma-ray radiation, originating from either core collapsing massive stars (e.g. \citealt{Woosley06}) or binary neutron star (BNS) mergers (e.g. \citealt{Abbott17}).
These sources can launch relativistic jets with opening angles of a few degrees.
According to the standard model (e.g. \citealp{Rees92,Meszaros97,Panaitescu98}), GRB afterglow emission takes place when the outflow from the GRB central engine impacts on the circumburst medium (CBM), resulting mainly in synchrotron radiation (for a review see e.g. \citealt{Piran04_rev,Meszaros06,Gao13}). The long-lasting afterglow emission can be detected days to months after the burst, and spans a broad range of electromagnetic spectrum (from gamma-ray to radio domain). It originates in two shock regions: a forward shock (FS) that propagates in the CBM (e.g. \citealt{GranotSari02}, hereafter GS02), and a reverse shock (RS) that propagates back into the flow itself and radiates at lower frequencies (e.g. \citealt{Meszaros99,Kobayashi00b,Kobayashi07a,Gao15}).

GRB afterglows encode a wealth of information on (1) the radiation mechanism, in particular the possible presence of large-scale magnetic fields ploughing the ejecta, which is still one of the main open issues in the field (e.g. \citealt{Jordana20}); (2) relativistic shock micro-physics; (3) energetics; (4) jet geometry.
All these issues can be addressed effectively and uniquely through observations at lower frequencies, especially in the radio band.
Observations of radio afterglows are key to diagnose the GRB physics (e.g. \citealt{Mundell07}), especially for the understanding of the RS component, which links directly to the nature of the outflow and, consequently, to the progenitor itself (e.g. \citealt{Kopac15}).
On the other hand, the detection of radio afterglows has proven challenging with current radio telescopes (e.g. \citealt{ChandraFrail12}) -- especially in single-dish mode \citep{Marongiu20b} -- mainly because of their mJy and sub-mJy nature.
To date, radio/mm followup campaigns in interferometric mode improved the observational coverage of the lower part of the emission spectrum (e.g. \citealt{Laskar13,Laskar15,Laskar18,Laskar19a}) through increasingly sensitive facilities -- such as the upgraded Giant Metre-wave Radio Telescope (GMRT, \citealt{Swarup90,Kapahi95,Gupta17})\footnote{\url{http://www.gmrt.ncra.tifr.res.in/}}, the Karl G. Jansky Very Large Array (VLA, \citealt{Thompson80})\footnote{\url{https://science.nrao.edu/facilities/vla}}, the Arcminute Microkelvin Imager Large Array (AMI-LA, \citealt{Zwart08})\footnote{\url{https://www.astro.phy.cam.ac.uk/research/research-projects/AMI}}, and the NOrthern Extended Millimeter Array (NOEMA, \citealt{Chenu16})\footnote{\url{http://iram-institute.org/EN/noema-project.php}}.

In addition to synchrotron radiation, the emission of GRB afterglows can be modelled via other radiation mechanisms (e.g. inverse Compton at high-energies; \citealt{MAGIC19b,Zhang20c}). 
Additionally, the jet collimation, energy injection, dust extinction and radio interstellar scintillation can further shape the observed afterglow.
Well-sampled GRB afterglows in time and frequency domains are usually modelled with fine-tuning to the standard model, from radio to gamma-ray frequencies (e.g. \citealt{Frail06,Laskar14,Perley14}), but especially ranging between optical and gamma-ray domain (e.g. \citealt{Lazzati02,Heyl03,Jakobsson05,Gendre06,CastroTirado07,Starling09,Zauderer13,Vanderhorst15}).
Sometimes additions for a fine-tuning of the model lack of broadband consistency check, suggesting that broadband available data (from radio to gamma-ray frequencies) could not be completely explained within the standard model (e.g. \citealt{Klotz08,Gendre10}); in this context, modelling and simulation of GRB afterglow evolution is a particularly challenging problem (e.g. \citealt{Granot07,Vaneerten18}), especially when radio observations are included in the analysis (e.g. \citealt{Frail00a, Frail00b,Frail03,Corsi05,Gendre10,Resmi12,Horesh15}).
In the radio domain there are other physical components that usually dominate the total emission, such as the RS (e.g. \citealt{Sari99,Kobayashi03a,Laskar13,Cucchiara15,Veres15,Laskar16,Alexander17,Laskar19b}), rebrightenings due to refreshed shocks, and flares caused by central-engine activity (e.g. \citealt{Bjornsson04,Zhang06,Melandri10,Chincarini10,Margutti10b}).

The ongoing technological evolution allowed to develop several computational packages to model GRB afterglows (e.g. \citealt{Rhoads99,Kobayashi99,Daigne00,Kumar03,Cannizzo04,Zhang09b,Vaneerten10a,Wygoda11,Vaneerten12b,DeColle12,Granot12,Laskar13,Leventis13,Rhodes20,Aksulu20,Ryan20,Ayache21}), but to date there is no computational tool that is able to fully describe the complex landscape of the GRB afterglows.

The richness of the data set collected for GRB\,160131A, in both time (from $430$~s to $\sim 163$~d) and frequency (from $6 \times 10^8$ to $7 \times 10^{17}$~Hz), makes it an ideal test bed for the standard GRB afterglow model.

This paper is organised as follows.
Observations are reported in Section~\ref{sec:obs}, and the modelling of broadband data is described in Section~\ref{par:multi_sed}.
After the presentation of our results in Section~\ref{par:results}, we discuss them in Section~\ref{sec:disc}, and finally we give our conclusions in Section~\ref{cap_GRB160131A:concls}.

In this paper we assume $\Lambda$CDM cosmological parameters of $\Omega_m = 0.32$, $\Omega_{\Lambda} = 0.68$, and $H_0 = 67$~km s$^{-1}$ Mpc$^{-1}$ \citep{cosmoPlanck18}.
We adopt the convention $F_{\nu} \propto t^{\alpha} \nu^{\beta}$ as adopted by GS02, where $\alpha$ and $\beta$ indicate the temporal decay index and the spectral index, respectively; we report the uncertainties at a $1\sigma$ confidence level unless stated otherwise.

\section{Observations and Data reduction}
\label{sec:obs}

GRB\,160131A was discovered by the {\it Neil Gehrels Swift Observatory} \citep{Gehrels04} on January $16$ at 08:20:31 UT, 2016 \citep{Page16}.
Discovered at redshift  $z = 0.972$ \citep{Malesani16,DeUgarte16b}, this very long GRB, with $T_{90} = 325\pm 72$~s \citep{Cummings16} has an isotropic-equivalent $E_{\gamma,{\rm iso}} = (8.3\pm0.7)\times 10^{53}$~erg in the $0.02-15$~MeV range \citep{Tsvetkova16}.
Prompt gamma-ray polarimetric measurements in the $100-300$~keV band indicated that GRB\,160131A is possibly highly polarised ($94 \pm 33$\,\%, although the confidence level is $< 3\sigma$, \citealt{Chattopadhyay19}): this suggests that the GRB is due to synchrotron emission within a time-independent, ordered magnetic field \citep{Nakar03,Granot03,Waxman03}, with an initial bulk Lorentz factor of $\Gamma_0 = 460 \pm 50$ and jet half-opening angle of $\theta_j = 3^{+3}_{-1.8}$~degrees, calculated from the jet breaks observed in Swift/XRT X-ray light curves\footnote{\url{http://www.swift.ac.uk/xrt_curves/}} \citep{Sari99c,Frail01}.
This constrain on $\theta_j$ corresponds to a beaming-corrected isotropic energy in the $\gamma$-ray band of $E_{\gamma} = E_{\gamma,iso} (1 - \cos{\theta_j}) = 6.0^{1.8}_{-0.5} \times 10^{51}$~erg, \citep{Chattopadhyay19}.
The study of the inhomogeneities in the optical light curves of GRB afterglows of \citet{Mazaeva18} shows that the early ($\lesssim 0.5$~d) optical afterglow of GRB\,160131A is characterised by a broken power-law with small scale deviations (wiggles), followed a steep decay, suggestive of a jet break at $t_j = 1.2 \pm 0.3$~d.

\subsection{X--ray observations with Swift/XRT}
\label{subsec:XRT}

\swift{}/XRT observed the region of GRB\,160131A in window timing (WT) mode from 60 to 595~s and in photon counting (PC) from 3820~s to 9~d after the BAT trigger and found a bright, uncatalogued X-ray source located at $\alpha=5^{\rm h}12^{\rm m}40.31^{\rm s}$, $\delta=-7^{\circ}02'59.4''$ (J2000), with an uncertainty of $1.4$~arcsec (radius, $90$\% containment)\footnote{\url{https://www.swift.ac.uk/xrt_positions/00672236/}}.
We obtained the observed $0.3$--10~keV light curve from the Leicester University repository\footnote{\url{https://www.swift.ac.uk/xrt_curves/00672236/}}, based on the time-average spectrum with a count-to-flux conversion factor of $3.55\times10^{-11}$~erg\,cm$^{-2}$~count$^{-1}$ (observed flux), and binned it up by imposing a minimum significance of $3\sigma$ per bin. The lack of evidence for a significant spectral evolution in the PC data justifies the adoption of a constant count-to-flux ratio.
We extracted the time-averaged spectrum ($4.0$ -- $131$~ks) using the Leicester web interface \citep{Evans09} based on {\sc heasoft} (v6.22).
We then grouped energy channels with the {\sc grppha} tool so as to ensure at least 20 counts per bin. 
The spectrum is well modelled by a highly absorbed power law using the {\sc xspec} model {\sc TBabs * zTBabs * powerlaw}, where the Galactic term was fixed to $N_{H,gal}=1.15\times10^{21}$~cm$^{-2}$ corresponding to the GRB direction \citep{Willingale13}\footnote{Derived using \url{https://www.swift.ac.uk/analysis/nhtot/}, taking the value $N_{H,tot}$.} and the redshift was fixed to $z=0.972$.
The best-fit photon index was $\Gamma_X = 2.04 \pm 0.06$ and the source-frame (intrinsic) hydrogen column $N_{H,int} = (5.0\pm0.1) \times 10^{21}$ cm$^{-2}$ ($\chi^2/{\rm dof}=171/178$).
We determined the instantaneous reference epoch for the XRT spectrum as follows: we preliminarily noticed that the light curve in the interested time interval can be modelled with a simple power-law $\propto t^{-\alpha_x}$ with $\alpha_x\simeq 1.2$.
Given that the observational coverage within this time window is reasonably uniform, the reference time $t_x$ was found by demanding that the instantaneous flux at $t_x$ equals the observed time-averaged one between $t_1=4$ and $t_2=131$~ks:
\begin{equation}
t_x\ =\ \Big[ \frac{1}{\alpha_x-1}\ \Big(\frac{t_1^{1-\alpha_x}-t_2^{1-\alpha_x}}{t_2-t_1}  \Big) \Big]^{-1/\alpha_x}\ =\ 33\ {\rm ks}\;.
\end{equation}
We followed a similar line of reasoning to find the reference energy for the $0.3$--$10$~keV average flux density light curve: using the two energy boundaries, $E_1=0.3$ and $E_2=10$~keV and the power-law index $\Gamma_X=2.04$, we calculated the energy $E_x$ at which the flux density equals the corresponding average flux density, finding $E_x=2.75$~keV ($6.65 \times 10^{17}$~Hz). This value is hereafter used as the reference energy for the average flux density light curve.

\longtab{
\small
\begin{longtable}{c|c|c|c}
\caption{Swift XRT Observations of GRB\,160131A.
$t-t_0$ indicates the epoch of observation, where $t_0$ is the GRB explosion date ($57418.3476$~MJD).
Flux densities refers to an energy of $2.75$~keV ($6.65 \times 10^{17}$~Hz).
\label{Tab:tot_data_x_online}
} \\
\hline
\hline
{$t - t_0$} & Flux density & Uncertainty & Frequency \\
{(days)} & (mJy) & (mJy) & (Hz) \\
\hline
\endfirsthead
\caption{Continued.} \\
\hline
{$t - t_0$} & Flux density & Uncertainty & Frequency \\
{(days)} & (mJy) & (mJy) & (Hz) \\
\hline
\endhead
\hline
\endfoot
\hline
\endlastfoot
3.83$\times 10^{-3}$ & 3.78$\times 10^{-2}$ & 5.63$\times 10^{-3}$ & 6.65$\times 10^{17}$  \\
3.85$\times 10^{-3}$ & 4.40$\times 10^{-2}$ & 6.55$\times 10^{-3}$ & 6.65$\times 10^{17}$  \\
3.87$\times 10^{-3}$ & 4.39$\times 10^{-2}$ & 6.69$\times 10^{-3}$ & 6.65$\times 10^{17}$  \\
3.89$\times 10^{-3}$ & 3.98$\times 10^{-2}$ & 6.34$\times 10^{-3}$ & 6.65$\times 10^{17}$  \\
3.91$\times 10^{-3}$ & 3.77$\times 10^{-2}$ & 5.62$\times 10^{-3}$ & 6.65$\times 10^{17}$  \\
3.94$\times 10^{-3}$ & 3.36$\times 10^{-2}$ & 5.35$\times 10^{-3}$ & 6.65$\times 10^{17}$  \\
3.96$\times 10^{-3}$ & 4.20$\times 10^{-2}$ & 6.26$\times 10^{-3}$ & 6.65$\times 10^{17}$  \\
3.98$\times 10^{-3}$ & 3.75$\times 10^{-2}$ & 5.59$\times 10^{-3}$ & 6.65$\times 10^{17}$  \\
\end{longtable}
}
We list the full table of X-ray data in the Appendix (Table~\ref{Tab:tot_data_x_online}).

\subsection{UVOIR observations}
\label{subsec:UVOT}

The {\it Swift} UltraViolet and Optical Telescope (UVOT; \citealt{Roming05}) observed the region of GRB\,160131A from $78$~s to $\sim 6$~d and found a source located at $\alpha=5^{\rm h}12^{\rm m}40.34^{\rm s}$, $\delta=-7^{\circ}02'59.1''$, with an uncertainty of $0.61$~arcsec (radius, $90 \%$ containment).
This position is $7.5$~arcsec from the center of the XRT error circle.
We analysed the UV band data using {\sc heasoft} (v. 6.22)\footnote{\url{https://heasarc.gsfc.nasa.gov/lheasoft/download.html}}, the dedicated software package for optical/X-ray astronomical spectral, timing, and imaging data analysis.
In particular, data were analysed for the six filters, {\it v, b, u, w1, w2} and {\it m2}, for which we extracted aperture photometry using a source region radius of $5''$, following the prescriptions by \citet{Brown09,Breeveld11}. Flux measurement having S/N $<3\sigma$ were replaced with the corresponding $3\sigma$ upper limits.

In the optical and near-infrared bands, GRB\,160131A was first observed in the Pan-STARRS {\it g', r', i', z', Y} filters with the 2-m Faulkes Telescope North (FTN; \citealt{Guidorzi16}) soon followed by the 2-m Faulkes Telescope South (FTS) and a 1-m unit in Siding Springs, all of which are operated by Las Cumbres Observatory Global Network (LCOGT; \citealt{LCO13}), starting from $\sim 74$ minutes to $6.6$~days (under proposal ARI2015A-001, PI: Kobayashi). We used the Spectral Camera (FOV $10.5' \times 10.5'$, resolution of $0.304''$/pixel) for the 2-m units, and the Sinistro Camera (FOV $26.5'\times 26.5'$, resolution of $0.467''$/pixel) for the 1-m unit. Individual exposures vary from a minimum of 30~s up to 120~s. Bias and flat-field corrections were applied using the specific LCOGT pipeline \citep{LCO13}. From February 3 to 6, 2016, we also used the 2-m Liverpool Telescope (LT; \citealt{Steele04,Guidorzi06}) at the Observatorio del Roque de Los~Muchachos (Canary Islands) and observed with the IO:O Camera (FOV $10'\times10'$, with a $2\times2$ binning, which corresponds to a resolution of $0.30''$/pixel) within the AB $r'$ and $i'$ filters. Bias and flat-field corrections were automatically applied using the LT pipeline.

The afterglow magnitudes were obtained through PSF-fitting photometry, after calibrating the zero-points with a dozen nearby Pan-STARRS catalogue stars\footnote{\url{https://panstarrs.stsci.edu/}} using the mean PSF AB magnitudes for the corresponding filters \citep{Tonry12}. Filter-dependent systematic errors, due to the zero-point scatter of the calibrating stars, were added to the statistical uncertainties of magnitudes, with the following average values in magnitude units: $0.02$, $0.01$, $0.04$, $0.02$, and $0.02$ for the {\it g', r', i', z'}, and {\it Y} filters, respectively.

The obtained calibrated magnitudes were corrected for the Galactic extinction along the line-of-sight of $\mathrm{E_{B-V}} = 0.09\,\mathrm{mag}$\footnote{We assumed the following extinctions in mag units: $A_{m2} = 0.90$, $A_{w2} = 0.79$, $A_{w1} = 0.64$, $A_{u} = 0.47$, $A_{b} = 0.39$, $A_{v} = 0.30$, $A_{g'} = 0.36$, $A_{r'} = 0.25$, $A_{i'} = 0.18$~mag, $A_{z'} = 0.14$, and $A_{Y}=0.12$.} \citep{Schlafly11}, and converted to flux densities \citep{Fukugita96}.

The full table of UVOIR data is available in the Appendix (Table~\ref{Tab:tot_data_optuv_online}).
\longtab{
\small
\begin{longtable}{c|c|c|c|c}
\caption{\label{Tab:tot_data_optuv_online} Optical and Swift UVOT observations of GRB\,160131A.
$t-t_0$ indicates the epoch of observation, where $t_0$ is the GRB explosion date ($57418.3476$~MJD).
All upper limits are at $3\sigma$.
} \\
\hline
\hline
{$t - t_0$} & Flux density & Uncertainty & Filter & Frequency \\
{(days)} & (mJy) & (mJy) & & (Hz) \\
\hline
\endfirsthead
\caption{Continued.} \\
\hline
{$t - t_0$} & Flux density & Uncertainty & Filter & Frequency \\
{(days)} & (mJy) & (mJy) & & (Hz) \\
\hline
\endhead
\hline
\endfoot
\hline
\endlastfoot
  0.093 &   1.191 &  0.054 & g'        &  6.47$\times 10^{14}$    \\
  0.094 &   1.180 &  0.053 & g'        &  6.47$\times 10^{14}$    \\
  0.094 &   1.170 &  0.053 & g'        &  6.47$\times 10^{14}$    \\
  0.095 &   1.138 &  0.051 & g'        &  6.47$\times 10^{14}$    \\
  0.103 &   1.057 &  0.029 & g'        &  6.47$\times 10^{14}$    \\
  0.104 &   1.047 &  0.029 & g'        &  6.47$\times 10^{14}$    \\
  0.105 &   1.028 &  0.028 & g'        &  6.47$\times 10^{14}$    \\
  0.106 &   0.991 &  0.036 & g'        &  6.47$\times 10^{14}$    \\
\end{longtable}
}

\subsection{Radio/mm observations}
\label{subsec:radio_vla}

VLA followup observations were carried out from February 1 to May 27, 2016, from $\sim 1$ to $\sim 117$~d post explosion \citep{Laskar16b,Laskar16c} under large Proposal VLA/15A-235 (PI: Berger)\footnote{\url{https://science.nrao.edu/science/science-program/large-proposals}}.
Data were taken in five spectral windows at C-band (with baseband central frequency of $6$~GHz), X-band ($10$~GHz), Ku-band ($15$~GHz), K-band ($22.25$~GHz), and Ka-band ($33.25$~GHz), with a nominal bandwidth of $\sim 0.4$~GHz.
3C48 and J0522+0113 were used as flux/bandpass and phase/amplitude calibrators, respectively.
To eventually observe multi-component behaviour in radio data, we split each radio band in eight parts, from $4.6$ to $37.4$~GHz, resulting in $\sim 300$ VLA flux densities.
The Common Astronomy Software Application ({\sc casa}, v. 5.1.1-4, \citealt{CASA})\footnote{\url{https://casa.nrao.edu/}} was used to calibrate, flag and image the data.
Images were formed from the visibility data using the CLEAN algorithm \citep{Hogbom74}.
The image size was set to ($240 \times 240$) pixels, the pixel size was determined as $1/5$ of the nominal beam width and the images were cleaned using natural weighting.

We also considered six observations (mainly upper-limits) from GMRT \citep{Chandra16b,Chandra16c}, AMI-LA \citep{Mooley16}, and NOEMA \citep{DeUgarte16c}.
The upper limits on the flux densities were calculated at a $3 \sigma$ confidence level.

All the $300$ radio/mm flux densities are reported in the Appendix (Table~\ref{Tab:tot_data_radiomm_online}).
%
\longtab{
\small
\begin{longtable}{c|c|c|c|c|c}
\caption{\label{Tab:tot_data_radiomm_online} Radio and millimetre observations of GRB\,160131A.
$t-t_0$ indicates the epoch of observation, where $t_0$ is the GRB explosion date ($57418.3476$~MJD).
All upper limits are at $3\sigma$.
} \\
\hline
\hline
{$t - t_0$} & Flux density$^a$ & Uncertainty & Frequency & Instrument & Reference \\
{(days)} & (mJy) & (mJy) & (Hz) & & (GCN, ref.) \\
\hline
\endfirsthead
\caption{Continued.} \\
\hline
{$t - t_0$} & Flux density$^a$ & Uncertainty & Frequency & Instrument & Reference \\
{(days)} & (mJy) & (mJy) & (Hz) & & (GCN, ref.) \\
\hline
\endhead
\hline
\endfoot
\hline
\endlastfoot
  1.360 &   1.940   &  0.070 &  9.25$\times 10^{10}$ &  NOEMA & 18976, \citep{DeUgarte16c} \\
  9.262 &   < 0.570 &  -     &  1.28$\times 10^{9}$  &  GMRT & 19009, \citep{Chandra16d} \\
  9.262 &   < 0.382 &  -     &  6.10$\times 10^{8}$  &  GMRT & 19009, \citep{Chandra16d} \\
 10.430 &   < 0.105 &  -     &  1.42$\times 10^{9}$  &  GMRT & 19010, \citep{Chandra16c} \\
 24.5   &   < 0.150 &  -     &  1.5$\times 10^{10}$  &  AMI-LA  & 19206, \citep{Mooley16} \\
 34.5   &   < 0.210 &  -     &  1.5$\times 10^{10}$  &  AMI-LA  & 19206, \citep{Mooley16} \\
  0.825 &   < 0.120 &  -     &  4.62$\times 10^{9}$  &  VLA & This work \\
  0.825 &   0.090   &  0.023 &  4.87$\times 10^{9}$  &  VLA & This work \\
\end{longtable}
}

\section{Data modelling}
\label{par:multi_sed}

We analyse the broadband observations in the context of synchrotron emission arising from relativistic shocks, following the standard afterglow model described by GS02.
The observed SED of each synchrotron component is described by three break frequencies (the characteristic frequency, $\nu_m$, the cooling frequency, $\nu_c$, and the self-absorption frequency, $\nu_{sa}$), and the flux density normalisation, $F_{\nu,m}$.
Depending on the order of $\nu_m$ and $\nu_c$, the synchrotron spectrum falls into two broad
categories: fast-cooling regime ($\nu_m > \nu_c$), where all the less energetic electrons cool rapidly, and slow-cooling ($\nu_m < \nu_c$) regime, where only the most energetic electrons cool rapidly (e.g., GS02, \citealt{Sari98,Gao13}).
The prompt phase of GRBs is expected to be in the fast-cooling regime \citep{Piran99}, whereas the transition to the slow-cooling regime is expected to take place during the early stages of the afterglow (\citealt{Meszaros97,Waxman97}, GS02).
During the afterglow phase, $\nu_{sa}$ is usually the smallest among the three frequencies.
When $\nu_{sa} > \nu_c$, the electron energy distribution may be significantly modified, resulting in inaccurate analytical models \citep{Gao13}.

The richness of our broadband data set allows us the modelling strategy combining two approaches to model the GRB afterglow emission: empirical approach (Sect.~\ref{par:emp_app}), and physical approach (Sect.~\ref{par:phys_app}).
In the empirical approach, we modelled SEDs (for each observing epoch) and light curves (for each observing frequency) with simple empirical functions; later, we analysed the best-fit results comparing them with the standard afterglow model (described by GS02), and the jet emission (e.g. \citealt{Panaitescu98,Rhoads99,Sari99b,Panaitescu02,Sari06,Granot07}).
This approach allows us to constrain the behaviour of the GRB afterglow emission -- in terms of the main observational features (breaking frequencies and possible jet break time) and the kind of CBM (ISM-like vs. wind-like) -- and then to apply the physical approach, where we modelled the data set of the GRB afterglow emission through a sophisticated modelling code -- fully self-consistent -- developed in Python, called {\sc sAGa} (Software for AfterGlow Analysis), briefly described in Section~\ref{app_saga}.

\subsection{Empirical approach}
\label{par:emp_app}

We start by adopting empirical functions for both SEDs and light curves in optical/X-rays domain (Sect.~\ref{par:breaks}).
The analysis of the radio data set (Sect.~\ref{par:radio}) better constrains the information inferred from the optical/X-ray analysis.
We assumed three kind of empirical functions, reported here for completeness:
\begin{itemize}
\item Single power-law (hereafter SPL):
\begin{equation}
F_{x} = F_0 \left(\frac{x}{x_0}\right)^{\gamma}
\label{eq:pl0}
\end{equation}
where $F_0$ is the flux density at the reference parameter $x$ ($x \equiv \nu$ with $x_0 \equiv \nu_0 = 1$~GHz for SEDs, and $x \equiv t$ with $x_0 \equiv t_0 = 1$~d for the light curves).
The slope index is $\gamma$, which corresponds to the spectral index $\beta$ for SEDs and the decay index $\alpha$ for the light curves.
\item Broken power-law (BPL):
\begin{equation}
F_{x,1b} = \begin{cases}
    F_b \left[\frac{1}{2} \left(\frac{x}{x_{b,1}}\right)^{-s\gamma_1} + \frac{1}{2} \left(\frac{x}{x_{b,1}}\right)^{-s\gamma_2}\right]^{-1/s} & \gamma_1 \geq \gamma_2 \\
    F_b \left[\frac{1}{2} \left(\frac{x}{x_{b,1}}\right)^{s\gamma_1} + \frac{1}{2} \left(\frac{x}{x_{b,1}}\right)^{s\gamma_2}\right]^{1/s} & \gamma_1 < \gamma_2 \\
    \end{cases}
\label{eq:pl1}
\end{equation}
where $F_{b,1}$ is the flux density at the reference break parameter $x_{b,1}$, corresponding to the break frequency $\nu_b$ for SEDs and the break time $t_b$ for the light curves, $s$ is the sharpness factor (we fixed $s = 5$), $\gamma_1$ and $\gamma_2$ are the slope indices before and after $x_b$, corresponding to the spectral index $\beta$ for SEDs and the decay index $\alpha$ for the light curves.
\item Double broken power-law (DBPL):
\begin{equation}
\begin{split}
F_{x,2b} = \begin{cases}
    F_{x,1b} \times \left[1 + \left(\frac{x}{x_{b,2}}\right)^{w(\gamma_2 - \gamma_3)}\right]^{-1/w} & \gamma_2 \geq \gamma_3 \\
    F_{x,1b} \times \left[1 + \left(\frac{x}{x_{b,2}}\right)^{w(\gamma_3 - \gamma_2)}\right]^{1/w} & \gamma_3 < \gamma_2 \\
    \end{cases}
\label{eq:pl2}
\end{split}
\end{equation}
where $s$ and $w$ are the sharpness factors (we fixed $s = w = 5$); $\gamma_1$, $\gamma_2$ and $\gamma_3$ are the slope indices among the break parameters $x_{b,1}$ and $x_{b,2}$, corresponding to the spectral index $\beta$ for SEDs and the decay index $\alpha$ for the light curves.
\end{itemize}

\subsection{Physical approach}
\label{par:phys_app}

Once we estimated the main observational features of the GRB afterglow, we modelled the data through {\sc sAGa}.
Built adopting a Bayesian statistics (e.g. \citealt{Sharma17,Marquette18}), our code adds up to other pre-existing broadband fitting tools in the literature (e.g. \citealt{Kobayashi99,Daigne00,Cannizzo04,Zhang09b,Vaneerten10a,Wygoda11,DeColle12,Laskar13,Leventis13,Rhodes20,Aksulu20,Ryan20,Ayache21}) and provides an independent check, emphasising the broadband study of GRB afterglows over the last two decades.
{\sc sAGa} performs simultaneously a broadband data analysis -- from radio to gamma-rays frequencies -- in a single iteration through a new approach that consists in the manipulation of all the data both at each observing epoch $t_{obs}$ and observing frequency $\nu_{obs}$, considering different radiation processes and other aspects, briefly described in Sect.~\ref{app_saga}.
This approach allows us to estimate in one fell swoop the micro-physics parameters of the afterglow and other physical information (the complete parameter space is listed in Table~\ref{tab:parameter_saga}).

{\sc sAGa} has been successfully tested on the broadband data of the afterglows of GRB\,120521C, GRB\,090423, and GRB\,050904, where the results obtained with {\sc sAGa} are consistent with those reported in the literature (especially in \citealp{Laskar14} -- hereafter L14 -- who make use of a similar approach for the characterisation of the GRB afterglow) within $\lesssim 2\sigma$.

We report a more detailed description -- test phase included -- of this Python package in a specific technical note \citep{Marongiu21a}.

\subsubsection{{\sc sAGa}: a physical/analytical approach for broadband modelling of GRB afterglows}
\label{app_saga}

The Python Package {\sc sAGa} models the data using the smoothly connected power-law synchrotron spectra for the FS (GS02, and the references therein), computing the break frequencies and normalisations as a function of the shock micro-physics parameters: the kinetic energy of the explosion ($E_{K,{\rm iso}}$), the CBM density ($n_0$ for ISM-like CBM; the normalised mass-loss rate $A_*$ for wind-like CBM), the power-law index of the electron energy distribution ($p$), the fractions of the blastwave energy delivered to relativistic electrons ($\epsilon_e$) and magnetic fields ($\epsilon_B$).
In addition to this standard model, {\sc sAGa} considers the inverse Compton (IC) radiation process by computing the Compton y-parameter from the FS parameters, and hence scaling the spectral break frequencies and flux densities of the synchrotron spectrum by the appropriate powers of $1 + y$ (\citealt{Sari01,Zhang07}, L14, GS02); if $y < 1$, the IC regime can be neglected, otherwise a high-energy component (of the order of $10$~MeV) appears in the spectrum and the cooling timescale is shortened by a factor $y$ \citep{Sari01,Piran04_rev}.
\begin{table*}
\caption{Free parameter space available, with relative range of definition (for further details, see Sect.~\ref{par:saga_results}), for {\sc sAGa} analysis.}
\label{tab:parameter_saga}
\small
\begin{tabular}{l | cl | l}
\hline
\hline
Parameter & Unit & Description & Parameter space \\
\hline
$p$                  & -                              & Power-law index of the electron energy distribution                       & $1.5$ -- $3.5$                      \\
$\epsilon_e$         & -                              & Fraction of the blastwave energy delivered to relativistic electrons      & $0$ -- $1/3$                        \\
$\epsilon_B$         & -                              & Fraction of the blastwave energy delivered to magnetic fields             & $0$ -- $1/3$                        \\
$E_{K,{\rm iso},52}$ & $10^{52}$~erg                  & Kinetic energy of the explosion (in units of $10^{52}$~erg)               & $10^{-2}$ -- $10^3$                 \\
$n_0$                & cm$^{-3}$                      & Density for ISM-like CBM                                                  & $10^{-3}$ -- $10^2$                 \\
$A_*$                & $5 \times 10^{11}$~g cm$^{-1}$ & Parameter connected with the wind-like density CBM                        & $10^{-3}$ -- $10^2$                 \\
$A_V$                & mag                            & Extinction in the host galaxy                                             & $0$ -- $10$                         \\
$t_j$                & d                              & Jet break time                                                            & According to the case               \\
$t_{ei,1}$           & d                              & Start time of the first injection                                     & According to the case               \\
$t_{ei,2}$           & d                              & Start time of the second injection                                    & According to the case               \\
$m$                  & -                              & Injection index                                                           & $0$ -- $3$ (ISM), $0$ -- $1$ (wind) \\
$m_2$                & -                              & Injection index (in case of two bumps during the energy injection regime) & $0$ -- $3$ (ISM), $0$ -- $1$ (wind) \\
\hline
\end{tabular}
\end{table*}

Moreover, {\sc sAGa} assumes:
\begin{itemize}
\item \textit{the uniform jet regime} (e.g. \citealt{Granot07,Zhang19b})\footnote{This jet regime is simpler than structured jet model, that assumes an angular distribution in energy and Lorentz factor, based on special relativistic hydrodynamics (e.g. \citealt{DeColle12,Granot18,Coughlin20_jet}), and other more complex regimes (e.g. \citealp{Huang04,Peng05,Wu05,Granot18}).}, based on purely geometrical or dynamical effects, that assumes a simplified conical jet blastwave, with a half opening angle $\theta_j$ and blastwave Lorentz factor $\Gamma$, where only the emission inside the $1/\Gamma$ cone is detectable due to relativistic beaming.
During the deceleration phase, $\Gamma$ decreases gradually until $1/\Gamma > \theta_j$ -- for an observer in the line-of-sight of the jet -- followed by an achromatic break in the light curve, at the jet break time $t_j$, measured both for ISM-like and wind-like CBM (see \citealp{Waxman97,Rhoads99,Sari99b,ChevalierLi00,Wang18}).
The light curve steepening can arise from two effects: the pure edge effect (e.g. \citealt{Panaitescu98,Granot07}) and the sideways expansion effect (e.g. \citealt{Rhoads99,Sari99b}).
In the pure edge effect, the blastwave dynamics does not change during the jet break transition, and hence the deceleration rate/dynamics of the jet (such as the breaking frequencies) is the same with the spherical blastwave.
On the other hand, the sideways expansion effect of a conical jet, implies that the conical jet exponentially decelerates; this feature translates in the change of the evolution of both the spectral break frequencies and flux densities at $t_j$.
{\sc sAGa} considers the uniform jet regime, based on the selection by the user (before launching the analysis) between the pure edge effect and the sideways expansion, through the modification of the evolution of the spectral break frequencies and flux densities at $t_{j}$ (\citealt{Sari99b,Panaitescu02,Sari06,Granot07}, GS02), smoothing over the transition with a fixed smoothing parameter ($s = 5$, \citealt{Granot01}).
\item \textit{the effect of non-relativistic/Newtonian (NR) ejecta} (e.g. \citealt{Wijers97,Zhang19b}), reached at the transition times $t_{\rm NR}$ (\citealt{Waxman97} for ISM-like CBM, and \citealt{ChevalierLi00} for wind-like CBM) when the relativistic blastwave, decelerated by the interaction with the CBM, is characterised by a bulk Lorentz factor $\gamma < \sqrt{2}$. Usually, this regime takes place in timescales of months/years (e.g. \citealt{Livio00,Zhang09b}), when the electrons should be in the slow cooling scenario ($\nu_m < \nu_c$).
{\sc sAGa} accounts for the NR regime modifying the evolution of the spectral break frequencies and flux densities at $t_{NR}$ (\citealt{Frail00a,Vaneerten10b,Leventis12}), smoothing over the transition with a fixed smoothing parameter ($s = 5$, \citealt{Granot01}).
\item \textit{the energy injection into the blastwave shock} (e.g. \citealt{Zhang02c,Granot06,Gao13b}), observed as one (or more) plateau/flattening in the light curves of GRB afterglows (e.g. \citealt{Nousek06,Liang07,Margutti10,Hascoet12}).
In general, the blastwave is fed by a long-lasting Poynting-flux-dominated wind, defined by the power-law decay $L(t) = L_0 \left(\frac{t}{t_0}\right)^{-q}$, where $t$ is the central engine time (corresponding to the observer time of GRB afterglow), $L_0$ is the luminosity at the reference time $t_0$, and $q \geq 0$\footnote{The same approach sometimes is based on $L(t) = L_0 (t/t_0)^q$ and $q \leq 0$ (e.g. \citealt{Misra07,Marshall11,Vaneerten14b,Laskar15}).}; this corresponds to the temporal evolution of the blastwave energy $E \propto t^{1 - q} = t^m$, where $m = 1 - q$ is the ``injection index''.
In the absence of energy injection, the standard hydrodynamic evolution requires that $m = 0$, $s = 1$ or $q = 1$ in the above expressions (e.g. \citealt{Gao13}).
{\sc sAGa} accounts for energy injection continuously adjusting the content -- in the time interval where this phenomenon takes place (between $t_{ei,i}$ and $t_{ei,f}$) -- of the kinetic energy in the standard afterglow regime ($E_{k,iso}(t)$, e.g. GS02) according to broken power-law functions described in \citealt{Laskar15}.
\item \textit{the interstellar scintillation effect} (ISS), caused by inhomogeneities in the electron density distribution in the Milky Way along the GRB line of sight, and observable through variations in measured flux density of the source at low frequencies ($\lesssim 10$~GHz) of radio domain \citep{Rickett90,Goodman97,Walker98,Frail97,Frail00a,Goodman06,Granot14,Misra19}; {\sc sAGa} accounts for ISS effect following the prescription described in \citep{Goodman06} and L14, to compute the modulation index $m_{scint}$ -- defined as the rms of the fractional flux density variation -- and the model-predicted flux density $F_{model}$ in the expected ISS contribution;
\item \textit{the dust extinction in the host galaxy along the sightline}, adopting the extinction curves of \citet{Pei92}, modelled using Milky Way (MW), or the dust models for Small and Large Magellan Clouds (SMC and LMC, respectively), to determine the extinction $A_V$, measured in the V band;
\item \textit{the UV absorption by neutral hydrogen} (from $z \gtrsim 1$), through a sight-line-averaged model for the optical depth of the intergalactic medium (IGM) as described by \citet{Madau95}, to compute the IGM transmission as a function of wavelength at the redshift of the GRB;
\item \textit{the photoelectric absorption for X-ray data}, through the related hydrogen-equivalent column density $N_H$ (in units of $10^{22}$ cm$^{-2}$), obtained by a polynomial fit of the effective absorption cross-section per hydrogen atom as a function of energy in the $0.03$--$10$~keV range assuming a given abundance pattern \citep{Morrison83}.
\end{itemize}

In {\sc sAGa} the best-fit solution is calculated through the maximisation of a likelihood function, using a Gaussian error model, described in L14.
The Bayesian approach adopted for the broadband modelling in {\sc sAGa} is performed through the Python {\sc emcee} package\footnote{\url{https://emcee.readthedocs.io/en/stable/}} \citep{Foreman13}, based on the Markov Chain Monte Carlo (MCMC) analysis; this tool leads to estimate uncertainties and correlations between the model parameters, and it is particularly useful in high-dimensional problems, like the current one.
These parameters are constrained through the definition of prior distributions that encode preliminary and general information.
{\sc sAGa} considers (1) uniform priors for the parameters that describe the exponential terms on the flux densities ($A_V$) and the power-law indices ($p$ and the injection index $m$), and (2) Jeffreys priors \citep{Jeffreys46}, for the parameters that span different orders of magnitudes ($E_{K,{\rm iso}}$, $n_0$, $A_*$, $\epsilon_e$, $\epsilon_B$ and $t_j$).
$\epsilon_e$ and $\epsilon_B$ are currently believed to be of the order of a few percent to tens of percent by energy \citep{Sironi13}; since generally they do not exceed their equipartition values of $1/3$ (e.g. L14)\footnote{This consists in the equal distribution of the internal energy among the magnetic field, the accelerated electrons and the baryons (protons/neutrons).}, the priors for these parameters are truncated at an upper bound of $1/3$.
These parameters are constrained through the parameter space derived from accurate modelling of the broadband GRB afterglows (e.g., \citealt{Schulze11,Laskar13,Santana14,Perley14,Sironi15b,Laskar16}), and are reported in Table~\ref{tab:parameter_saga}.

\section{Results}
\label{par:results}

\subsection{Preliminary SED analysis}
\label{par:prelim_analysis}

\begin{figure*} 
\centering
{\includegraphics[width=120mm]{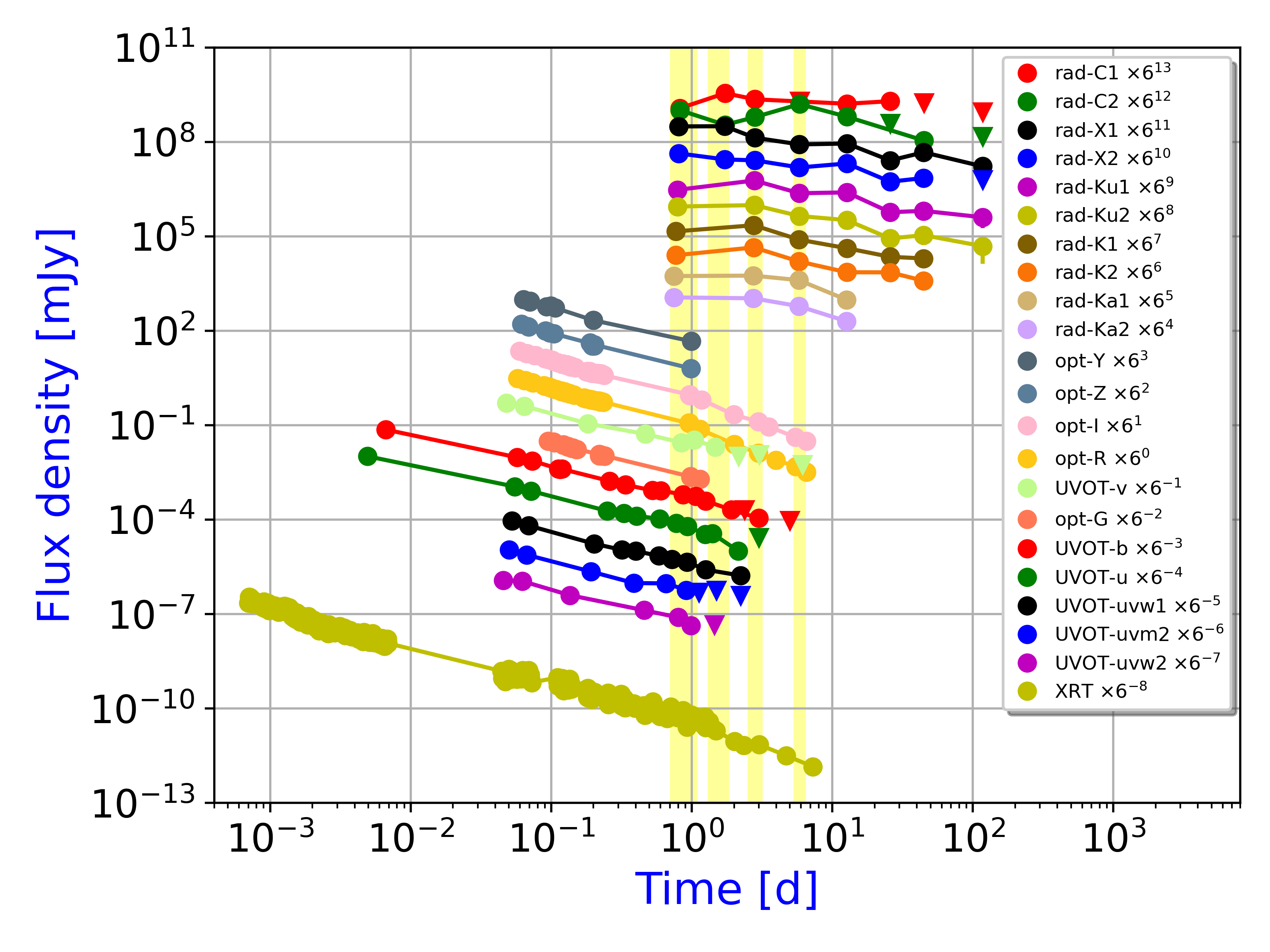}}
\caption{GRB\,160131A light curves from radio to X-rays.
Yellow shaded areas show the time intervals (centred to $0.8$~d, $1.7$~d, $2.7$~d, and $5.8$~d) where SEDs have been empirically analysed.
Filled circles indicate detections (uncertainties are smaller than the corresponding symbol sizes), connected with each other through a segment}, and upside down triangles indicate $3\sigma$ upper limits.
\label{fig:multi_sed_tutto}
\end{figure*}

\begin{figure*} 
\centering
{\includegraphics[width=77mm]{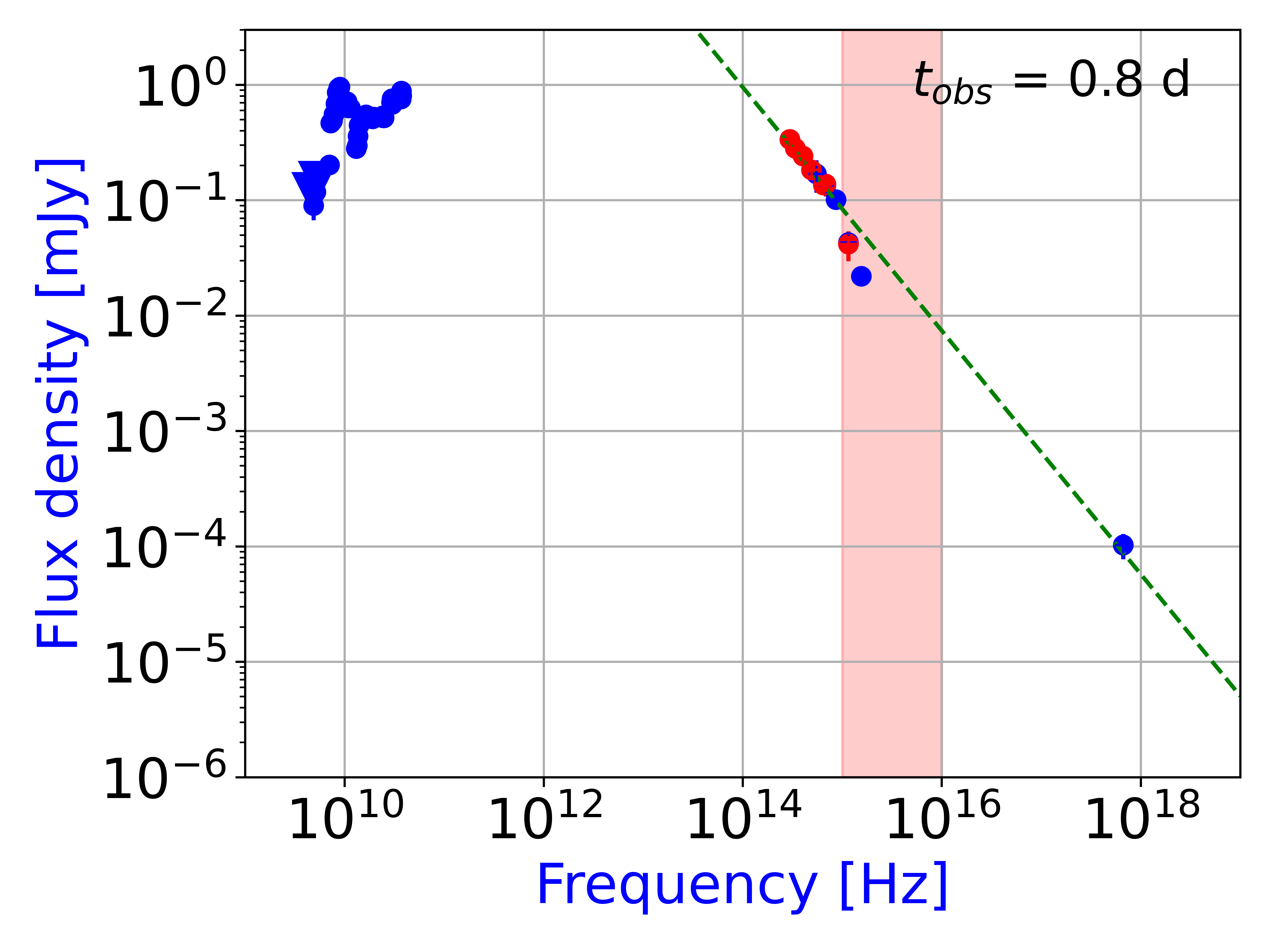}} \quad
{\includegraphics[width=77mm]{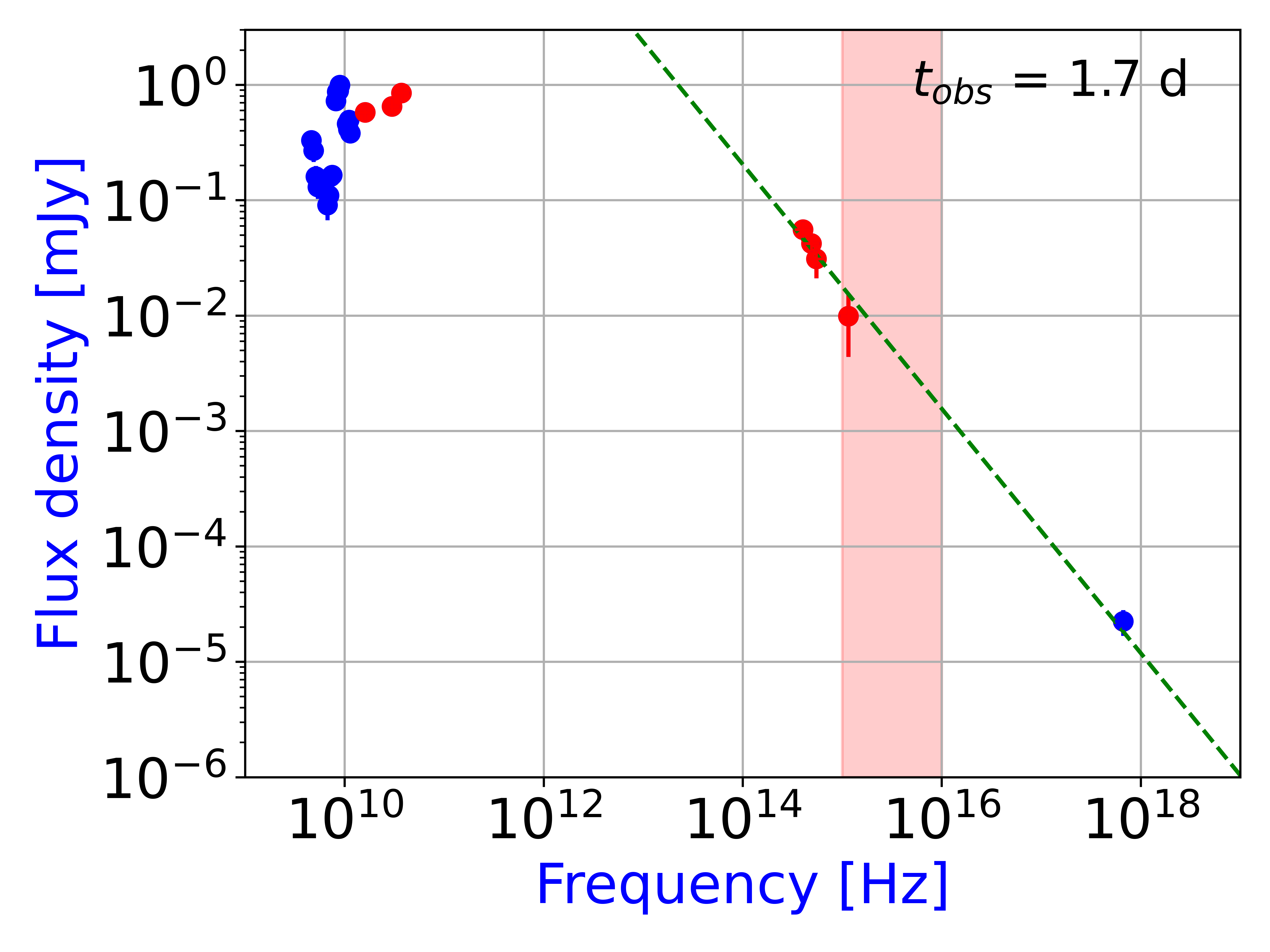}} \\
{\includegraphics[width=77mm]{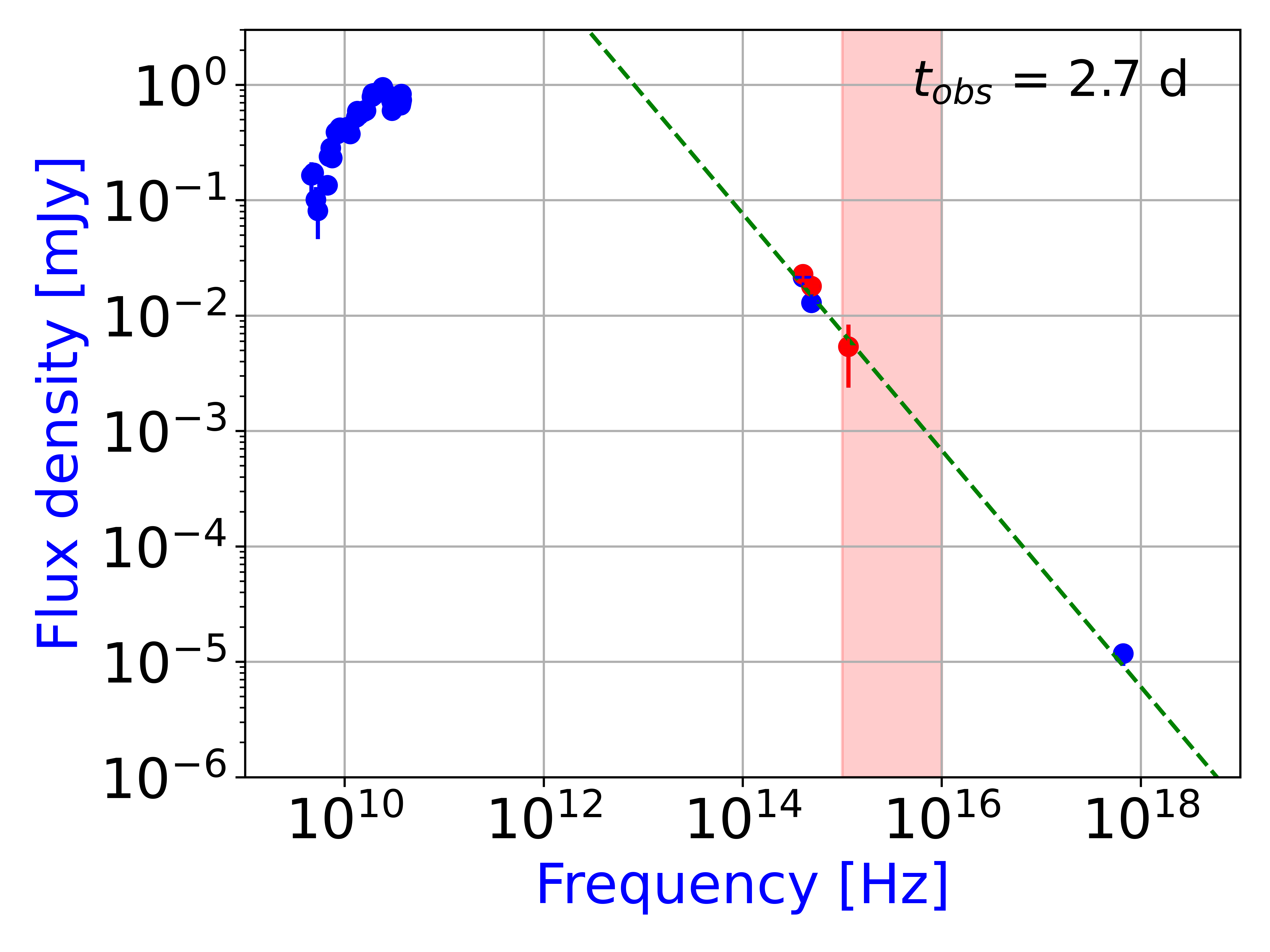}} \quad
{\includegraphics[width=77mm]{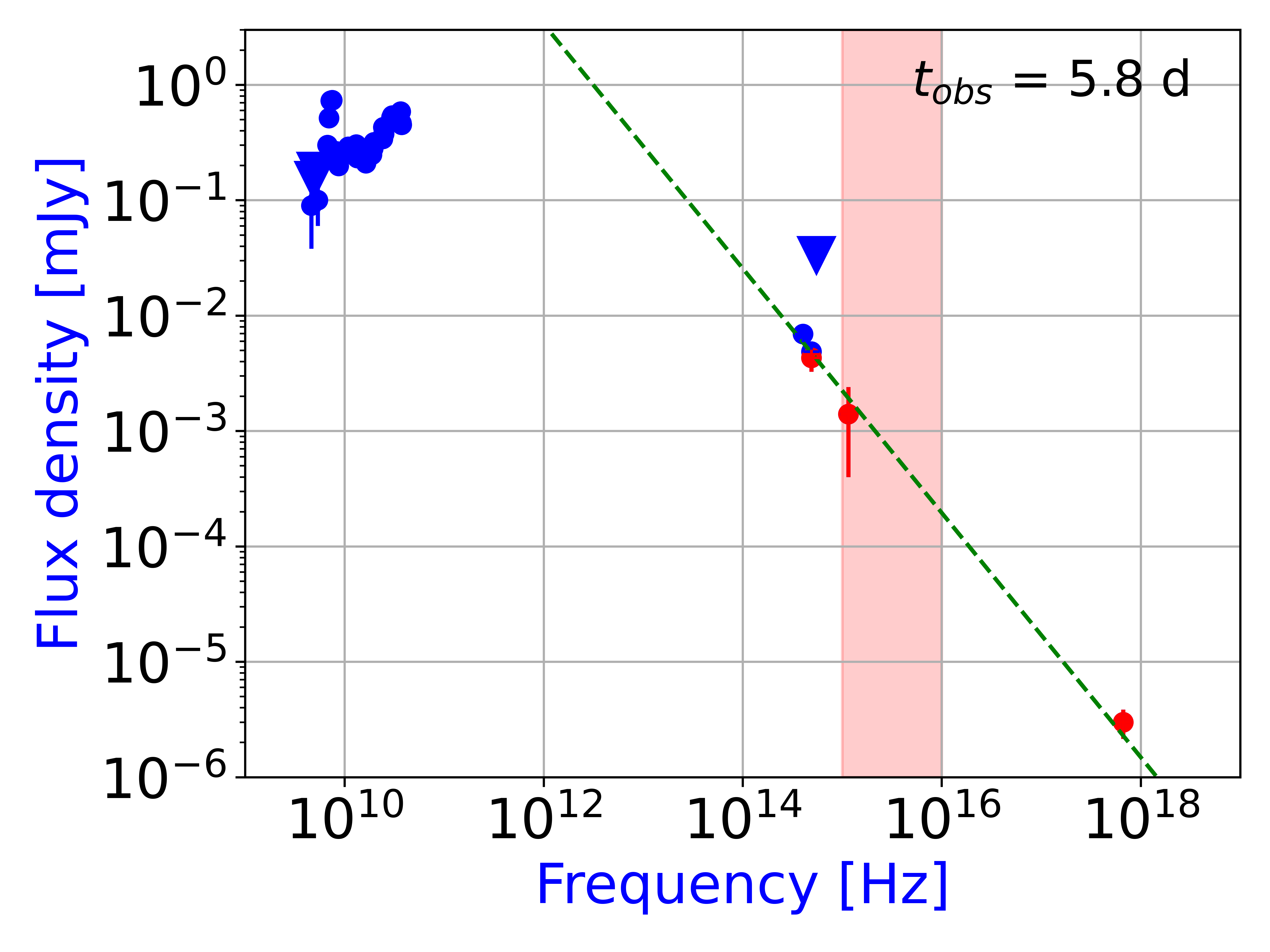}} \\
\caption{Broadband SEDs of GRB\,160131A at $0.8$~d (top left), $1.7$~d (top right), $2.7$~d (bottom left), and $5.8$~d (bottom right).
Blue (red) points are measured (linearly interpolated in a log-log plot) data.
These SEDs display radio peaks (at $0.8$~d, $1.7$~d, and $5.8$~d) and dust extinction (red shaded regions, especially at $0.8$~d).
Green dashed line shows the resulting modelling of the high-energy data (optical/X-ray).
Filled circles indicate detections, and upside down triangles indicate $3\sigma$ upper limits.
}
\label{fig:multi_sed_interp}
\end{figure*}

From the multi-frequency light curves (from radio to X-rays) displayed in Fig.~\ref{fig:multi_sed_tutto} we extract SEDs at four time-intervals (centred to $0.8$, $1.7$, $2.7$, and $5.8$~d), characterised by a richness of broadband data.
To investigate the relation between radio and optical/X-rays, we linearly (in a log-log plot) interpolated data (Fig.~\ref{fig:multi_sed_interp}, red points) at those epochs, where needed.

The high-energy side of the SEDs (Fig.~\ref{fig:multi_sed_interp}) is well-fitted by a power law with a mean value of $\beta_{he} = -1.09 \pm 0.04$\footnote{This value has been obtained neglecting (only in this specific case) the data in the range $10^{15}-10^{16}$~Hz, heavily affected by dust extinction.}, corresponding to a photon index $\Gamma = 1 - \beta_{he} = 2.09 \pm 0.04$, compatible with $\Gamma_X$ obtained from XRT data (Sect.~\ref{subsec:XRT}).
This constrains the behaviour of the break frequencies (especially $\nu_c$ and $\nu_m$), as well as the possible jet break, the time evolution of the blastwave, and the kind of environment (ISM vs. wind).

\subsection{Optical/X-ray data set: $\nu_{m}$ -- $\nu_c$ location, CBM density profile, and jet break}
\label{par:breaks}

\begin{figure} 
\centering
{\includegraphics[width=90mm]{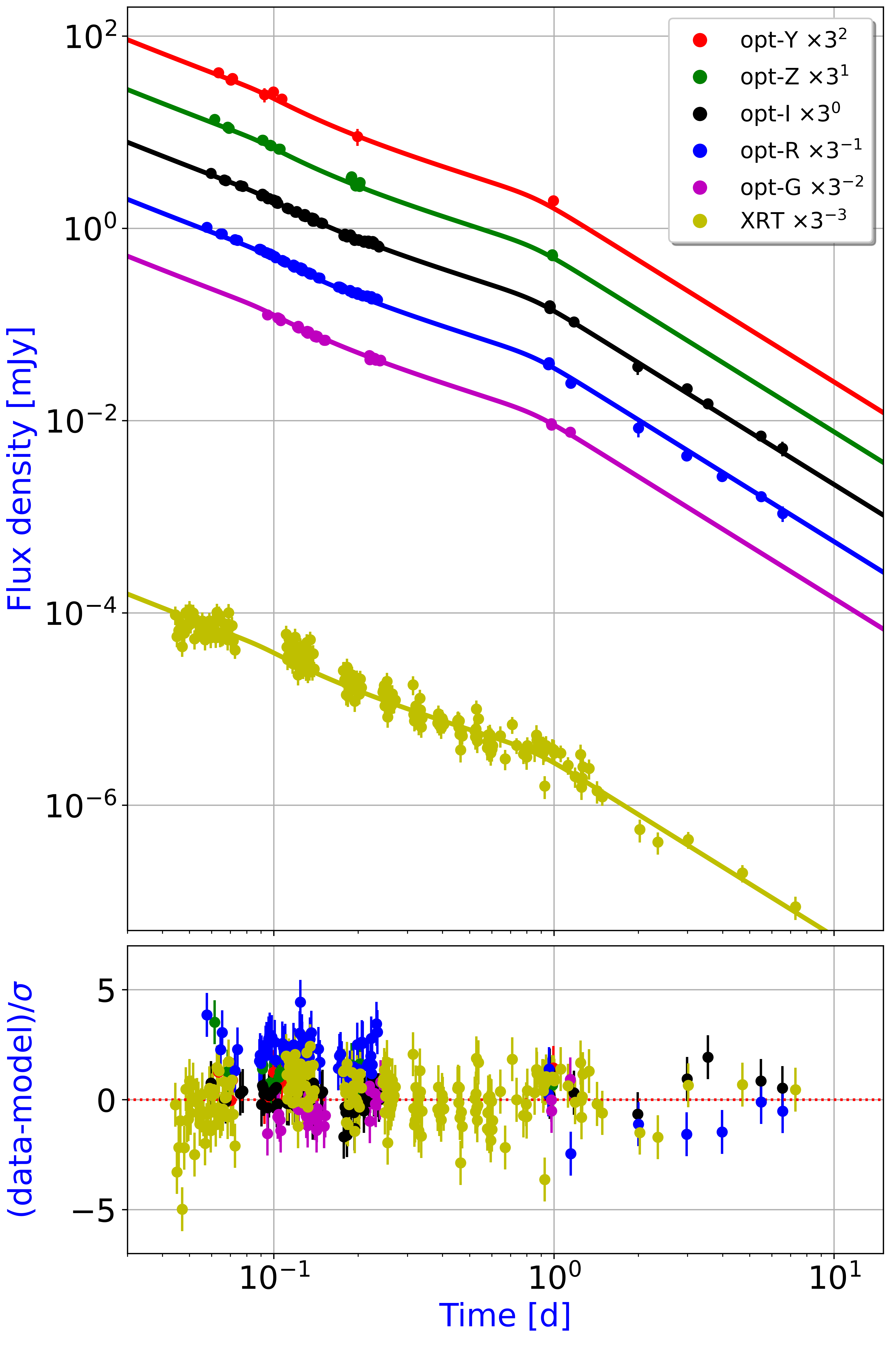}}
\caption{Light curves for GRB\,160131A of visible and X-ray data modelled with DBPL (Eq.~\ref{eq:pl2}).
We observe the plateau, probably ascribable with energy injection between $\sim 10^4$ and $\sim 7 \times 10^4$~s ($\sim 0.1$ and $0.8$~d), and the achromatic break at $\sim 9 \times 10^4$~s ($\sim 1$~d) interpreted in terms of jetted emission.
The bottom panel shows to the residuals of the fit.}
\label{fig:ene_in}
\end{figure}

As we can see in Fig.~\ref{fig:ene_in}, the optical/X-ray fluxes decay with temporal index $\alpha_{he} \sim -1.25$ up to $\sim 0.1$~d, followed by a plateau (more pronounced in the optical data) in the temporal range $\sim 0.1$ -- $0.8$~d ($\alpha_{X,ei} \sim -1$), possibly suggesting energy injection (Sect.~\ref{par:ene_inj_grb}); after the plateau the flux decay steepens to $\alpha_{he} \sim -1.8$ and can be interpreted in terms of a jet break (Sect.~\ref{par:radiolc}).

In the context of the standard afterglow model, the absence of any break frequencies between optical and X-rays domains suggests that $\nu_m$ and $\nu_c$ must lie either below or above the optical/X-ray frequencies $\nu_{opt,X}$ at the first epoch of observations $t_{obs,0}$ ($\sim 10^{-3}$~d).
In the following, we explore the different possibilities:
\begin{description}
\item[\textbf{Fast cooling regime.}] $\nu_{opt,X} < \nu_c < \nu_m$ is incompatible with this regime because the optical/X-ray spectra are expected to show only positive values of $\beta$ ($1/3 \lesssim \beta \lesssim 2$ for any possible spectrum).
Moreover, $\nu_c < \nu_{opt,X} < \nu_m$ is incompatible with this regime because the optical/X-ray spectra are expected to show $\beta \sim -0.5$ instead the observed $\beta_{he} = -1.08$.
Finally, $\nu_c < \nu_m < \nu_{opt,X}$ case is compatible with fast cooling regime because, following the indices $\alpha$ and $\beta$ calculated for different spectral regimes in GS02, it requires an electron energy index $p = -2\beta_{he} \sim 2.18$ and a decay rate $\alpha = (2-3p)/4 \sim -1.14$ (regardless of the CBM), compatible with $\alpha_{he}$; this suggests that $\nu_m$ is just below optical frequencies at $t_{obs,0}$.
\item[\textbf{Slow cooling regime.}] $\nu_{opt,X} < \nu_m < \nu_c$ is incompatible with this regime because the optical/X-ray spectra are expected to show only positive values of $\beta$ ($1/3 \lesssim \beta \lesssim 2$ for any possible spectrum).
Moreover, $\nu_m < \nu_{opt,X} < \nu_c$ is incompatible with this regime, because it requires $p = 1-2\beta_{he} \sim 3.18$ and $\alpha \sim - 1.64$ for an ISM-like CBM ($\alpha \sim - 2.14$ for a wind-like CBM) in GS02, too steep for real light curves.
Finally, $\nu_m < \nu_c < \nu_{opt,X}$ case is compatible with slow cooling regime, because it requires $p = -2\beta_{he} \sim 2.18$ and $\alpha = (2-3p)/4 \sim -1.14$ (the same regime of fast cooling case), suggesting that $\nu_m$ is well below optical frequencies at $t_{obs,0}$.
\end{description}

This picture constrains $\nu_m$ and $\nu_c$ below $\nu_{opt} = 3 \times 10^{14}$~Hz at $t_{obs,0}$.
Moreover, the absence of any break in these light curves until $\sim 0.1$~d (after which energy injection and jet break occur) suggests a decreasing evolution of $\nu_c$, thus favouring an ISM-like CBM over wind-like CBM in the standard afterglow model.

From the upper limit on $\nu_{opt}$ and using the temporal scaling for both $\nu_m$ ($t^{-3/2}$) and $\nu_c$ ($t^{-1/2}$ for ISM), we constrain the passage of $\nu_m$ and $\nu_c$ in the radio frequencies.
The passage of $\nu_m$ in Ka-band is constrained at $t_{obs} < 2.1$~d, in K-band at $t_{obs} < 2.8$~d, in Ku-band at $t_{obs} < 3.6$~d, in X-band at $t_{obs} < 4.7$~d, and in C-band at $t_{obs} < 6.7$~d.
Moreover, $\nu_c$ is expected to cross the radio domain at late-time (Ku-band at $t_{obs} < 4 \times 10^5$~d), and hence virtually unobservable.

Assuming the classical results by \citet{Sari99b}, the decay of the light curve after the break ($t_j \sim 1$~d) is $-p$ for $\nu_m < \nu < \nu_c$  and $\nu > \nu_c$ (corresponding to our picture).
This post-jet decay ($\alpha_{post,j} = -p \sim -2.2$) is steeper than expected for the optical/x-ray decay ($\alpha_{he} \sim -1.8$), and hence we assumed a milder jet break model (pure edge effect, Sect.~\ref{app_saga}), characterised by a post-jet decay $\alpha_{post,j} = \alpha_{pre,j} - (3-k)/(4-k)$ \citep{Granot07}: assuming ISM-like CBM (and hence $k = 0$), we obtain $\alpha_{post,j} = -1.25 - 0.75 = -2$, compatible with the observed value ($\alpha \sim -1.8$).

Summing up, the optical/X-ray data suggest that (1) the CBM is preferably described by ISM, (2) the transition between fast and slow cooling regime is not constrained by optical/X-ray observations, (3) $p \sim 2.2$, (4) both $\nu_m$ and $\nu_c$ lie below $\nu_{opt} = 3 \times 10^{14}$~Hz already at $t_{obs,0}$, and (5) a milder jet break model (pure edge effect) is in accordance with the optical/X-ray data.
A more accurate identification of the break frequencies requires a comprehensive data analysis within a self-consistent broadband modelling (Sect.~\ref{par:saga_results}).

\subsection{VLA data set}
\label{par:radio}

We analyse both the radio SEDs at each epoch from $0.8$~d to $117$~d and the light curves from $4.6$~GHz to $37.4$~GHz.

\subsubsection{Radio SEDs: the $\nu_{sa}$ location, and the multi-component approach}
\label{par:radiosed}

\begin{figure*} 
\centering
{\includegraphics[width=72mm]{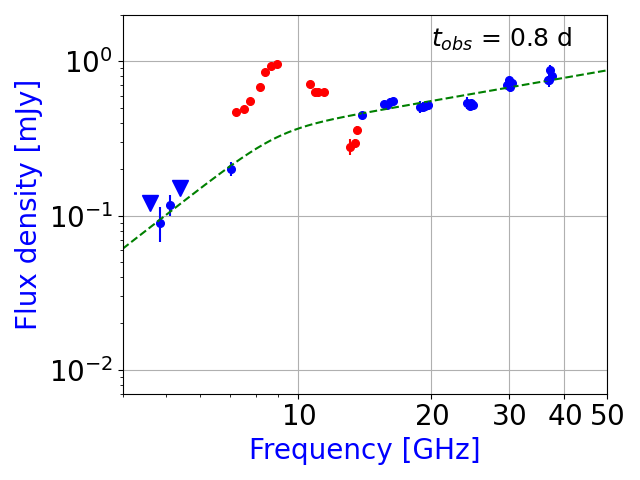}} \quad
{\includegraphics[width=72mm]{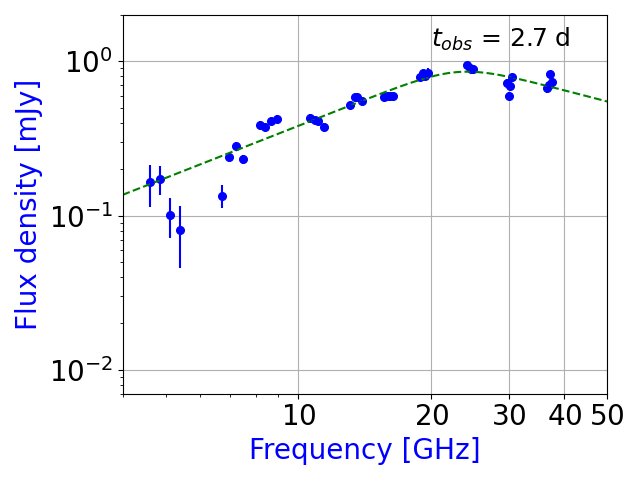}} \\
{\includegraphics[width=72mm]{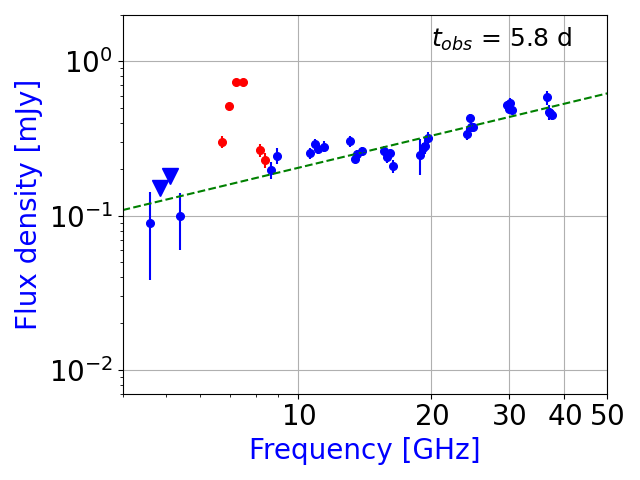}} \quad
{\includegraphics[width=72mm]{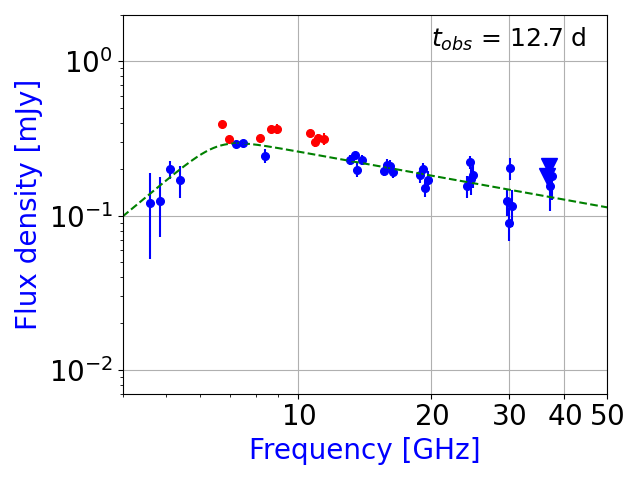}} \\
{\includegraphics[width=72mm]{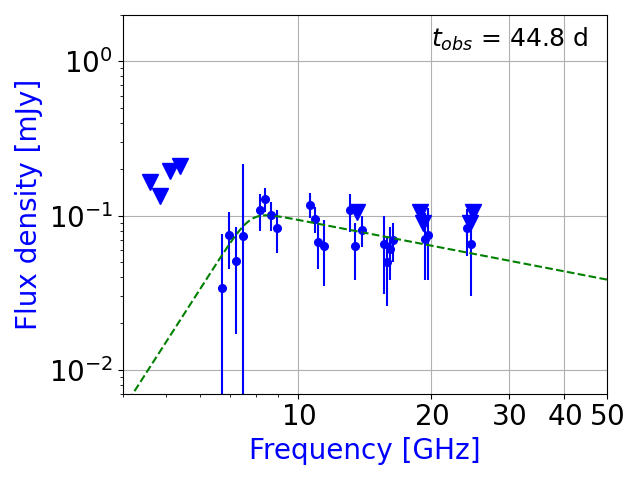}} \quad
\caption{Radio SEDs of GRB\,160131A from $0.8$ to $44.8$~d.
Top left: data together with a BPL (Eq.~\ref{eq:pl1}) at $0.8$~d; red points identify the bump and were ignored by the fit.
Top right: radio SED at $2.7$~d fitted with a BPL.
Middle left: data together with an empirical SPL (Eq.~\ref{eq:pl0}) at $5.8$~d; red points identify the bump $\sim 8$~GHz and were ignored by the fit.
Middle right: radio SED at $12.7$~d fitted with a BPL.
Bottom: data together with a BPL (Eq.~\ref{eq:pl1}) at $44.8$~d.
Green dashed lines show the resulting modelling.
Filled circles indicate detections, and upside down triangles indicate $3\sigma$ upper limits.
}
\label{fig:radio sed_1}
\end{figure*}

\begin{figure*} 
\centering
{\includegraphics[width=72mm]{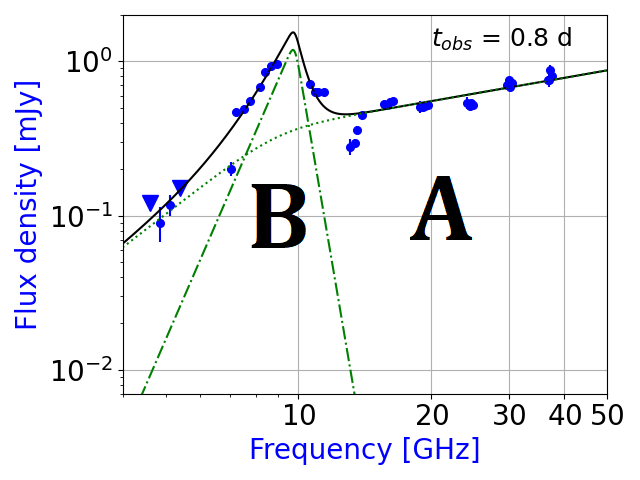}} \quad
{\includegraphics[width=72mm]{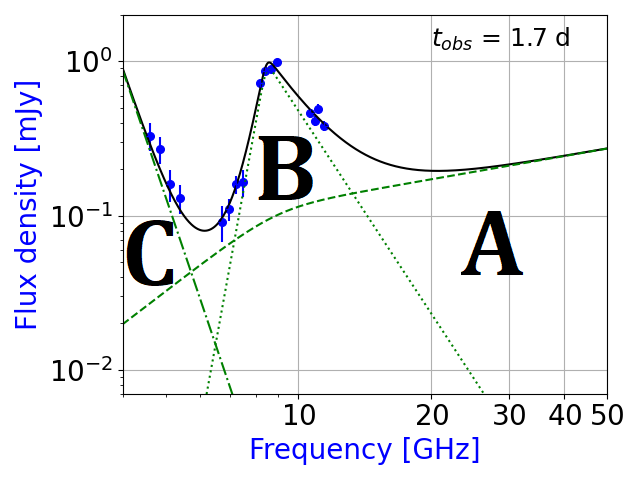}} \\
{\includegraphics[width=72mm]{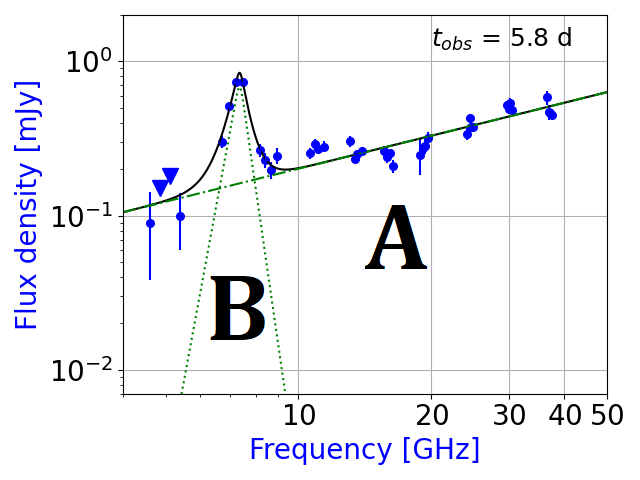}} \quad
{\includegraphics[width=72mm]{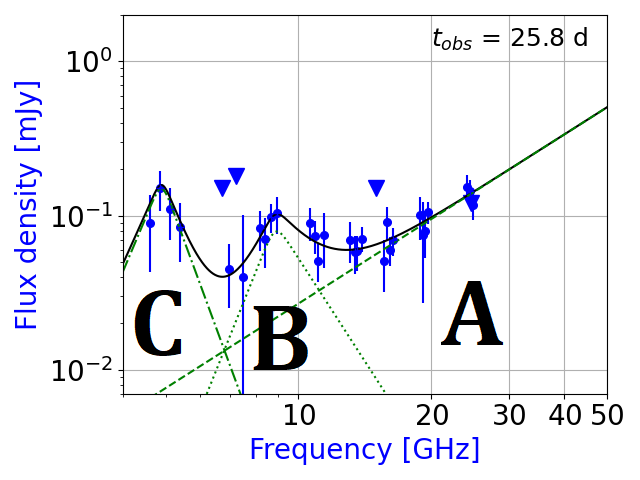}} \\
\caption{Radio SEDs of GRB\,160131A from $0.8$ to $25.8$~d in multi-component approach.
Top left: data together with a sum between two BPLs at $0.8$~d.
Top right: radio data at $1.7$~d together with a sum between a SPL and a two BPLs.
Bottom left: data together with a sum between a SPL and a BPL at $5.8$~d.
Bottom right: radio SED at $25.8$~d fitted with a sum between a SPL and two BPLs.
Black lines show the resulting modelling, and green dash-dot or dotted lines indicate each component.
Filled circles indicate detections, and upside down triangles indicate $3\sigma$ upper limits.
}
\label{fig:radio sed_multi}
\end{figure*}

One of the most impressive features in radio SEDs is the presence of spectral bumps or peaks at several epochs (Fig.~\ref{fig:radio sed_1}, red circles).
We preliminarily modelled these radio SEDs ignoring the peaks with either a power-law or a broken power-law (Fig.~\ref{fig:radio sed_1}), to compare the resulting spectral indices with those expected from the synchrotron emission of GRB afterglows.
Then, we analyse the radio SEDs including all the data set in a multi-component approach (Fig.~\ref{fig:radio sed_multi}).
\begin{itemize}
\item \textbf{$0.8$~d radio SED.} This SED shows a peak at $\sim 9$~GHz and width $\Delta\nu\sim 2$~GHz (Fig.~\ref{fig:radio sed_1}, top left).
Neglecting this peak, this SED is described by a BPL (Eq.~\ref{eq:pl1}; Table~\ref{tab:seds_radio}).
The constraints on $\nu_m$ described in Sect.~\ref{par:breaks} suggest that for this epoch $\nu_m < 150$~GHz; the comparison between the values of $\beta$ showed in Table~\ref{tab:seds_radio} and in Fig.~1 of GS02 suggests that $\nu_{sa}$ crossed the radio band in slow cooling regime (scenario 1, $\nu_{sa} < \nu_m < \nu_c$, GS02).
Unfortunately, the presence of the extra-component peaking at $\sim 9$~GHz prevents us from better constraining $\nu_{sa}$.
\item \textbf{$2.7$~d radio SED.} This SED is characterised by a broad peak at $\sim 25$~GHz, which can be modelled with a BPL (Eq.~\ref{eq:pl1}; Fig.~\ref{fig:radio sed_1}, top right; Table~\ref{tab:seds_radio}).
The constraints described in Sect.~\ref{par:breaks} suggest that for this epoch $\nu_m < 22$~GHz.
This SED is compatible with slow cooling regime (scenario 1, $\nu_{sa} < \nu_m < \nu_c$, GS02): $\beta_{2,bpl}$ in Table~\ref{tab:seds_radio} is steeper than $1/3$ for this regime, suggesting probably the proximity of $\nu_b \sim \nu_m$ with $\nu_{sa}$.
\item \textbf{$5.8$~d radio SED.} This SED, characterised by a strong and narrow peak at $\sim 7$~GHz, is modelled with a SPL (Eq.~\ref{eq:pl0}; Fig.~\ref{fig:radio sed_1}, middle left; Table~\ref{tab:seds_radio}); for this epoch, $\nu_m < 7.3$~GHz (Sect.~\ref{par:breaks}) suggests slow cooling regime, but the value of $\beta$ is incompatible with regimes described in the standard afterglow model.
\item \textbf{$12.7$~d radio SED.} This SED, showing a peak at $\sim 7$~GHz, can be modelled with a BPL (Eq.~\ref{eq:pl1}; Fig.~\ref{fig:radio sed_1}, middle right; Table~\ref{tab:seds_radio}).
Since at this epoch we expect that $\nu_m < 2.3$~GHz (Sect.~\ref{par:breaks}), this behaviour is compatible with slow cooling regime (scenario 2 of GS02, $\nu_m < \nu_{sa} < \nu_c$), where it is $\nu_{sa} = \nu_b$, $\beta_{1,bpl} = 2.5$ and $\beta_{2,bpl} = (1-p)/2$ (suggesting $p = 2.04 \pm 0.10$).
\item \textbf{$44.8$~d radio SED.} This SED is similar to the $12.7$~d one, except that it is just dimmer.
It can be modelled with a BPL (Eq.~\ref{eq:pl1}; Fig.~\ref{fig:radio sed_1}, bottom; Table~\ref{tab:seds_radio}).
Since at this epoch it is $\nu_m < 0.35$~GHz (Sect.~\ref{par:breaks}), this behaviour could still be compatible with slow cooling regime (scenario 2), although $\beta_{2,bpl}$ is steeper than expected; in this scenario it is $\nu_{sa} = \nu_b$, $\beta_{1,bpl} = 2.5$ and $\beta_{2,bpl} = (1-p)/2$ (suggesting $p = 2.1 \pm 0.6$).
\end{itemize}
\begin{table*}
\caption{Best-fit parameters obtained by empirically fitting the radio SEDs of GRB\,160131A from $0.8$ to $44.8$~days after the GRB trigger (see Fig.~\ref{fig:radio sed_1}).
``SPL'' and ``BPL'' indicate a power-law (Eq.~\ref{eq:pl0}) and a broken power-law model (Eq.~\ref{eq:pl1}), respectively.
$\nu_b$ is the break frequency and $F_b$ the flux density at $\nu=\nu_b$; $\beta_{1,bpl}$ and $\beta_{2,bpl}$ are the two BPL spectral indices, while $\beta_{pl}$ is the SPL index.
The reduced chi square is denoted with $\chi^2_r$.}
\label{tab:seds_radio}
\centering
\begin{tabular}{l | ccccc}
\hline
\hline
$t_{obs}$                    & $0.8$~d          & $2.7$~d          & $5.8$~d          & $12.7$~d         & $44.8$~d         \\
\hline
Model                        & BPL              & BPL              & SPL              & BPL              & BPL              \\
$\nu_b$\tablefootmark{(a)}   & $8.9 \pm 0.6$    & $23.1 \pm 0.5$   & -                & $6.6 \pm 0.3$    & $7.8 \pm 0.3$    \\
$F_b$\tablefootmark{(b)}     & $0.32 \pm 0.02$  & $0.86 \pm 0.10$  & -                & $0.32 \pm 0.01$  & $0.09 \pm 0.01$  \\
$\beta_{pl}$                 & -                & -                & $0.69 \pm 0.04$  & -                & -                \\
$\beta_{1,bpl}$              & $2.2 \pm 0.4$    & $1.13 \pm 0.03$  & -                & $2.39 \pm 0.34$  & $4.46 \pm 1.90$  \\
$\beta_{2,bpl}$              & $0.50 \pm 0.05$  & $-0.75 \pm 0.11$ & -                & $-0.52 \pm 0.05$ & $-0.55 \pm 0.26$ \\
$\chi^2_r$                   & $0.79$           & $1.60$           & $1.27$           & $0.74$           & $0.88$           \\
\hline
\end{tabular}
\tablefoot{
\tablefoottext{a}{In units of GHz.}  \\
\tablefoottext{b}{In units of mJy.}  \\
}
\end{table*}

The relatively large uncertainties on flux density in the SEDs at $\nu \lesssim 6$~GHz inevitably affect the ability to constrain $\nu_{sa}$.
Assuming $\nu_b \sim \nu_{sa}$ in the radio SEDs at $0.8$~d and $12.7$~d (Table~\ref{tab:seds_radio}), we obtain that $\nu_{sa}$ could evolve approximately as $t^{-0.1}$, compatibly with being constant over time, as expected for the ISM (GS02).

Including now the peaks in the radio SEDs, we consider all the radio data set in a multi-component approach.
In addition to the continuum associated with FS emission (Sect.~\ref{par:radiosed}, hereafter component A), radio SEDs suggest other two distinct emission components (Fig.~\ref{fig:radio sed_multi}).
\begin{description}
\item[\textbf{Component B}] appears at four epochs ($0.8$, $1.7$, $5.8$, and $25.8$~d) and is characterised by a faint peak around $9$~GHz (Fig.~\ref{fig:radio sed_multi}).
We fit this component with a BPL, obtaining the results showed in Table~\ref{tab:seds_radio_multi}.
\item[\textbf{Component C}] shows up in the $25.8$~d radio SED, and it partially appears at $1.7$~d (Fig.~\ref{fig:radio sed_multi}), when the lack of radio data at $\lesssim 5$~GHz does not allow us to resolve its peak.
We fit this component with a SPL ($1.7$~d) and a BPL ($25.8$~d), obtaining the results showed in Table~\ref{tab:seds_radio_multi}.
\end{description}
In the multi-component approach, we briefly focus on the radio SED at $1.7$~d (Fig.~\ref{fig:radio sed_multi}, top right), well-fitted by a combination of a SPL at $\lesssim 5$~GHz (a possible part of the component C) and a BPL peaking at $\sim 9$~GHz (component B, Table~\ref{tab:seds_radio_multi}).
As opposed to the other SEDs, the absence of data at high frequencies prevents us from constraining component A associated with the FS emission of GRB afterglow.
In Fig.~\ref{fig:radio sed_multi} (top right) we added the component A with a BPL, characterised by the same spectral indices of the case of $0.8$~d radio SED, and flux density $0.1$~mJy (Table~\ref{tab:seds_radio_multi}).
\begin{table}
\caption{Parameters for empirical fits to radio SEDs of GRB\,160131A from $0.8$ to $25.8$~d in multi-component approach (see Fig.~\ref{fig:radio sed_multi}).
The letter into brackets in the "Type" row indicates the associated component.
See the caption of Table~\ref{tab:seds_radio} for a full description of the fit parameters.
}
\label{tab:seds_radio_multi}
\small
\begin{tabular}{l | c c c c}
\hline\hline
$t_{obs}$                       & 0.8~d                     & 1.7~d                     & 5.8~d                     & 25.8~d           \\
\hline
$N_{comp}$                      & 2                         & 3                         & 2                         & 3                \\
\hline
Type                            & -                         & SPL (C?)                  & SPL (A?)                  & SPL (A)          \\
$\beta_{pl}$                    & -                         & $-9.9 \pm 0.3$            & $0.68$\tablefootmark{(c)} & $1.8 \pm 0.4$    \\
\hline
Type                            & BPL (A)                   & BPL (A)                   & BPL (B)                   & BPL (C)          \\
$\nu_{peak}$\tablefootmark{(a)} & $8.9$\tablefootmark{(c)}  & $8.9$\tablefootmark{(c)}  & $7.4 \pm 0.2$             & $4.9 \pm 0.1$    \\
$F_{peak}$\tablefootmark{(b)}   & $0.32$\tablefootmark{(c)} & $0.1$\tablefootmark{(c)}  & $0.68 \pm 0.07$           & $0.15 \pm 0.01$  \\
$\beta_{1,bpl}$                 & $2.2$\tablefootmark{(c)}  & $2.2$\tablefootmark{(c)}  & $15.5 \pm 1.3$            & $6.8 \pm 0.2$    \\
$\beta_{2,bpl}$                 & $0.5$\tablefootmark{(c)}  & $0.5$\tablefootmark{(c)}  & $-19.9 \pm 2.0$           & $-7.8 \pm 0.2$   \\
\hline
Type                            & BPL (B)                   & BPL (B)                   & -                         & BPL (B)          \\
$\nu_{peak}$\tablefootmark{(a)} & $9.8 \pm 0.5$             & $8.5 \pm 0.2$             & -                         & $8.9 \pm 0.5$    \\
$F_{peak}$\tablefootmark{(b)}   & $1.2 \pm 0.5$             & $0.9 \pm 0.1$             & -                         & $0.08 \pm 0.01$  \\
$\beta_{1,bpl}$                 & $6.6 \pm 0.3$             & $15.8 \pm 1.3$            & -                         & $7.1 \pm 1.2$    \\
$\beta_{2,bpl}$                 & $-16.8 \pm 1.8$           & $-4.4 \pm 0.2$            & -                         & $-4.4 \pm 1.2$   \\
\hline
$\chi^2_r$                      & $1.8$                     & $1.03$                    & $1.7$                     & $1.1$            \\
\hline
\end{tabular}
\tablefoot{
\tablefoottext{a}{In units of GHz.}  \\
\tablefoottext{b}{In units of mJy.}  \\
\tablefoottext{c}{Fixed.}            \\
}
\end{table}

Summing up, the radio SEDs suggest that (1) the slow cooling regime occurs at $t \lesssim 0.8$~d, (2) at $5.8$~d the features are incompatible with the standard GRB afterglow model, (3) at $12.7$~d $\nu_{sa} \sim 7$~GHz, and (4) radio data set is composed by 3 spectral components (A, B, and C), of which only the first one (A) is connected with a known physical effect (the continuum associated with FS emission).
We delve deeper into them in Sect.~\ref{sec:disc}.

\subsubsection{Radio light curves: evidence for a jet}
\label{par:radiolc}

Radio data help constrain both the FS emission and the jet opening angle.
In this context we analysed the radio light curves ignoring the peaks ascribed to additional components (Sect.~\ref{par:radiosed}) and data below $8$~GHz because of the high variability, probably caused by strong interstellar scintillation (ISS, Sects.~\ref{par:reverse} and \ref{app_saga}), which prevent from well constraining the rise and decline rates.

\begin{figure*} 
\centering
{\includegraphics[width=72mm]{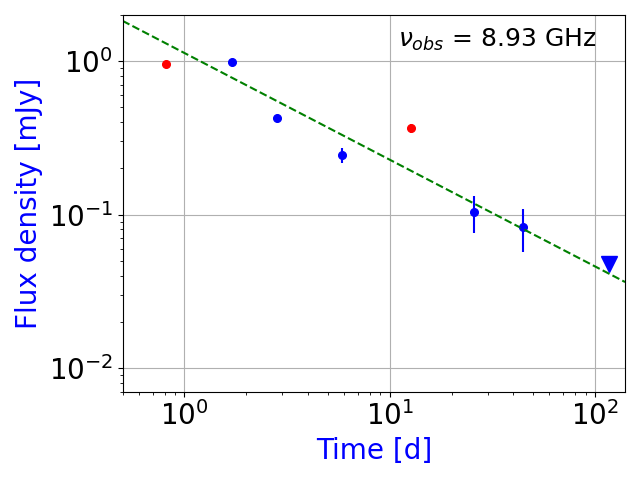}} \quad
{\includegraphics[width=72mm]{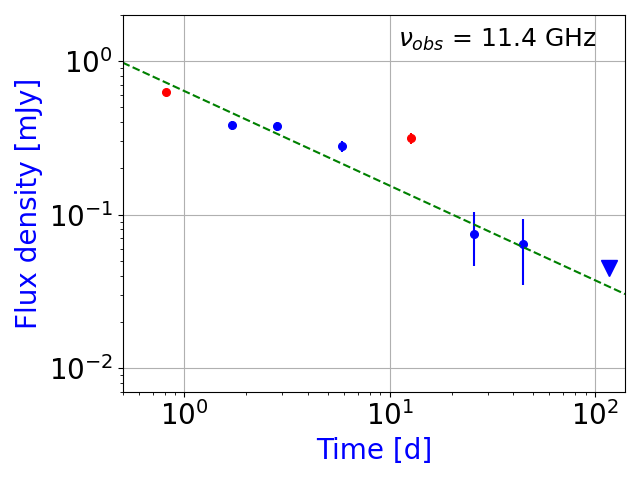}} \\
{\includegraphics[width=72mm]{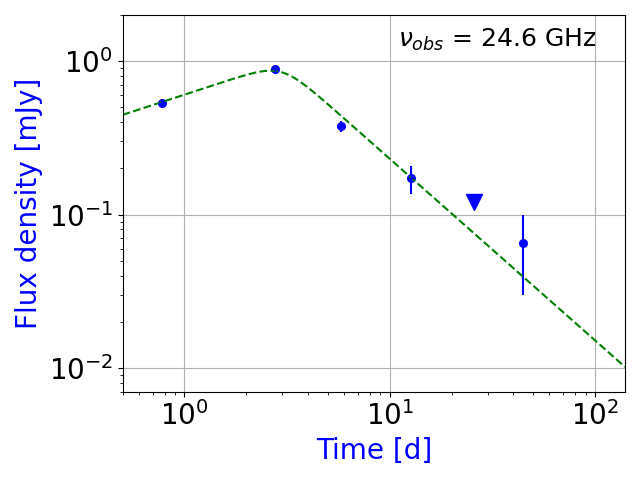}} \quad
{\includegraphics[width=72mm]{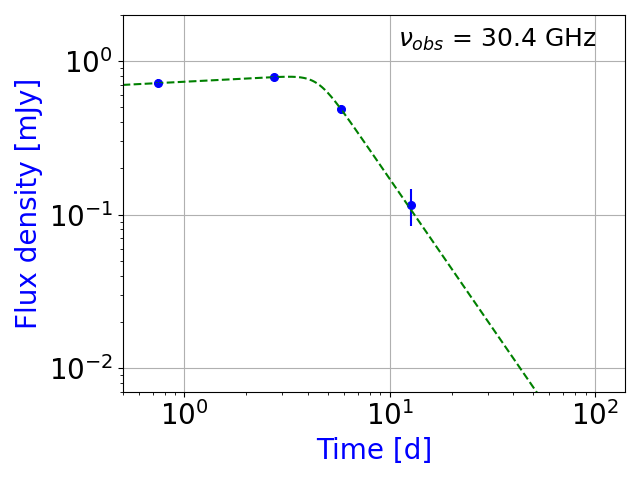}} \\
{\includegraphics[width=72mm]{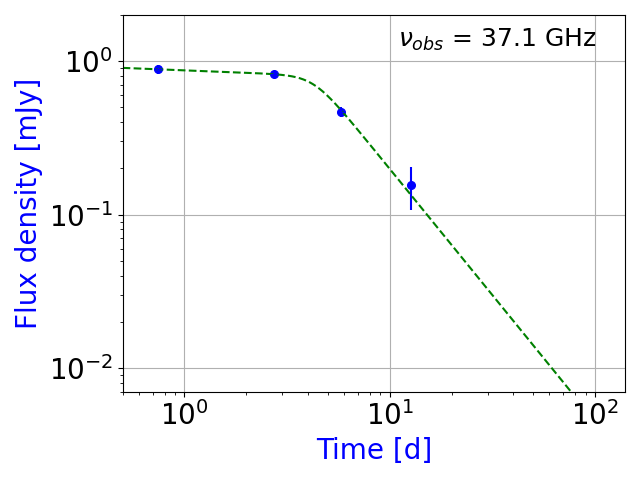}} \quad
\caption{Radio light curves of GRB\,160131A in the range $9-37$~GHz.
$8.93$~GHz (top left) and $11.4$~GHz (top right) fitted with a SPL (Eq.~\ref{eq:pl0}); the other light curves ($24.6$~GHz, middle left; $30.4$~GHz, middle right; $37.1$~GHz, bottom) are fitted with a BPL (Eq.~\ref{eq:pl1}).
Blue filled circles indicate detections, and upside down triangles indicate $3\sigma$ upper limits; red circles indicate the ignored points corresponding to the peaks observed in radio SEDs (Fig.~\ref{fig:radio sed_1}), and green lines show the resulting model.
}
\label{fig:radio lc_1}
\end{figure*}

In the standard afterglow model, a jet break arises at the time $t_{j}$ when the bulk Lorentz factor $\Gamma$ decreases below the inverse opening angle of the jet $\theta_j^{-1}$ and its edges become visible to an observer (Sect.~\ref{app_saga}).
Once $\nu_m$ has crossed the observing frequency, the flux density decays steeply following a jet break.
In this regime, the steepening in the radio light curves is expected to follow that of the steepening in the optical/X-ray light curves, depending on the time it takes for $\nu_m$ to cross the radio band \citep{Laskar15}.
The identification of $\nu_b \sim 23$~GHz with $\nu_m$ observed in the SED at $t_{obs} = 2.7$~d (Fig.~\ref{fig:radio sed_1} and Table~\ref{tab:seds_radio}) indicates that the light curve at $\nu_{obs} \sim \nu_b$ would peak at $t_{obs}$.
We observed this behaviour in the light curve at $24.6$~GHz (Fig.~\ref{fig:radio lc_1}, middle left), well-fitted by a BPL (Eq.~\ref{eq:pl1}); the best-fit results (Table~\ref{tab:lc_radio}) show that $\alpha_{2,bpl}$ is also compatible with $\alpha_{he}$ obtained for optical/X-ray light curves (Sect.~\ref{par:breaks}), and hence with the passage of $\nu_m$ in the light curves of standard GRB afterglow model \citep{Sari98}.
The radio light curves above $24.6$~GHz show a steep decay of the flux densities at $t_b$ ranging between $\sim 3$ and $\sim 5$~d, compatible with jet break; modelling with BPL (Eq.~\ref{eq:pl1}) shows $-0.1 \lesssim \alpha_1 \lesssim 0.1$ and $-2 \lesssim \alpha_2 \lesssim -1.6$ (Fig.~\ref{fig:radio lc_1}, middle left and bottom; Table~\ref{tab:lc_radio}).
At $t = t_j \sim 1$~d, as inferred from optical/X-ray light curves, $\nu_m$ lies close to $\sim 10^{11}$~GHz, that is well below the optical/X-ray domain: this is consistent with the steep decline observed around the same epoch in these bands.

For completeness, we obtain further information about break frequencies of synchrotron emission from the decreasing temporal decay indices $\alpha$ in the light curves between $8$~GHz and $24.6$~GHz.
In particular:
\begin{itemize}
    \item the value $\alpha \sim -0.6$ (Table~\ref{tab:lc_radio}) obtained modelling the light curves between $8$~GHz and $14$~GHz (Fig.~\ref{fig:radio lc_1}, top left and top right) with a SPL (Eq.~\ref{eq:pl0}), suggests that -- in agreement with what inferred from the high-energy data analysis (Sect.~\ref{par:breaks}) -- (1) $\nu_c$ crosses these frequencies after $45$~d, and (2) the passage of $\nu_m$ occurs at $t \lesssim 3$~d \citep{Sari98};
    \item the decreasing temporal indices in the light curves between $14$~GHz and $24.6$~GHz, evolving from $\sim -0.8$ at $14$~GHz to $\sim -1.2$ at $24$~GHz, are suggestive of the passage of $\nu_c$ in these light curves above $\sim 120$~d, and the passage of $\nu_m$ at these observing frequencies is very close to $3$~d \citep{Sari98}.
\end{itemize}
\begin{table*}
\caption{Parameters for empirical fits to VLA radio light curves of GRB\,160131A from $4.6$ to $37.4$~GHz (see Fig.~\ref{fig:radio lc_1}).
$t_b$ indicates the break time corresponding to the flux density $F_b$, $\alpha_{1,bpl}$ and $\alpha_{2,bpl}$ indicate the temporal decay indices for broken power-law, and $\alpha_{pl}$ indicates the temporal decay index for a power-law.
See the caption of Table~\ref{tab:seds_radio} for a full description of the other fit parameters.
}
\label{tab:lc_radio}
\centering
\begin{tabular}{l | ccccc}
\hline
\hline
$\nu_{obs}$\tablefootmark{a} & $8.93$           & $11.4$           & $24.6$           & $30.4$                  & $37.1$                   \\
\hline
Model                        & SPL              & SPL              & BPL              & BPL                     & BPL                      \\
$t_b$\tablefootmark{b}       & -                & -                & $2.96 \pm 0.04$  & $4.47 \pm 0.14$         & $4.16 \pm 0.11$          \\
$F_b$\tablefootmark{c}       & -                & -                & $0.85 \pm 0.13$  & $0.71 \pm 0.05$         & $0.71 \pm 0.07$          \\
$\alpha_{pl}$                & $-0.64 \pm 0.04$ & $-0.62 \pm 0.02$ & -                & -                       & -                        \\
$\alpha_{1,bpl}$             & -                & -                & $0.44 \pm 0.05$  & $0.07$\tablefootmark{d} & $-0.05$\tablefootmark{d} \\
$\alpha_{2,bpl}$             & -                & -                & $-1.18 \pm 0.02$ & $-1.93 \pm 0.23$        & $-1.59 \pm 0.13$         \\
$\chi^2_r$                   & $1.8$            & $1.3$            & $1.1$            & $0.99$                  & $1.02$                   \\
\hline
\end{tabular}
\tablefoot{
\tablefoottext{a}{In units of GHz.}   \\
\tablefoottext{b}{In units of days.}  \\
\tablefoottext{c}{In units of mJy.}   \\
\tablefoottext{d}{Fixed.}   \\
}
\end{table*}

\subsection{Physical approach: modelling with {\sc sAGa}}
\label{par:saga_results}

The complexity of the broadband spectral and temporal properties, in particular the spectral radio peaks (Fig.~\ref{fig:multi_sed_interp}), imposes an iterative analysis (optical, optical/X-ray, optical/X-ray/radio) to probe the physical characteristics of the afterglow of GRB\,160131A, and to oversee when the broadband model of GRB afterglow starts losing validity.
We considered in this analysis a jetted (edge-regime) FS emission with dust extinction and energy injection in ISM-like CBM; we also considered ISS effect, typical of radio domain, following the procedure described in \citet{Misra19}.
The modelling ignored the data at $t_{obs} < T_{90} = 4 \times 10^{-3}$~d, when the prompt emission was not over yet.

From the analysis reported in Sections~\ref{par:breaks} and \ref{par:radio}, we adopted as starting points the following values for the micro-physics parameters (Sect.~\ref{app_saga} and Table~\ref{tab:parameter_saga}): $p = 2.2$, $\epsilon_B = 0.01$, $n_0 = 1$~cm$^{-3}$, $E_{k,iso,52} = 50$, $A_V = 0.1$, $t_j =1$~d, and $m = 0.2$.
Moreover, according to a method to constrain $\epsilon_e$ through the identification of the radio peaks (observed in the radio light curves) connected with the passage of $\nu_m$ \citep{BeniaminivanderHorst17}, we used the peak (with a flux density $F \sim 0.9$~mJy) observed in the $24.6$~GHz light curve at $t_{obs} \sim 3$~d (Sect.~\ref{par:radiolc}) to estimate $\epsilon_e \sim 0.1$ as a starting point.

\subsubsection{From optical to X-rays}
\label{par:opt_res}

\begin{figure*} 
\centering
{\includegraphics[width=150mm]{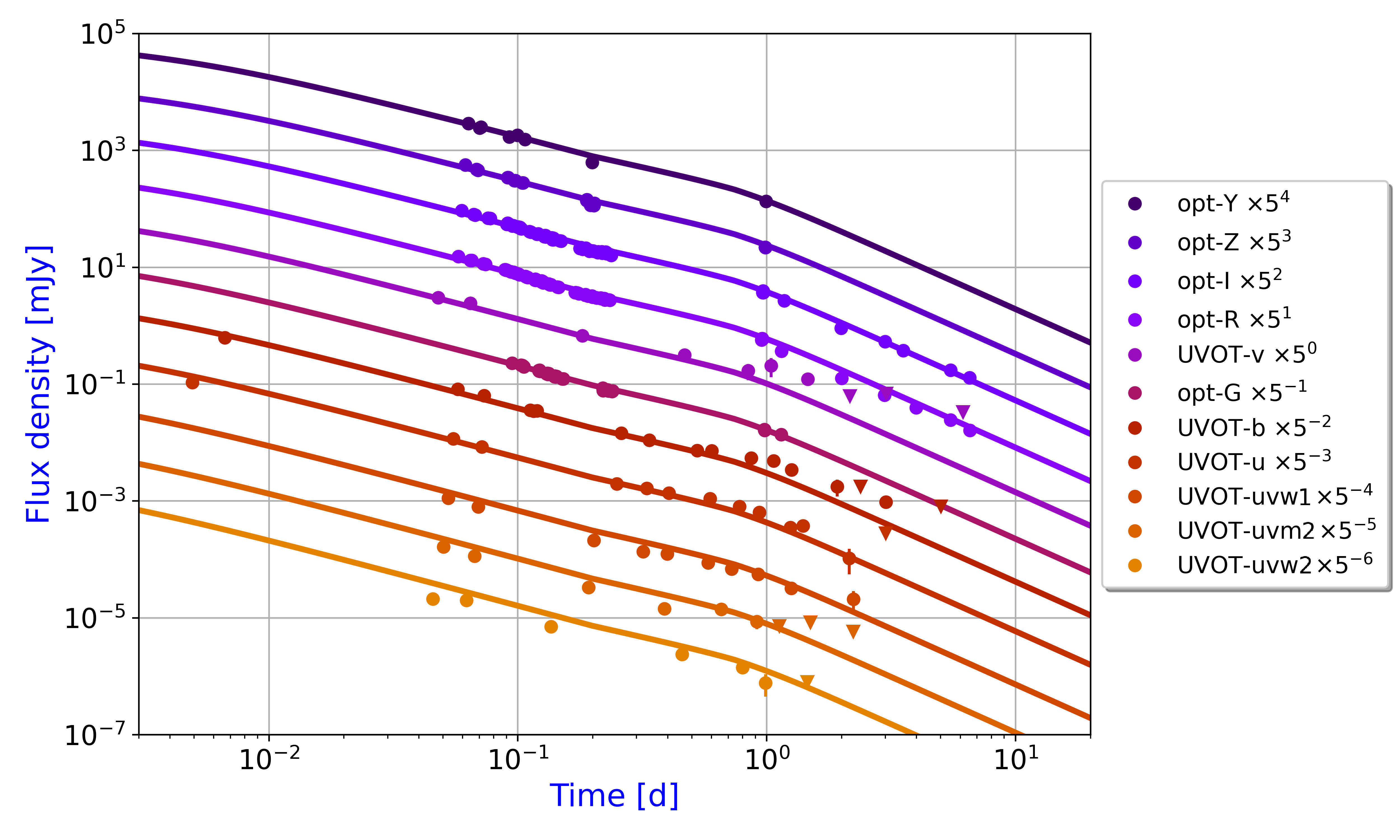}}
\caption{Broadband modelling (UVOIR frequencies; Table~\ref{tab:optjet_res_ism}, first column) of GRB\,160131A for a FS model with a ISM-like CBM (GS02); we considered in this analysis a jetted (edge-regime) emission with dust extinction and energy injection.
Filled circles indicate detections, and downward triangles indicate $3\sigma$ upper limits.}
\label{fig:lc_uvoir_saga}
\end{figure*}

\begin{figure*} 
\centering
{\includegraphics[width=150mm]{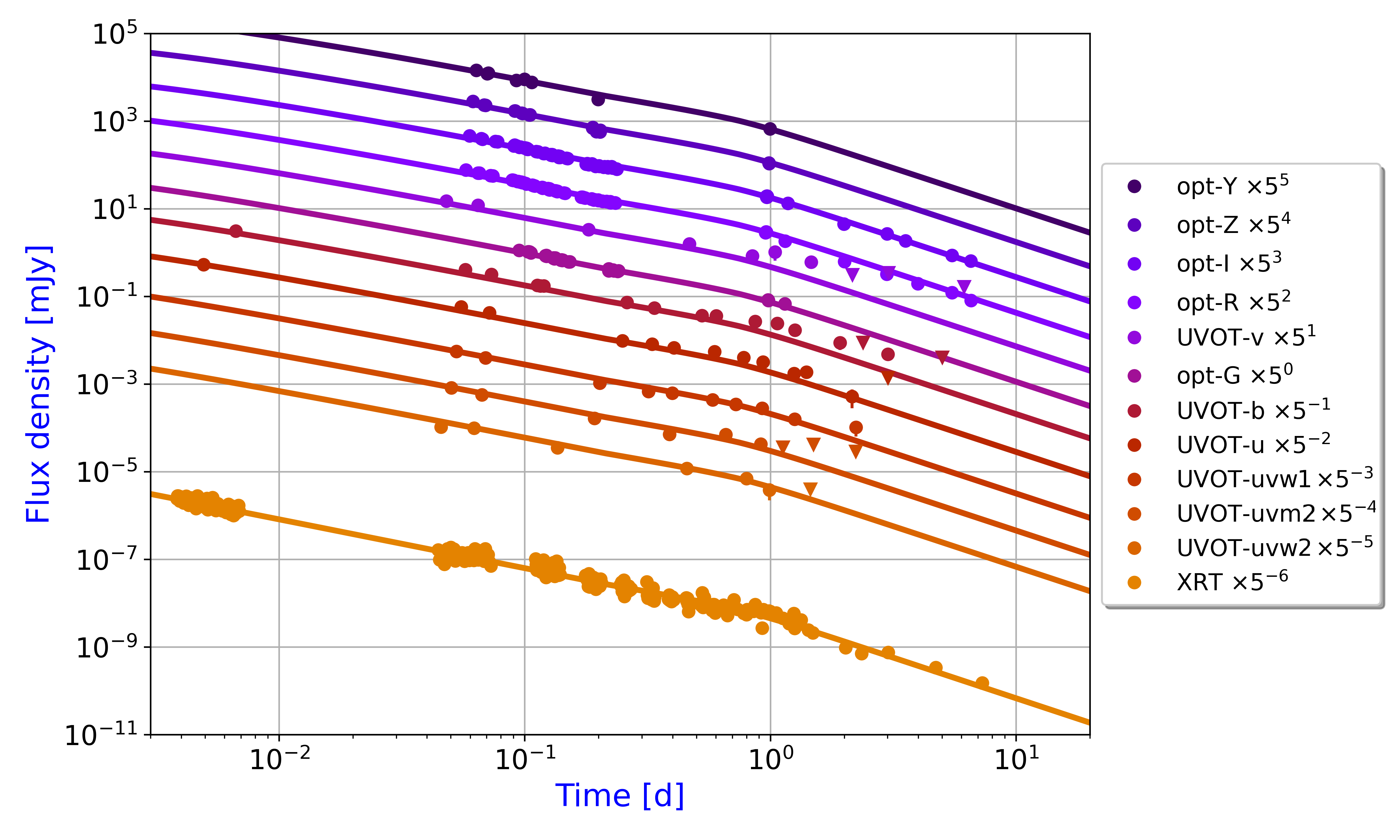}}
\caption{Broadband modelling (from optical to X-ray frequencies; Table~\ref{tab:optjet_res_ism}, second column) of GRB\,160131A.
See the caption of Fig.~\ref{fig:lc_uvoir_saga} for a full description of the modelling.
Filled circles indicate detections, and downward triangles indicate $3\sigma$ upper limits.}
\label{fig:lc_uvoirx_saga}
\end{figure*}

The iterative process of modelling from $3 \times 10^{14}$ to $6.6 \times 10^{17}$~Hz, the results of which are reported in Table~\ref{tab:optjet_res_ism} (in the first two columns), shows a good best-fit model ($\chi^2_r \sim 1$), as displayed in the broadband light curves (Fig.~\ref{fig:lc_uvoir_saga} for optical frequencies, and Fig.~\ref{fig:lc_uvoirx_saga} for optical/X-ray domain).
\begin{figure*} 
\centering
{\includegraphics[width=72mm]{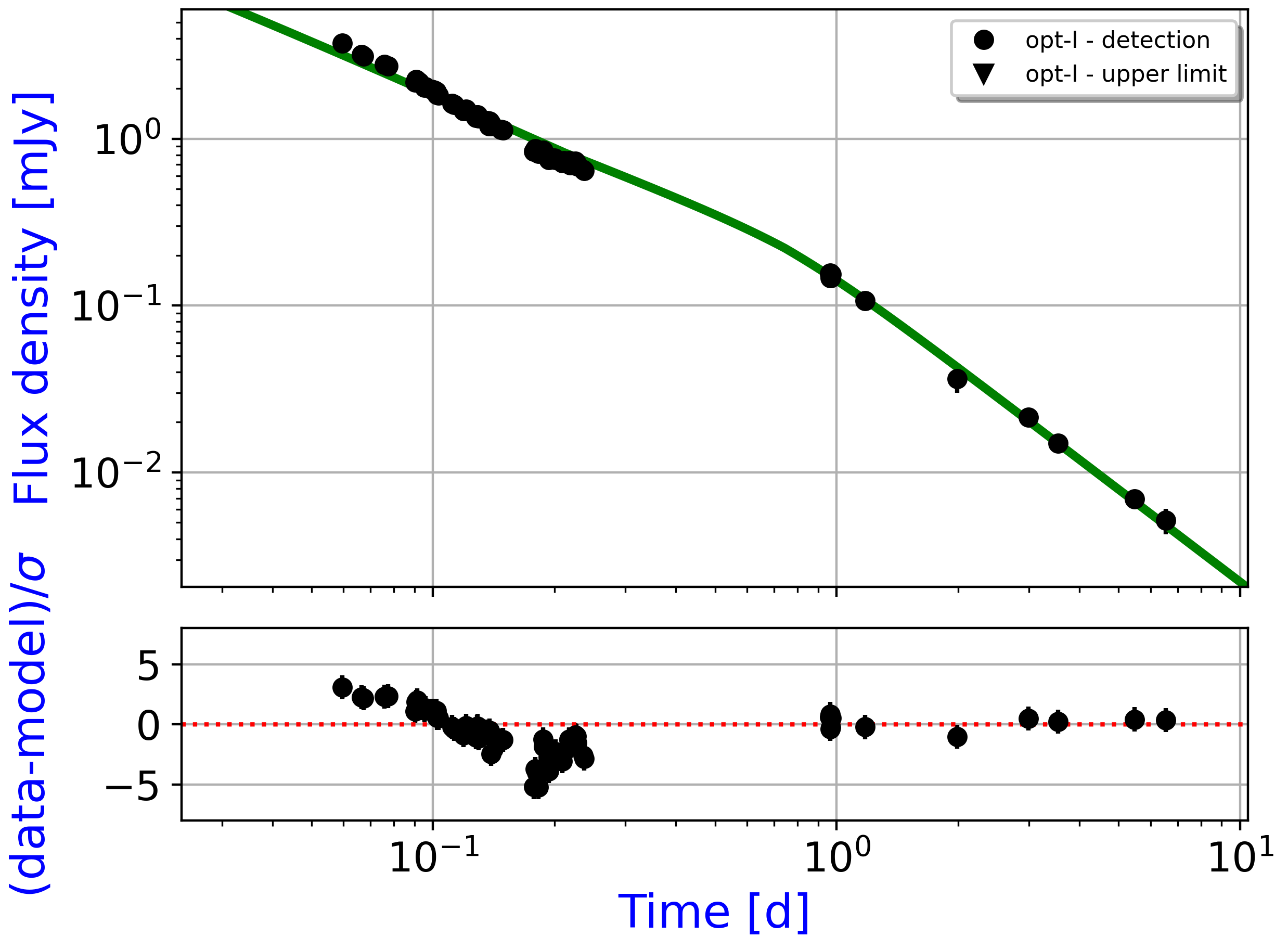}} \quad
{\includegraphics[width=72mm]{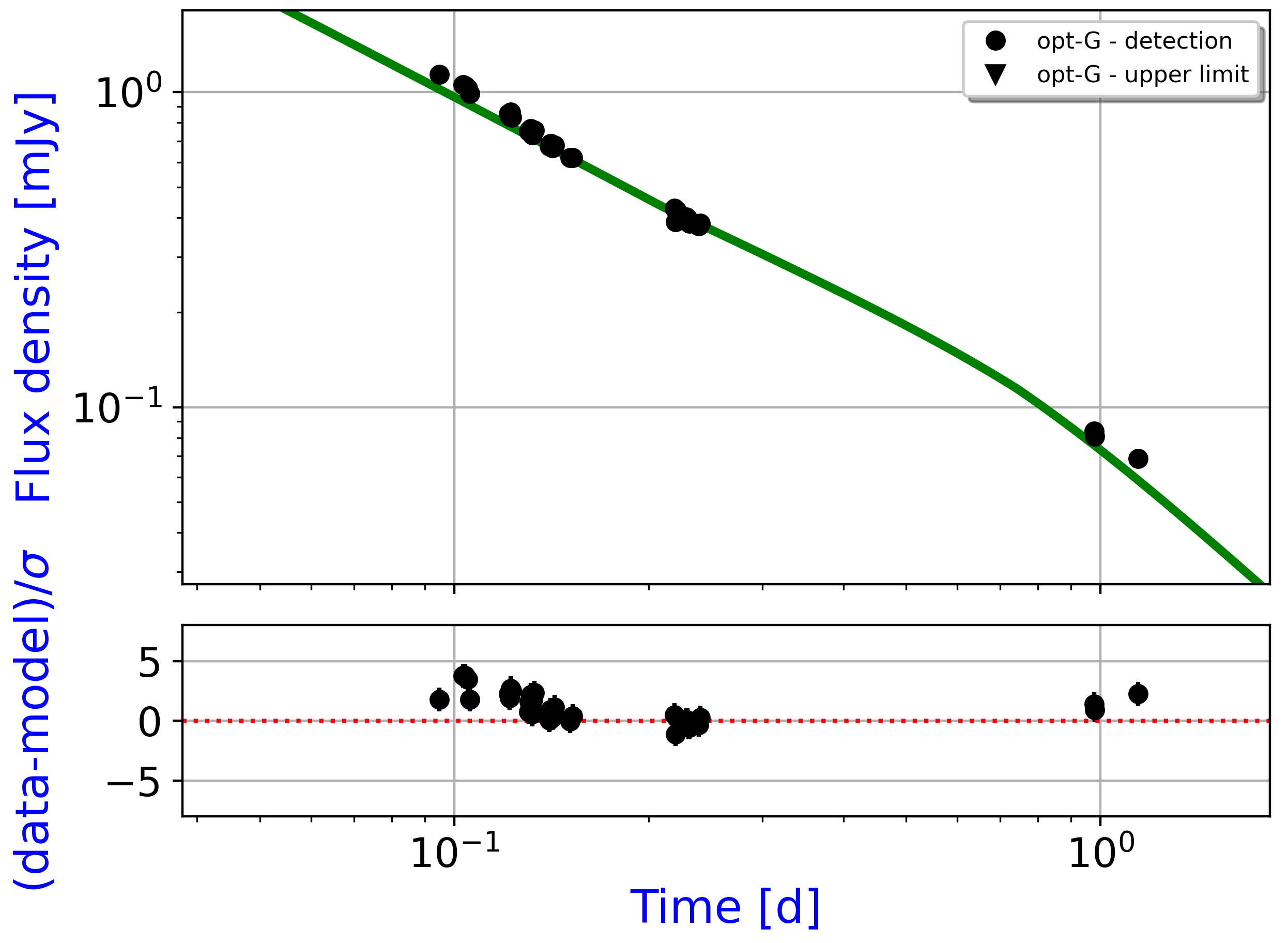}} \\
{\includegraphics[width=72mm]{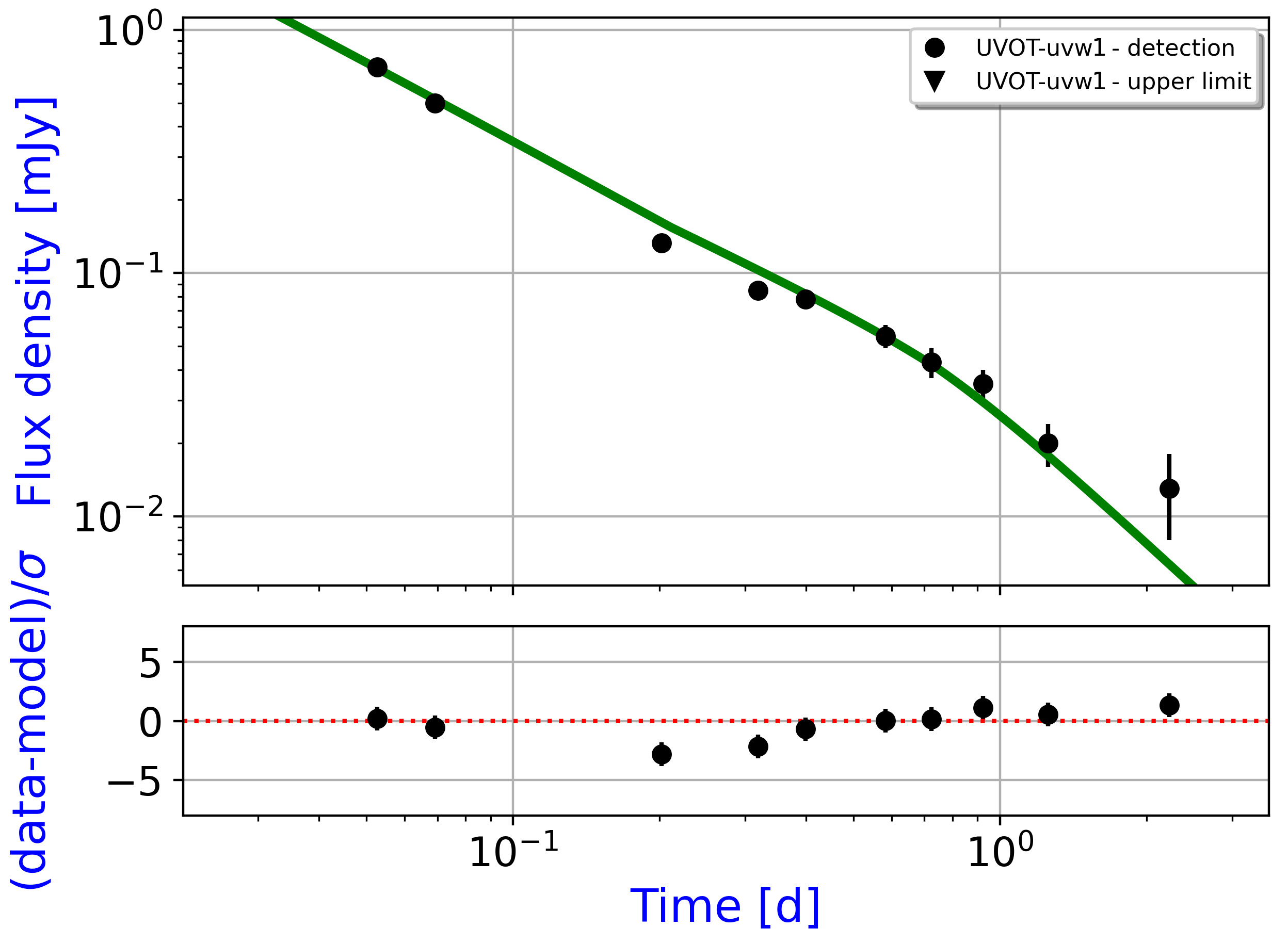}} \quad
{\includegraphics[width=72mm]{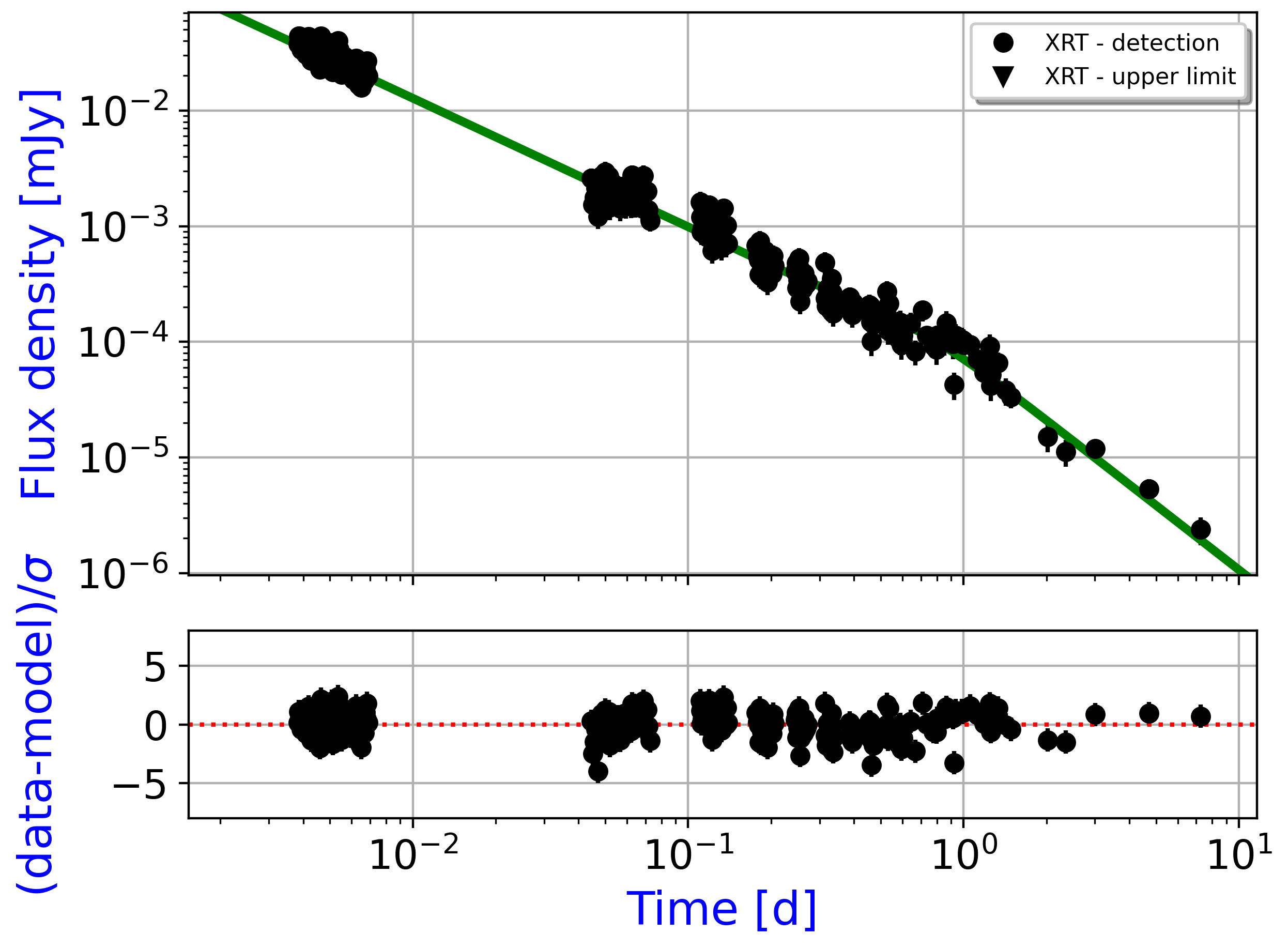}} \\
\caption{Light curves of GRB\,160131A in the UVOIR/X-rays domain at i'-filter ($4.03 \times 10^{14}$, top left), g'-filter ($6.47 \times 10^{14}$, top right), UV/uvw1-filter ($1.15 \times 10^{15}$, bottom left), and X-ray frequency ($6.65 \times 10^{17}$, bottom right), referred to the broadband modelling from optical to X-ray frequencies (Table~\ref{tab:optjet_res_ism}, second column), displayed in Fig.~\ref{fig:lc_uvoirx_saga}.
The bottom panel of each light curve corresponds to the residuals of the fit.
See the caption of Fig.~\ref{fig:lc_uvoir_saga} for a full description of the modelling.
Filled circles indicate detections, upside down triangles indicate $3\sigma$ upper limits, and green lines show the resulting model.}
\label{fig:zoom_saga_uvoirx}
\end{figure*}
\begin{figure*} 
\centering
{\includegraphics[width=150mm]{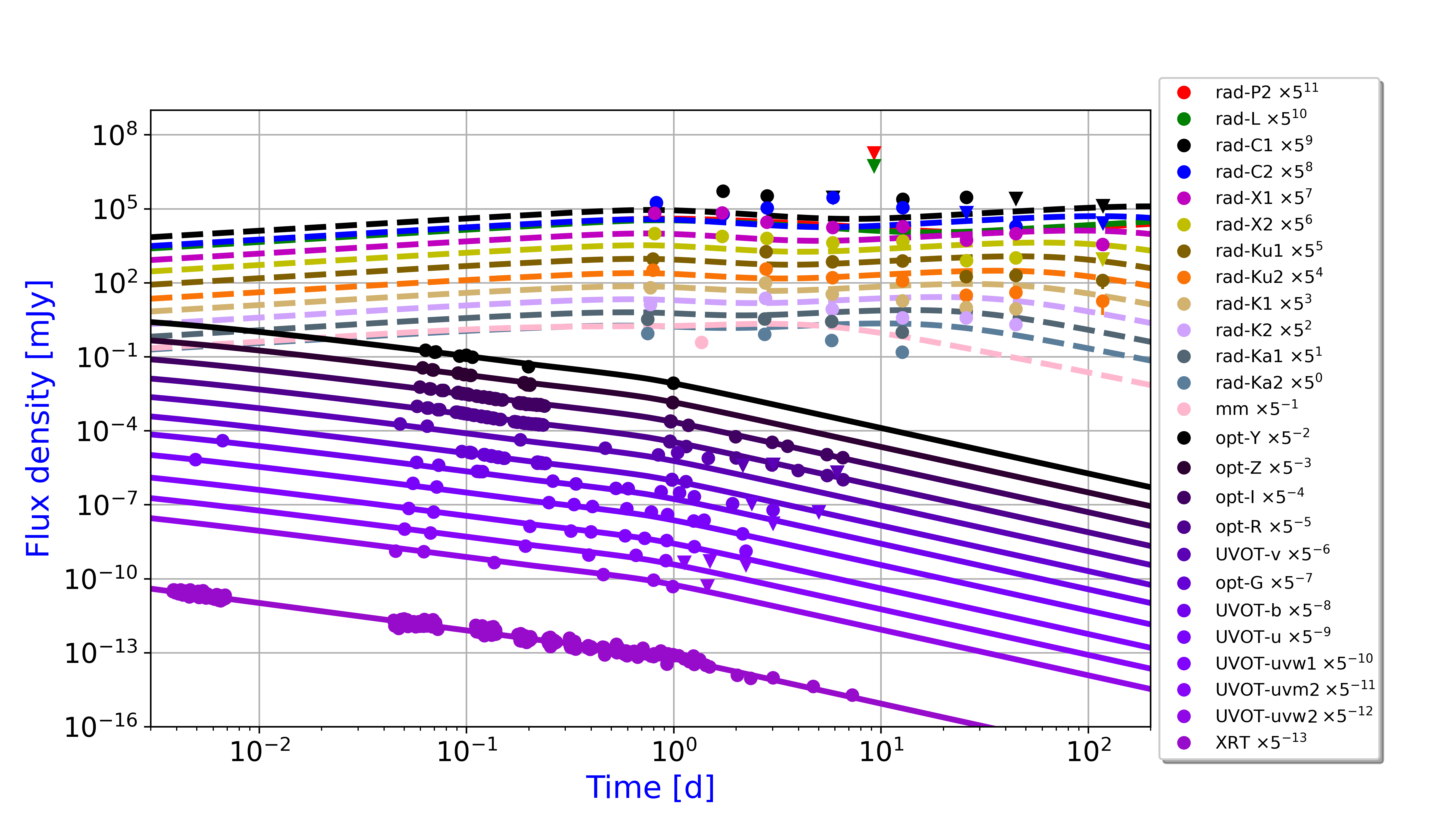}}
\caption{Broadband modelling (from optical to X-ray frequencies; Table~\ref{tab:optjet_res_ism}, second column) of GRB\,160131A.
See the caption of Fig.~\ref{fig:lc_uvoir_saga} for a full description of the modelling.
Filled circles indicate detections, and downward triangles indicate $3\sigma$ upper limits.
For completeness we include all the radio data (dashed lines, not modelled in this approach) and relative light curves (derived from optical/X-ray modelling).}
\label{fig:lc_optuvx_saga}
\end{figure*}

Our results (Table~\ref{tab:optjet_res_ism}) show that the spectrum is in fast cooling until $t_{trans} \sim 0.02$~d and the NR regime occurs at $\sim 300$~d; the cooling due to IC scattering is negligible because of the very low Compton y-parameter ($0.02$).

\begin{figure*} 
\centering
{\includegraphics[width=110mm]{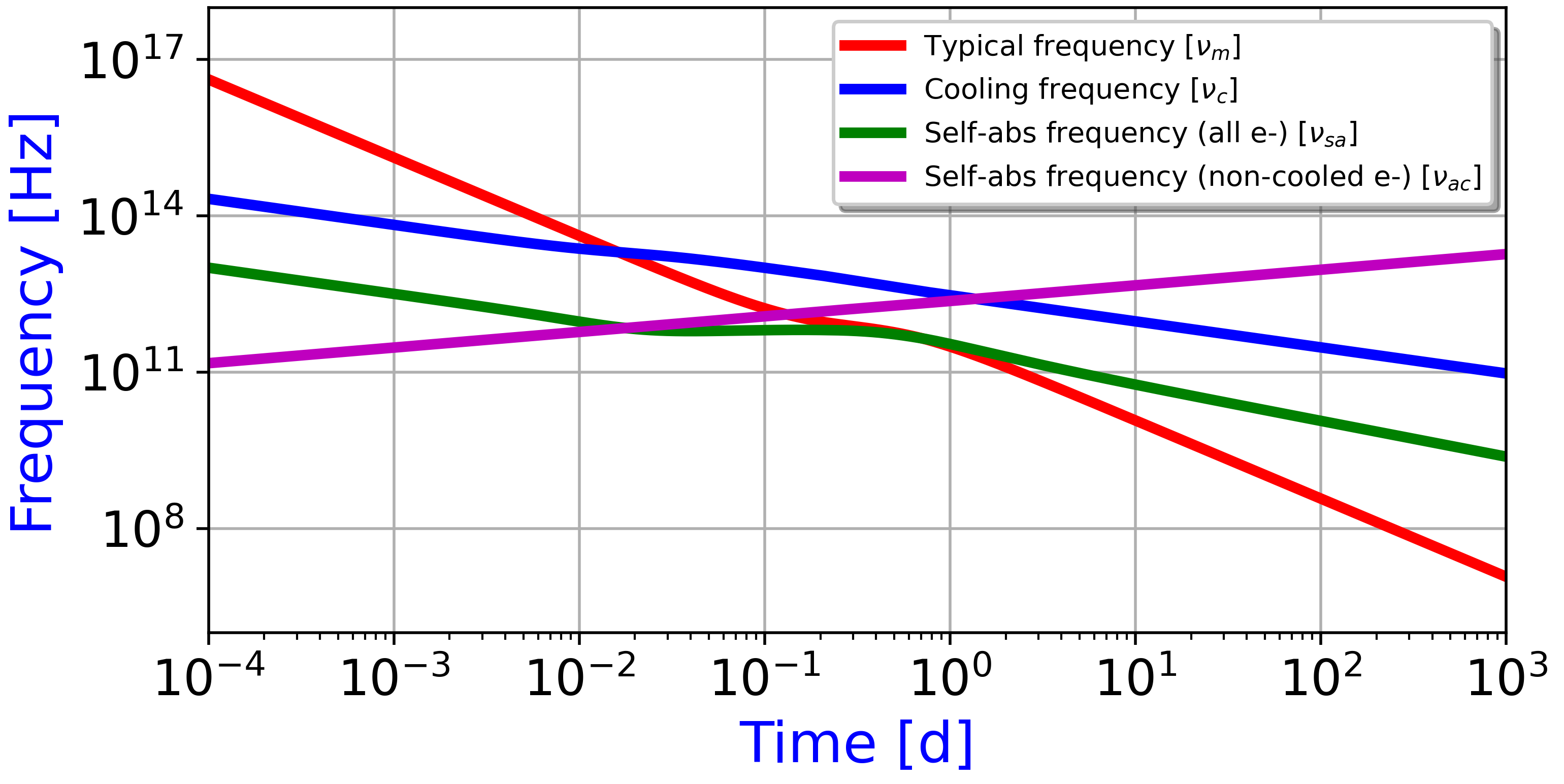}}
\caption{Temporal evolution of the synchrotron break frequencies for afterglow emission of GRB\,160131A, based on analysis of UVOIR/X-ray data (Table~\ref{tab:optjet_res_ism}, second column).
See the caption of Fig.~\ref{fig:lc_uvoir_saga} for a full description of the modelling.
The self-absorption frequency produced by noncooled electrons $\nu_{ac}$ makes sense only in fast-cooling regime ($\lesssim 0.02$~d).}
\label{fig:break_nu_optx_saga}
\end{figure*}

{\sc sAGa} also estimates the behaviour of the synchrotron break frequency over time (Fig.~\ref{fig:break_nu_optx_saga}).
With reference to the lines of reasoning argued in Sect.~\ref{par:breaks} ($\beta_{he} = -1.09$ suggests that $\nu_m$ and $\nu_c$ must lie in the same spectral regime below $\nu_{opt,X}$ at $t_{obs,0} \sim = 10^{-3}$~d), the temporal evolution of $\nu_c$ and $\nu_m$ are in accordance with {\sc sAGa} results (Fig.~\ref{fig:break_nu_optx_saga}).
On the other hand, with reference to what was argued in Sect.~\ref{par:radiolc} (radio SEDs suggest that $\nu_{sa} \sim 7$~GHz until $t_{obs} \sim 13$~d), the temporal evolution of $\nu_{sa}$ (Fig.~\ref{fig:break_nu_optx_saga}) is incompatible with {\sc sAGa} results ($\nu_{sa} \sim 100$~GHz at $t_{obs} \sim 13$~d), caused by the lack of radio data in the optical/X-ray analysis.

For completeness, Fig.~\ref{fig:zoom_saga_uvoirx} shows light curve at some observing frequency, and Fig.~\ref{fig:lc_optuvx_saga} shows also all the radio data (dashed lines, not included in this part of the modelling) with the predicted SEDs in this domain obtained from modelling of optical/X-ray data; these data do not match the high-energy sample, as we show and discuss in the next section.
\begin{table*}
\caption{Summary statistics from MCMC analysis obtained with {\sc sAGa} applied to the visible and UV data of GRB\,160131A for a model based on a jetted (edge-regime) FS emission with optical absorption and energy injection, in ISM-like CBM.
$t_{trans,51}$ and $t_{trans,12}$ indicate the transition time between FS spectral regimes ($5 \rightarrow 1$ and $1 \rightarrow 2$, respectively) as described in GS02; $\chi^2_r$ indicates the reduced chi-squared of the best-fit model.
}
\label{tab:optjet_res_ism}
\centering
\begin{tabular}{l | c | ccc}
\hline\hline
Parameter                   & Unit          & UVOIR                                   & UVOIR/X-Ray                          & Radio/X                                 \\
\hline
$p$                         & -             & $2.20^{+0.07}_{-0.04}$                  & $2.14^{+0.02}_{-0.01}$               & $2.20$\tablefootmark{a}                 \\
$\epsilon_e$                & -             & $(1.1 \pm 0.2) \times 10^{-2}$          & $(1.3^{+0.3}_{-0.2}) \times 10^{-2}$ & $(3.4^{+0.5}_{-0.2}) \times 10^{-2}$    \\
$\epsilon_B$                & -             & $(1.5^{+1.2}_{-0.9}) \times 10^{-1}$    & $(9.3^{+9.0}_{-5.1}) \times 10^{-2}$ & $(1.5 \pm 0.3) \times 10^{-3}$          \\
$n_0$                       & cm$^{-3}$     & $8.4^{+20.5}_{-5.9}$                    & $10.7^{+12.8}_{-6.4}$                & $(3.6^{+2.7}_{-0.8}) \times 10^1$       \\
$E_{52}$                    & $10^{52}$~erg & $(4.4^{+1.1}_{-0.8}) \times 10$         & $(4.9^{+0.9}_{-0.8}) \times 10$      & $(1.2^{+0.1}_{-0.2}) \times 10$         \\
$A_v$                       & mag           & $(1.1^{+0.5}_{-0.6}) \times 10^{-1}$    & $(1.8 \pm 0.4) \times 10^{-1}$       & $0.2 \pm 0.1$                           \\
$t_j$                       & d             & $0.9^{+0.2}_{-0.1}$                     & $0.82^{+0.03}_{-0.02}$               & $0.9 \pm 0.1$                           \\
$\theta_j$                  & deg           & $5.6^{+0.9}_{-0.8}$                     & $5.6^{+0.6}_{-0.7}$                  & $7.7^{+0.7}_{-0.3}$                     \\
$t_{NR}$                    & d             & $(3.0^{+0.9}_{-0.8}) \times 10^{2}$     & $(2.7^{+1.1}_{-0.7}) \times 10^2$    & $(1.2^{+0.1}_{-0.3}) \times 10^{2}$     \\
$t_{b,0}$                   & d             & $(1.99^{+0.03}_{-0.07}) \times 10^{-1}$ & $(2.09 \pm 0.01) \times 10^{-1}$     & $(2.10^{+0.01}_{-0.03}) \times 10^{-1}$ \\
$m$                         & -             & $0.181 \pm 0.002$                       & $0.120 \pm 0.002$                    & $(5.03^{+0.05}_{-0.02}) \times 10^{-2}$ \\
\hline
$\nu_m$\tablefootmark{b}    & Hz            & $4.2 \times 10^{11}$                    & $3.0 \times 10^{11}$                 & $1.7 \times 10^{11}$                    \\
$\nu_c$\tablefootmark{b}    & Hz            & $1.8 \times 10^{12}$                    & $3.1 \times 10^{12}$                 & $5.9 \times 10^{14}$                    \\
$\nu_{sa}$\tablefootmark{b} & Hz            & $4.0 \times 10^{11}$                    & $3.4 \times 10^{11}$                 & $1.5 \times 10^{11}$                    \\
$\nu_{ac}$\tablefootmark{b} & Hz            & $1.4 \times 10^{12}$                    & $2.3 \times 10^{12}$                 & $2.9 \times 10^{12}$                    \\
$t_{trans,51}$              & d             & $5.6 \times 10^{-2}$                    & $1.8 \times 10^{-2}$                 & $9.2 \times 10^{-5}$                    \\
$t_{trans,12}$              & d             & $0.8$                                   & $0.5$                                & $1.3$                                   \\
\hline
$\chi^2_r$                  & -             & $1.22$                                  & $0.97$                               & $10.97$                                 \\
\hline
\end{tabular}
\tablefoot{
\tablefoottext{a}{Fixed.}                        \\
\tablefoottext{b}{Measured at $t_{obs} = 1$~d.}  \\
}
\end{table*}

\subsubsection{From radio to X-ray frequencies}
\label{par:tutto_res}

The addition of the radio/mm data set from $0.6$ to $92.5$~GHz does not include the data points affected by the bumps (Sect.~\ref{par:radio}), because the best-fit model with all the radio data set was very bad ($\chi^2_r > 30$).

To verify the stability and robustness of the best-fit solution, we repeated the analysis assuming three different starting values for $p$ ($2.1$, $2.4$, $2.9$); we obtained $p \sim 2$, lower than that estimated from the high-energy approach (Sect.~\ref{par:breaks}), but compatible with the analysis of the radio SEDs in empirical approach (Sect.~\ref{par:radiosed}).
The bad modelling of these three analyses ($\chi^2_r > 20$) contributed to consider a fixed value of $p$ ($2.2$, according to the high-energy approach; Sect.~\ref{par:breaks}) as a compromise.

\begin{figure*} 
\centering
{\includegraphics[width=150mm]{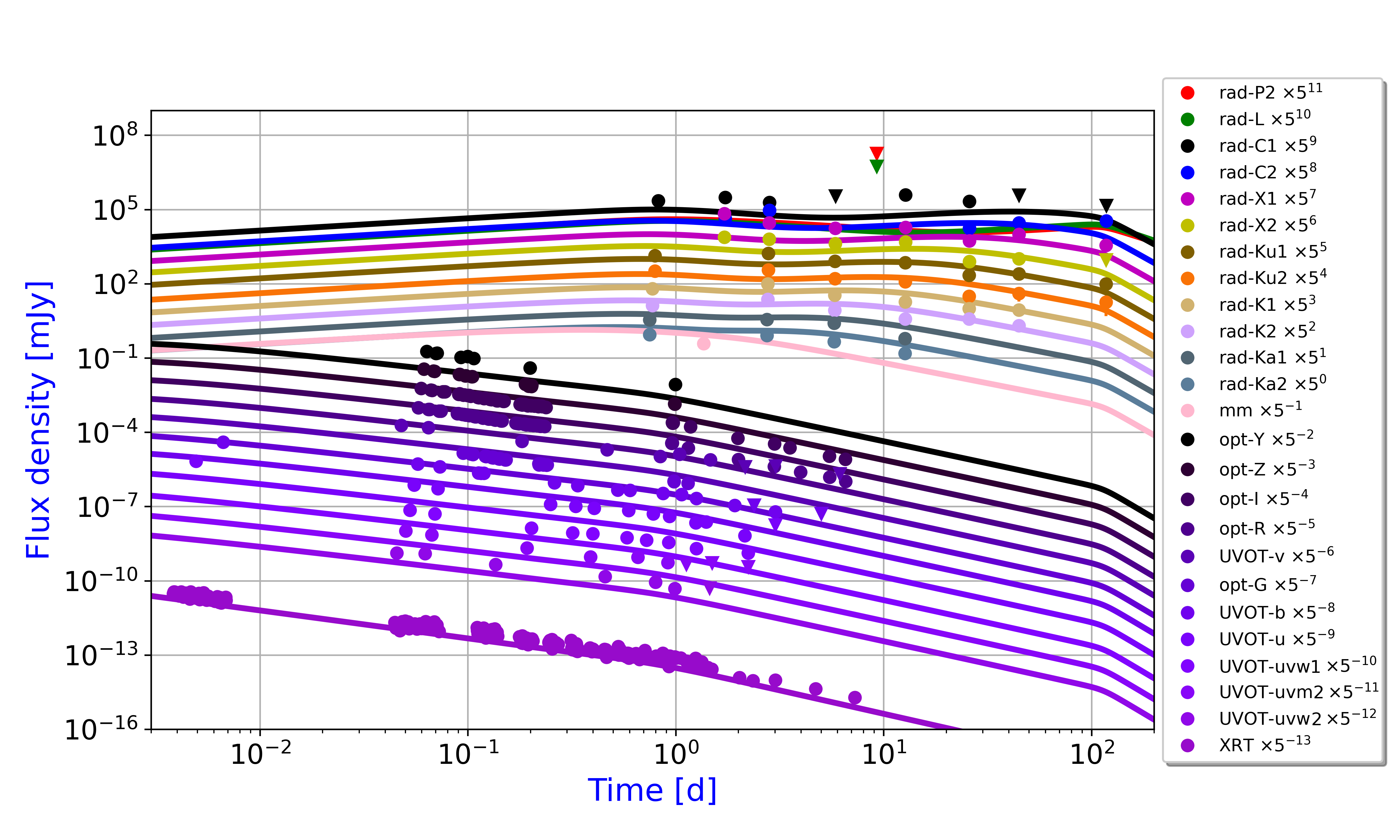}}
\caption{Broadband modelling of GRB\,160131A from radio to X-ray frequencies (Table~\ref{tab:optjet_res_ism}, third column).
See the caption of Fig.~\ref{fig:lc_uvoir_saga} for a full description of the modelling.
Filled circles indicate detections, and downward triangles indicate $3\sigma$ upper limits.
}
\label{fig:lc_radoptuvx_saga}
\end{figure*}

\begin{figure*} 
\centering
{\includegraphics[width=11cm]{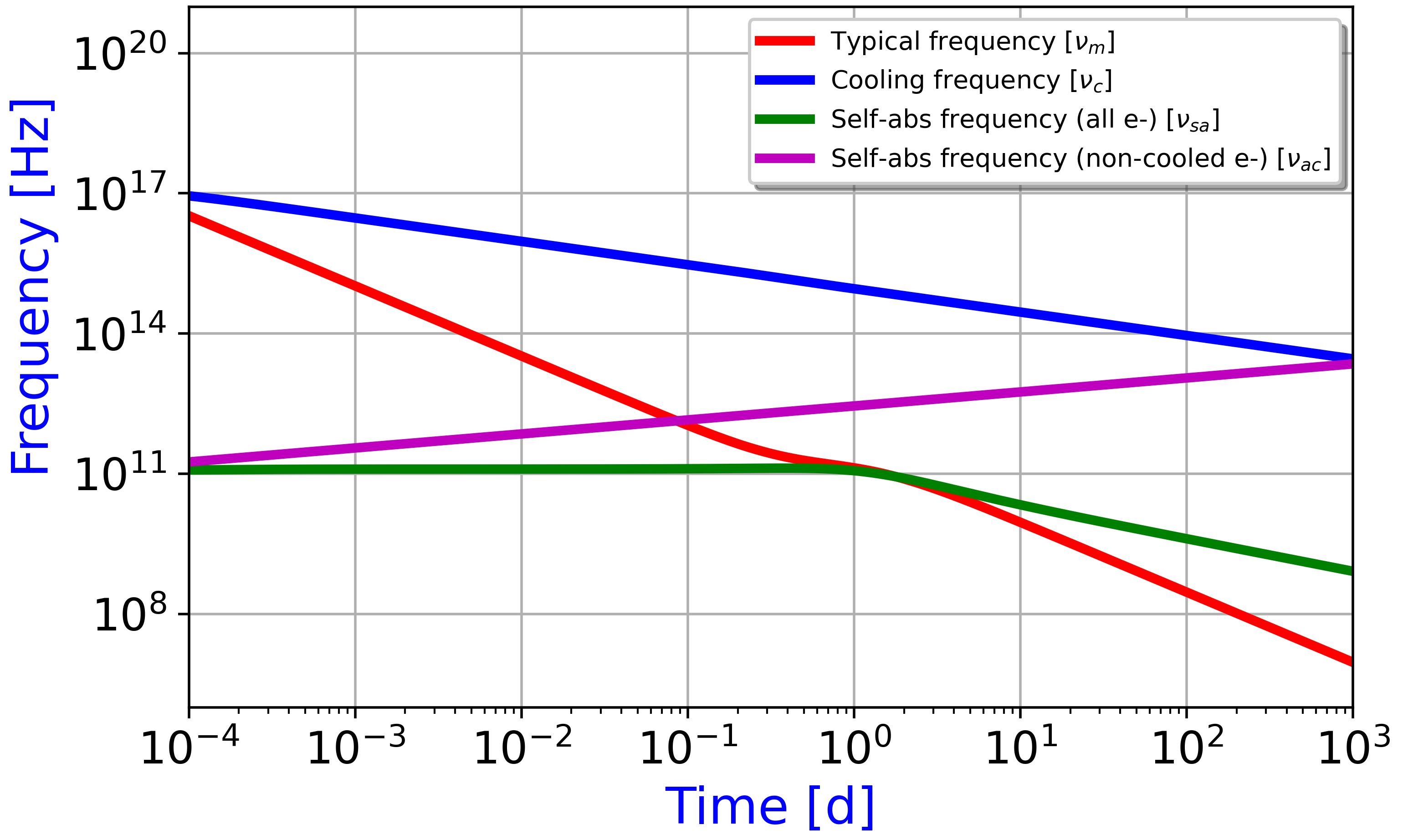}}
\caption{Temporal evolution of the synchrotron break frequencies for afterglow emission of GRB\,160131A, based on analysis of broadband data (from radio to X-ray frequencies; Table~\ref{tab:optjet_res_ism}, third column).
See the caption of Fig.~\ref{fig:lc_uvoir_saga} for a full description of the modelling.
The self-absorption frequency produced by noncooled electrons $\nu_{ac}$ makes sense only in fast-cooling regime ($\lesssim 9 \times 10^{-5}$~d).
}
\label{fig:break_nu_all_saga}
\end{figure*}

Unsurprisingly, the best-fit model has a very high $\chi^2_r$ ($\sim 10$; Table~\ref{tab:optjet_res_ism}, third column).
This is indicative of the problems faced by the standard GRB afterglow model, common in cases when a rich data set at low frequencies is available (Fig.~\ref{fig:lc_radoptuvx_saga}).
Our results (Table~\ref{tab:optjet_res_ism}, third column) show that the jet break time of $0.9$~d translates into a jet opening angle $\theta_{j} \sim 8$~degrees, $t_{trans} \sim 9 \times 10^{-5}$~d, and the NR regime occurs at $\sim 120$~d.
Moreover, Fig.~\ref{fig:lc_radoptuvx_saga} shows that the model is well suited only for radio (except for $\nu \lesssim 10$~GHz) domains, partially well at X-ray frequencies, and poorly in the optical band.
This behaviour suggests other radiation mechanisms responsible for the afterglow emission for GRB\,160131A.
As in the case of the analysis of optical/X-ray data (Sect.~\ref{par:opt_res}), the Compton y-parameter is $0.02$, indicating that cooling due to IC scattering is negligible.
The temporal evolution of the cooling frequency $\nu_c$ (Fig.~\ref{fig:break_nu_all_saga}) suggests that it lies above the X-rays (as opposed to $\nu_{sa}$ and $\nu_m$), in contrast with the behaviour expected from empirical considerations based on the optical/X-ray spectra (Sect.~\ref{par:breaks}).

\section{Discussion}
\label{sec:disc}

The addition of radio data set in the afterglow modelling considerably complicates the broadband analysis, challenging the standard GRB afterglow model.

We point out three problematic features at radio frequencies:
\begin{enumerate}
\item the presence of the same rather constant peak at $\sim 8$~GHz in SEDs up to $\sim 25$~d, whose width $\Delta \nu / \nu$ evolves from $\sim 0.5$ at $1.7$~d to $\sim 0.1$ at $\sim 25$~d, with a temporary disappearance at $\sim 2.7$~d (Fig.~\ref{fig:radio sed_1}).
\item the SED at $5.8$~d evolves with $\beta \sim 0.7$ (Table~\ref{tab:seds_radio} and Fig.~\ref{fig:radio sed_1}), a value which is incompatible with slow cooling regimes for FS emission (Sect.~\ref{par:breaks}).
\item flux densities at low frequencies ($\lesssim 7$~GHz) seem to be constant over time (Figs.~\ref{fig:radio sed_1} and \ref{fig:radio sed_multi}).
\end{enumerate}

Our results suggest that radio data could hardly be accounted for along with the optical/X-ray data within the framework of the standard GRB afterglow model.
This is not unprecedented: for example, \citet{Kangas19} reported the lack of detectable jet breaks in the radio light curves of a sample of $15$ GRB afterglows, whereas X-rays seem to support it.
However, we underline that they (1) considered only one spectral regime (5-1-2) of afterglow emission in GS02, (2) assumed the sideways expansion for jetted emission, and (3) ignored any observed rise period of the light curve and any early features attributed to flares, plateau or RS in the literature.
They interpret the long-lasting single power-law decline of the radio emission in terms of a two-component jet.

There are other possible assumptions that might not necessarily hold true for the afterglow of GRB\,160131A: (1) constant micro-physics parameters, in the light of evidence of the temporal evolution of the micro-physics parameters in the afterglow of GRB\,190114C \citep{Misra19}, (2) unique CBM, as in the case of evidence of the transition from a wind-like to ISM-like CBM in the afterglow of GRB\,140423A \citep{Li20}, and (3) uniform jet model, in the light of evidence of other jet models used to interpret the broadband data for several GRB afterglows, such as the structured jet model (e.g. \citealp{DeColle12,Granot18,Alexander18,Coughlin20_jet}), two-component jet (e.g. \citealp{Berger03,Peng05,Racusin08,Liu11,Holland12}) and other more complex regimes (e.g. \citealp{Huang04,Wu05,Granot18}).
In the latest years growing evidence has been found that favours the structured jet\footnote{Recently, the open-source Python package {\sc afterglowpy} became available for on-the-fly computation of structured jet afterglows with arbitrary viewing angle \citep{Ryan20}.}, as in the case of the GRB\,170817A associated to GW\,170817 \citep{Alexander18}.

\subsection{Energy injection}
\label{par:ene_inj_grb}

A flattening in the optical/X-rays light curves prior to $0.8$~d of GRB\,160131A could call for energy injection.
Nothing can be inferred in this regard from radio data, which were taken starting from $\sim 1$~d.

\begin{figure} 
\centering
{\includegraphics[width=9cm]{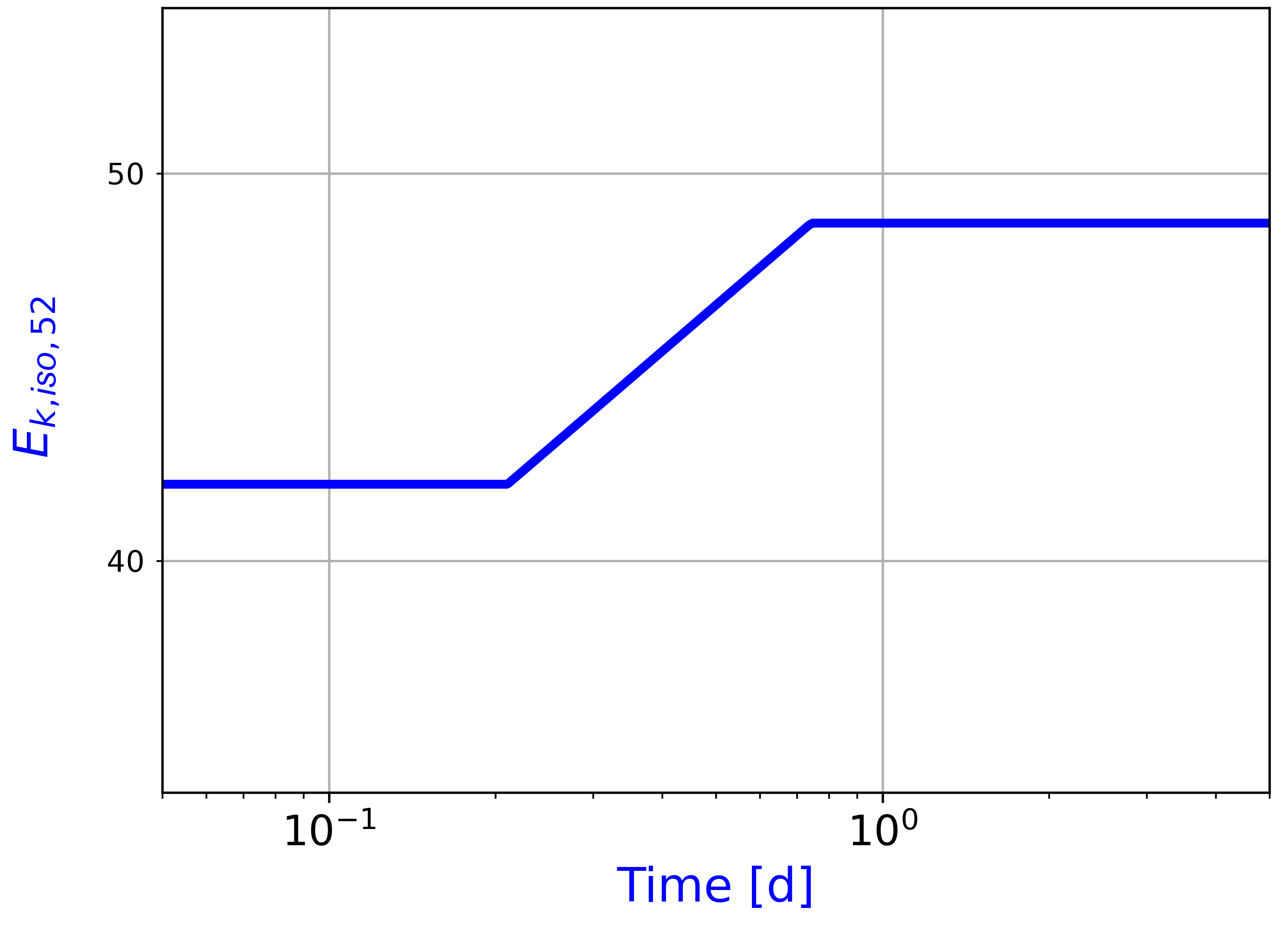}}
\caption{Isotropic equivalent kinetic energy $E_{k,iso,52}$ (in units of $10^{52}$~erg) as a function of time, as determined from modelling of the optical/X-ray data set (Table~\ref{tab:optjet_res_ism}, second column).}
\label{fig:kin_ene}
\end{figure}

In the energy injection approach (Sect.~\ref{app_saga}), the inferred value $p \sim 2.2$ (Sect.~\ref{par:results}) suggests $\nu_c<\nu_X$, where the flux density is $F_{\nu > \nu_c} \propto E_{k,iso,52}^{(2+p)/4} t^{(2-3p)/4}$ (GS02); in this regime we obtain $F_{\nu > \nu_c} \propto E_{k,iso,52}^{1.1} t^{-1.3}$.
The temporal evolution of the injected energy is parameterised as $E \propto t^m$, and hence $F_{\nu > \nu_c} \propto t^{1.1m - 1.3}$.
Fitting the X-ray light curve with a power-law from $\sim 0.2$~d to $\sim 0.8$~d, roughly corresponding to the flattening, we obtain $\alpha_{X,ei} = -1.0 \pm 0.2$; apparently, this temporal decay index apparently does not call for the energy injection effect in the modelling, but the addition of the UVOIR data set in the broadband modelling necessarily invokes this effect.
In the energy injection approach, the value of $\alpha_{X,ei}$ implies $m = 0.27 \pm 0.20$, or, equivalently, $q = 1 - m = 0.73 \pm 0.20$.
This conclusion is perfectly compatible with our optical/X-ray modelling (Sect.~\ref{par:opt_res} and Table~\ref{tab:optjet_res_ism}, first and second column), where we adopted the energy injection approach (Sect.~\ref{app_saga}); in particular, we obtained an increasing $E_{k,iso,52}$ from $\sim 4.2 \times 10^{53}$~erg to $\sim 4.9 \times 10^{53}$~erg (Fig.~\ref{fig:kin_ene}).
A similar energy injection process was discussed for GRB\,100418A, for which it was found $m \sim 0.7$ \citep{Marshall11,Laskar15}.

As we can see in Fig.~\ref{fig:ene_in}, X-rays light curve shows a less pronounced flattening with respect to optical light curves.
This unusual light curve was also observed with GRB\,090102 \citep{Gendre10}, where the optical flattening could then be interpreted as (1) a change of the CBM (e.g. \citealp{RamirezRuiz01b,Chevalier04}), and (2) a normal fireball expanding in an ISM, with a RS component (the lack of radio data does not corroborate this assumption).
Another similar feature is present in GRB\,060908 \citep{Covino10}, where it is possible to model the optical and X-ray afterglow independently, but the multi-frequency spectral and temporal data challenge available theoretical scenarios.
The broadband modelling of the afterglow of the ultra-long duration GRB\,111209A \citep{Kann18} shows a strong chromatic rebrightening in the optical domain, modelled with a two-component jet; the late afterglow also shows several smaller, achromatic rebrightenings, which are likely to be energy injections.

\subsection{The possible role of ISS in the multi-component radio SEDs}
\label{par:reverse}

The evidence of the multi-component SEDs at radio frequencies (A, B, and C; Sect.~\ref{par:radiosed}) suggests further radiation mechanisms for the GRB afterglow in addition to the continuum associated with FS emission.

\begin{figure*} 
\centering
{\includegraphics[width=75mm]{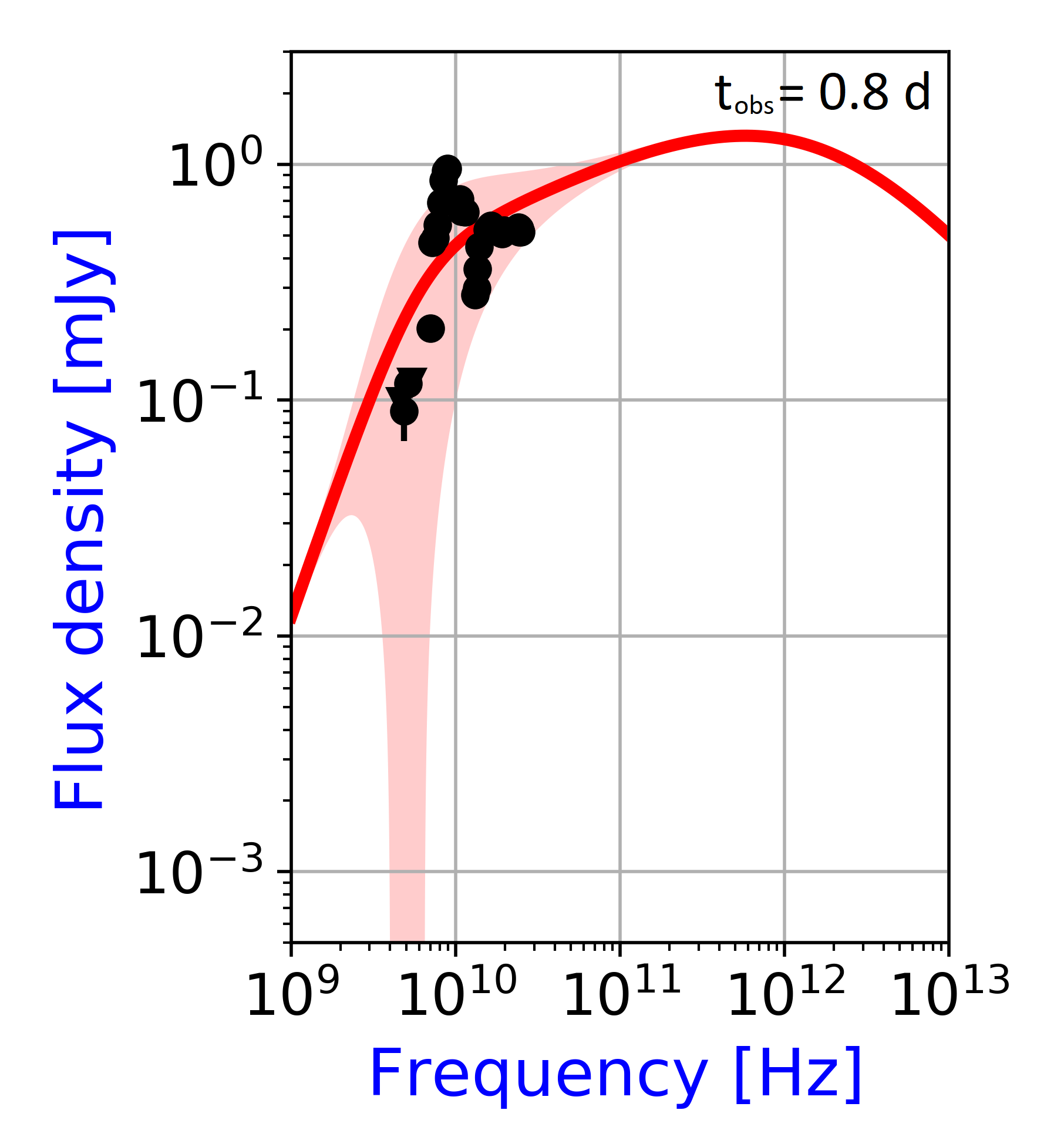}} \quad
{\includegraphics[width=75mm]{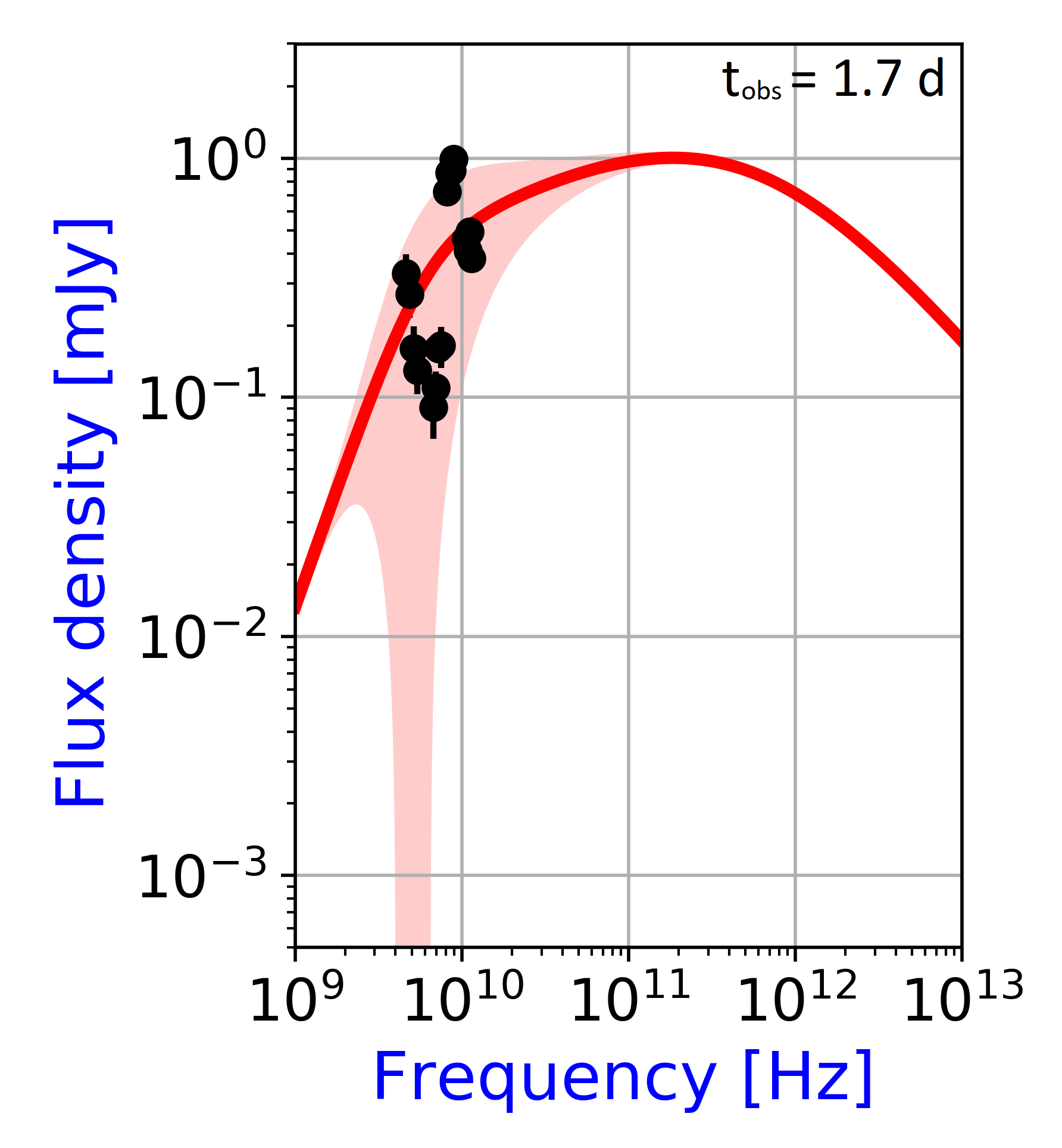}} \\
{\includegraphics[width=75mm]{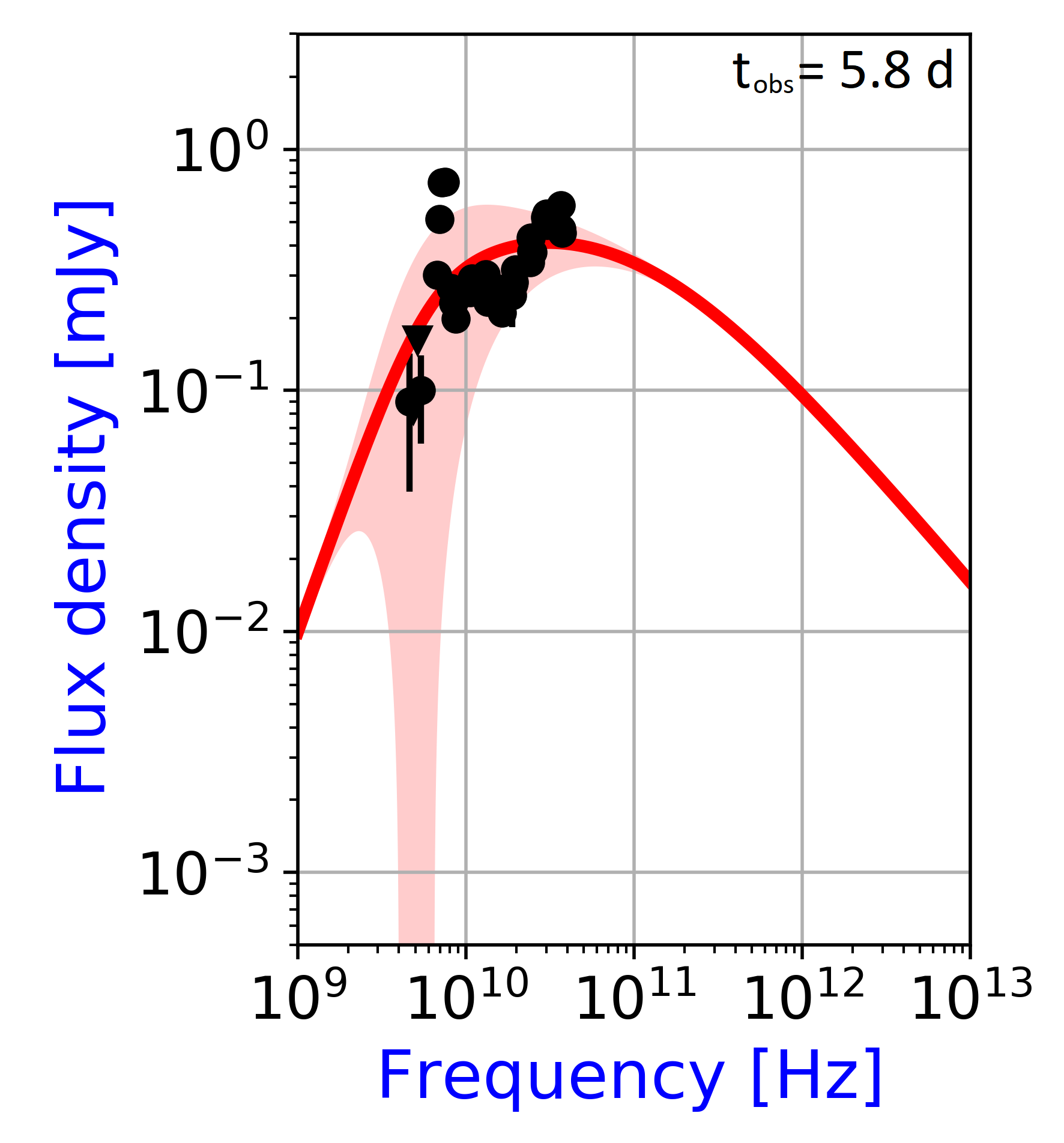}} \quad
{\includegraphics[width=75mm]{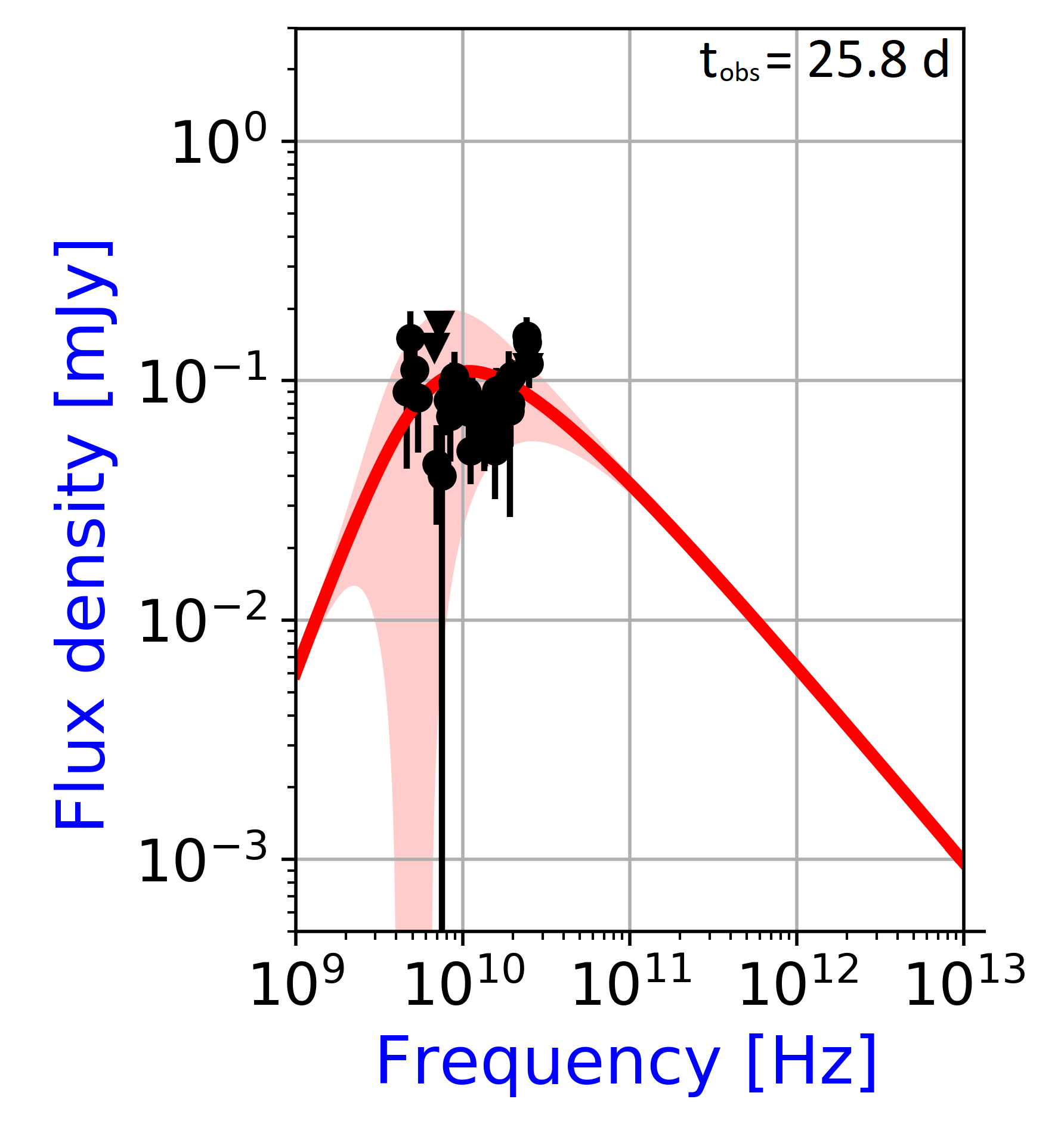}} \\
\caption{Radio SEDs of GRB\,160131A at $0.8$~d (top left), $1.7$~d (top right), $5.8$~d (bottom left), and $25.8$~d (bottom right), obtained through a radio modelling for a FS model in ISM; we considered a jetted (edge-regime) emission with ISS effect.
Filled circles indicate detections, and upside down triangles indicate $3\sigma$ upper limits; the red shaded regions represent the expected variability due to ISS effect, obtained through the prescription described in \citet{Misra19}.}
\label{fig:radio_scint_sp}
\end{figure*}

\begin{figure*} 
\centering
{\includegraphics[width=78mm]{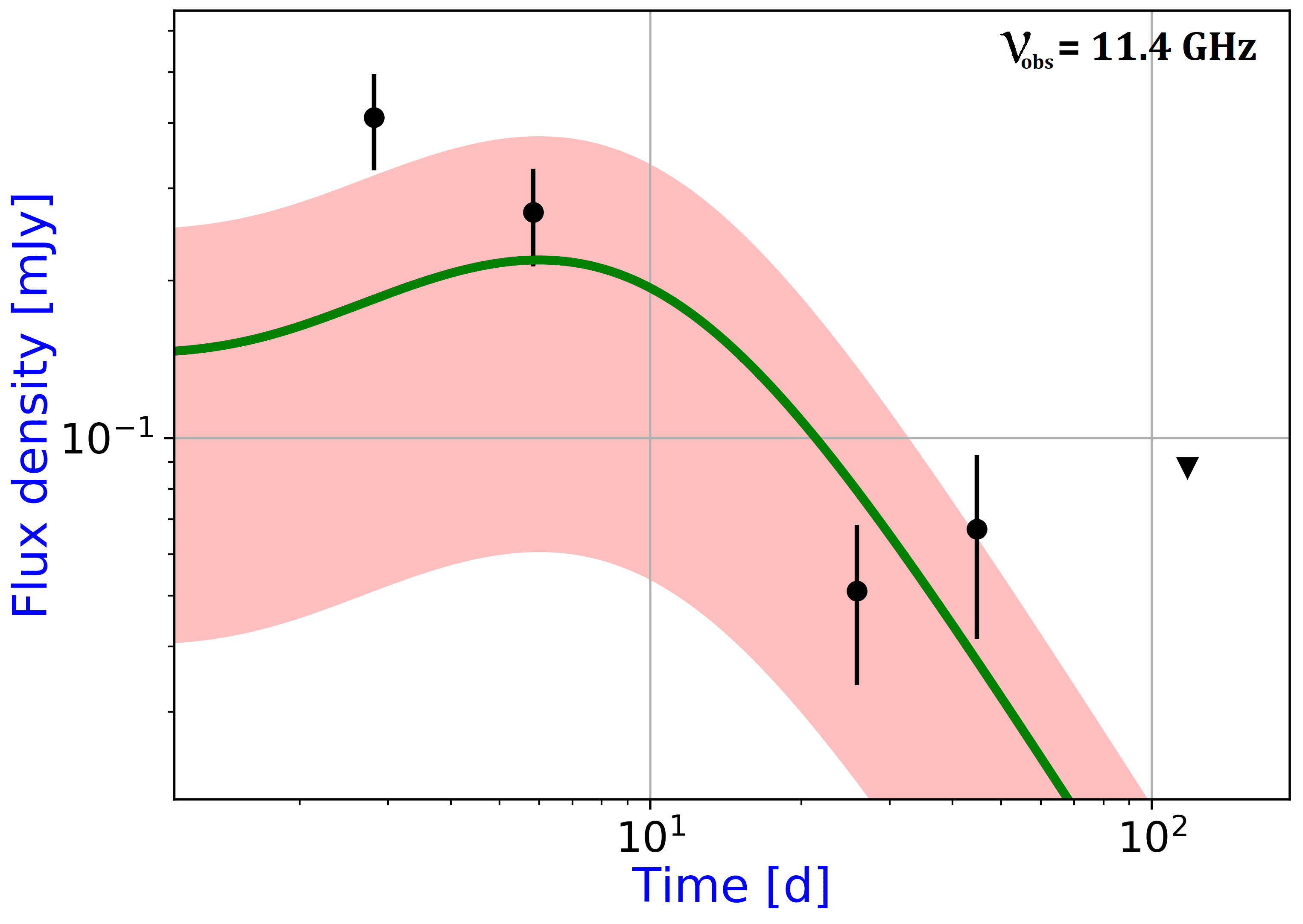}} \quad
{\includegraphics[width=78mm]{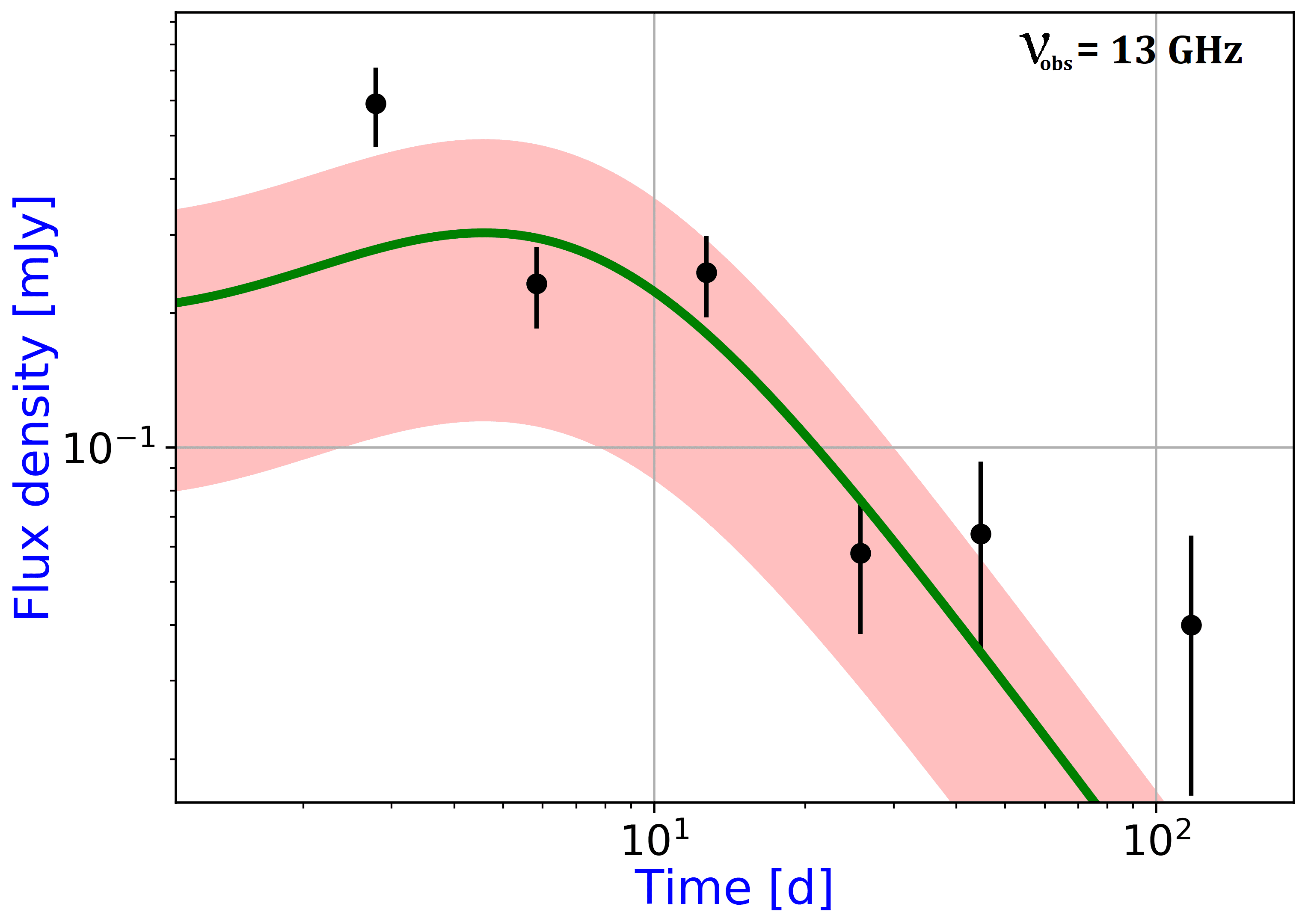}} \\
{\includegraphics[width=78mm]{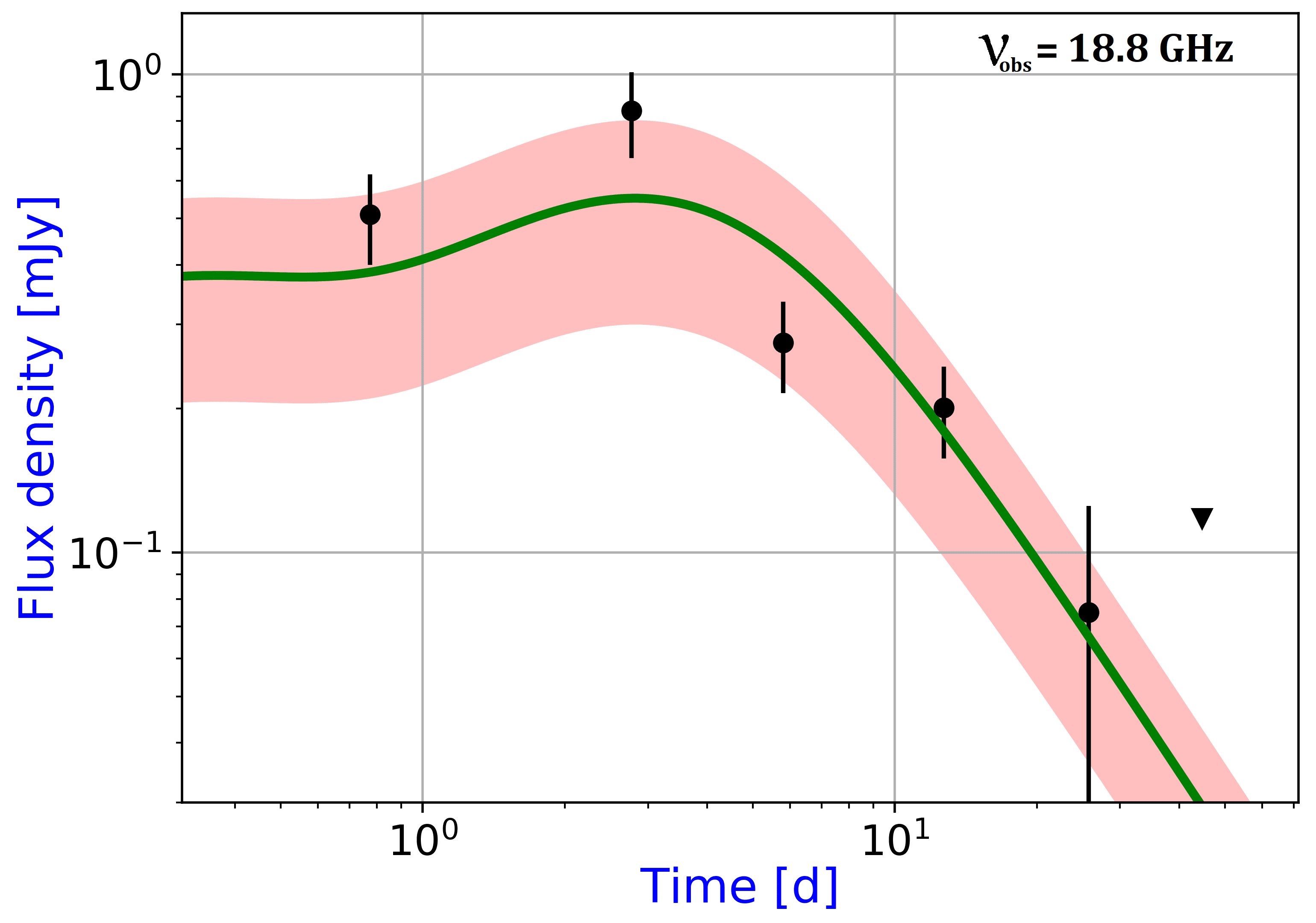}} \quad
{\includegraphics[width=78mm]{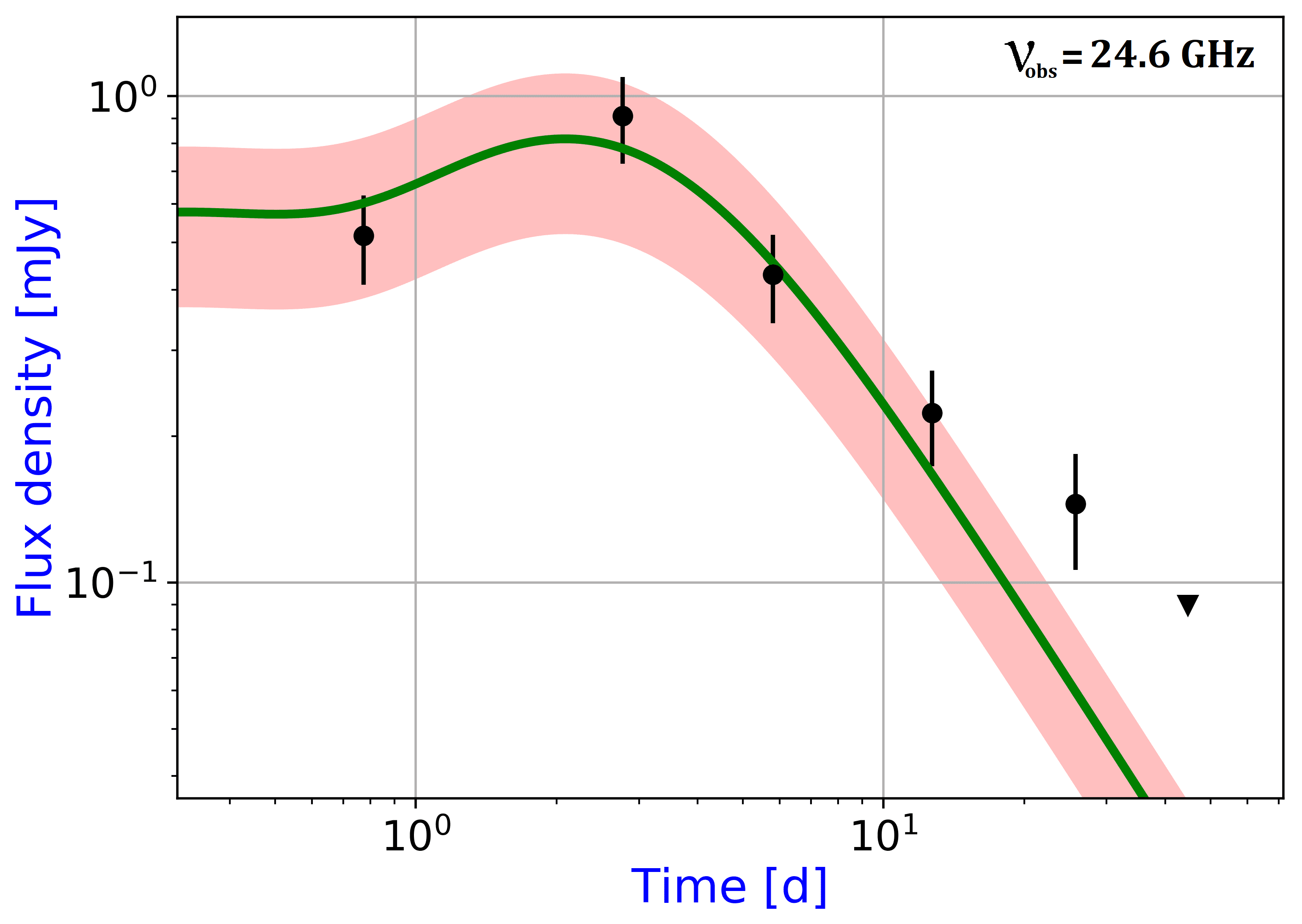}} \\
\caption{Radio light curves of GRB\,160131A at $11.4$~GHz (top left), $13$~GHz (top right), $18.8$~GHz (bottom left), and $24.6$~GHz (bottom right), obtained through a radio modelling (from radio to X-ray frequencies) for a FS model in ISM; we considered a jetted (edge-regime) emission with ISS effect, dust extinction and energy injection.
Filled circles indicate detections, and upside down triangles indicate $3\sigma$ upper limits; the red shaded regions represent the expected variability due to ISS effect, obtained through the prescription described in \citet{Misra19}.}
\label{fig:radio_scint_lc}
\end{figure*}

The presence of peaks in radio SEDs had already been observed in other sources, and the main candidate to explain this pronounced radio variability is the ISS (or other extreme scattering effects); in particular, VLA SED at $\sim 2$~d of GRB\,130925A \citep{Horesh15} shows a peak at $\sim 8$~GHz, with $\Delta \nu / \nu\sim 0.7$, compatible with our values ($\Delta \nu / \nu \sim 0.1$ -- $0.5$, as observed in Sect.~\ref{sec:disc}).
\citet{Horesh15} suggest that these peaks are well modelled with the ISS emission model in which the emission originates from either mono-energetic electrons or an electron population with an unusually steep power-law energy distribution.
Moreover, thanks to a simple modelling of the radio data set with {\sc sAGa}, we obtained SEDs (Fig.~\ref{fig:radio_scint_sp}) and light curves (Fig.~\ref{fig:radio_scint_lc}) well modelled with the expected variability due to ISS effect (red shaded regions).

Another interpretation for this radio excess at early-times should have been ascribable to the presence of RS in addition to a FS (e.g., \citealt{Gomboc08,Melandri10,Japelj14,Alexander17,Laskar18}), because (1) the RS emission is expected to peak at lower frequencies than the FS, and (2) the RS spectrum is expected to cut off steeply above the RS cooling frequency \citep{Kobayashi00b}.
A recent work \citep{Laskar19b} showed for the first time that within a SED it is possible to disentangle the contributions of RS and of FS in the radio band.
Moreover, the first case of a SED instantaneously and clearly decomposed into RS and FS components (GRB\,181201A, \citealt{Laskar18c}) suggests that an early-time radio peak is consistent with emission from a refreshed RS produced by the violent collision of two shells with different Lorentz factors emitted at different times.
Nevertheless, the peak at lower frequency bands observed in radio SEDs of \citet{Laskar19b}, characterised by $\Delta \nu / \nu \sim 3$, is much broader of our ones ($\Delta \nu / \nu \sim 0.1$ -- $0.5$, as observed in Sect.~\ref{sec:disc}), calling for something else that comes into play in addition to the RS prescription.
This incompatibility is strengthened by the lower limit on $n_0$ estimated with {\sc sAGa} ($n_0 \gtrsim 5$~cm$^{-3}$) and the strong observed correlation -- highlighted in several analyses (e.g., GRB\,160509A in \citealt{Laskar16}, GRB\,161219B in \citealt{Laskar18}, and GRB\,181201A in \citealt{Laskar19b}) -- between broadband detections of RS emission and CBM characterised by low densities (typically $n_0 \lesssim 10^{-2}$~cm$^{-3}$ in ISM-like CBM, and $A_* \lesssim 10^{-2}$ in wind-like CBM).
In the hindsight, these features could have possibly been observed in more sparse radio data sets from past GRBs as well, and erroneously interpreted as RS evidence.

We further rule out the presence of RS emission analysing these peaks in the radio SEDs according to the prescription taken up by \citet{Laskar18c}.
They assume $\nu_{c,rs}$ to be located near each observed radio spectral peak, in order to compute a conservative lower limit to the optical light curve\footnote{Once the RS has crossed the ejecta (timescale of days), the flux above $\nu_{c,rs}$ declines rapidly because no electron is newly accelerated within the ejecta.}.
At radio frequencies, the first spectral peak takes place at $F \approx 0.9$~Jy in X-band ($\sim 9$~GHz) at $0.8$~d (Fig.~\ref{fig:radio sed_1}); following the reasoning about the evolution of $\nu_{c,rs}$, we assume $\nu_{c,rs} \approx 9$~GHz and $F_{\nu,pk} \approx 0.9$~Jy at this epoch.
\begin{itemize}
    \item In the relativistic RS regime, the Y-band ($\sim 3 \times 10^{14}$~Hz) would be crossed by a relativistic RS (ISM) at $t_{pk} \sim 8.5 \times 10^{-4}$~d with $F_{\nu,pk} \sim 730$~Jy ($t_{pk} \sim 3.1 \times 10^{-3}$~d and $F_{\nu,pk} \sim 465$~Jy for wind).
    Unfortunately, there are no optical data at those epochs, and hence we scale $F_{\nu,pk}$ at $t_{pk}$ knowing that the observed Y-band light curve evolves as $\sim t^{-1.25}$ (Sect.~\ref{par:breaks}), obtaining $F_{\nu,pk} \sim 920$~Jy for ISM-like CBM ($F_{\nu,pk} \sim 180$~Jy at $\sim 3.1 \times 10^{-3}$~d for wind-like CBM), incompatible with the relativistic RS regime.
    \item In the Newtonian RS approach, for the same spectral peak we obtain the passage of $\nu_{c,rs}$ in Y-band (1) in the range $\approx (1.7$ -- $0.5) \times 10^{-3}$~d (corresponding to $F_{\nu,pk} \sim 450$ -- $728$~Jy) for ISM-like CBM, and (2) in the range $\approx (8.6$ -- $1.7) \times 10^{-3}$~d (corresponding to $F_{\nu,pk} \sim 260$ -- $450$~Jy) for wind-like CBM.
    Also in this case, there are no optical data at those epochs to verify this assumption; the observed Y-band light curve evolves as $\sim -1.25$, resulting in $F_{\nu,pk} \sim 390 - 1930$~Jy for ISM-like CBM ($F_{\nu,pk} \sim 50 - 390$~Jy for wind-like CBM); this behaviour seems to be compatible with the predicted Y-band light curve.
\end{itemize}
The radio peak clearly observed in the $1.7$~d SED at the same frequency (Fig.~\ref{fig:radio sed_multi}, top right) is incompatible with the temporal evolution of $\nu_{c,rs}$ for RS emission because, considering the observed peak at $\sim 9$~GHz in the $0.8$~d-radio SED, at $1.7$~d we would observe $\nu_{c,rs} \sim 3$~GHz in ISM-like CBM ($\nu_{c,rs} \sim 2$~GHz in wind-like CBM);
this suggests that the RS is unlikely to play a dominant role in radio data of GRB\,160131A.

Other possible explanations for the radio spectral bumps could be (1) the two-component jet, one in which the optical/X-ray emission arises from a narrower, faster jet than that producing the radio observations (e.i. \citealt{Peng05,Racusin08,Holland12}), or (2) the presence of a population of thermal electrons, not accelerated by the FS passage into a relativistic power-law distribution \citep{Eichler05}, characterised by Lorentz factor much lower than the minimum Lorentz factor of the shock-accelerated electrons (``cold electron model'', \citealt{Ressler17}).

\section{Conclusions}
\label{cap_GRB160131A:concls}

We presented our results on the broadband modelling of the afterglow of GRB\,160131A, whose observations span from $\sim 330$~s to $\sim 160$~d post explosion at 26 frequencies from $6 \times 10^8$~Hz to $7 \times 10^{17}$~Hz.

In the data modelling we considered a jetted (edge-regime) FS emission with energy injection, ISS effect, dust extinction and absorption effects, in ISM-like CBM.
Our results on the UVOIR/X-ray data alone show the following results: $p \sim 2.2$, $\epsilon_e \sim 0.01$, $\epsilon_B \sim 0.1$, $n_0 \gtrsim 10$~cm$^{-3}$, $E_{K,iso} \gtrsim 5 \times 10^{53}$~erg, $A_V \sim 0.2$~mag, and $t_j \sim 0.9$~d.
The constrain on $t_j$ leads to an estimate of the jet half opening angle of $\theta_j \sim 6^{\circ}$, corresponding to a beaming-corrected kinetic energy of the explosion $E_K = E_{K,iso} (1 - \cos{\theta_j}) \gtrsim 3 \times 10^{51}$~erg, in agreement with the typical values of long GRBs (Figs.~21 and 22 of \citealt{Laskar15}).
The spectrum is in fast cooling until $\sim 0.02$~d, the non-relativistic regime sets in at $\sim 100$~d, and the energy injection is characterised by $m \sim 0.15$.
The radio data set -- when it is as rich as in this case -- show the presence of spectral bumps in several SEDs, incompatible with a simple standard GRB afterglow model and probably ascribable with either ISS (or other extreme scattering effects) or a more complex multi-component structure.
This incompatibility is corroborated by the broadband modelling from radio to high energies, where the model works well at radio domain (except for $\nu \lesssim 10$~GHz), partially well at X-ray frequencies, and poorly in the optical band.
These results challenge the standard GRB afterglow model, and highlight the key role and as-yet poorly understood physics that manifests itself especially when a rich data set (from radio to high-energy domain) -- as in the case of GRB\,160131A -- is included in the modelling.

Future broadband followup of GRB afterglows, particularly at radio frequencies with the latest and forthcoming generation facilities -- especially in interferometric mode -- such as the Very Large Baseline Array (VLBA\footnote{\url{https://science.nrao.edu/facilities/vlba}}), LOw Frequency ARray (LOFAR, \citealt{LOFAR13}) or the next generation Square Kilometer Array (SKA, e.i. \citealt{ASKAP}), are essential to reach an exhaustive comprehension of the GRB afterglow physics, particularly within the newborn era of multi-messenger astronomy.

\begin{acknowledgements}

We thank the anonymous referee for helping us improve the paper.
Support for this work was provided by Universit\`a degli Studi di Ferrara through grant FIR~2018 ``A Broad-band study of Cosmic Gamma-Ray Burst Prompt and Afterglow Emission" (PI Guidorzi).
M.~Marongiu gratefully acknowledges the University of Ferrara for the financial support of his PhD scholarship.
M.~Marongiu is very grateful to R.~Martone for useful conversations about GRB science; moreover, M.~Marongiu thanks P.~Bergamini and G.~Angora for the useful discussion about Python programming language and data analysis.
G.~Stratta acknowledges support from PRIN-MIUR 2017 (grant 20179ZF5KS).
A.~Gomboc acknowledges the financial support from the Slovenian Research Agency (grants P1-0031, I0-0033, J1-8136, J1-2460) and networking support by the COST Actions CA16104 GWverse and CA16214 PHAROS.
N.~Jordana and C.G.~Mundell acknowledge financial support from Mr Jim Sherwin and Mrs Hiroko Sherwin.
D.~Kopac acknowledges the financial support from the Slovenian Research Agency (research core funding No. P1-0188).
The National Radio Astronomy Observatory is a facility of the National Science Foundation operated under cooperative agreement by Associated Universities, Inc..

This is a pre-print of an article accepted for publication in A\&A.
The final authenticated version will be available online.

\end{acknowledgements}

\section*{ORCID iDs}

M.~Marongiu: \url{https://orcid.org/0000-0002-5817-4009} \\
C.~Guidorzi: \url{https://orcid.org/0000-0001-6869-0835} \\
G.~Stratta: \url{https://orcid.org/0000-0003-1055-7980} \\
A.~Gomboc: \url{https://orcid.org/0000-0002-0908-914X} \\
N.~Jordana-Mitjans: \url{https://orcid.org/0000-0002-5467-8277} \\
S.~Dichiara: \url{https://orcid.org/0000-0001-6849-1270} \\
S.~Kobayashi: \url{https://orcid.org/0000-0001-7946-4200} \\
D.~Kopa{\v c}: \url{https://orcid.org/0000-0001-8099-230X} \\
C.~G.~Mundell: \url{https://orcid.org/0000-0003-2809-8743} \\

\bibliographystyle{aa}            
\bibliography{alles_grbs}         


\end{document}